\documentclass[a4paper,12pt]{article}

\usepackage{epsfig,graphicx,amsmath,amssymb}
\usepackage{url}
\usepackage{subcaption}
\usepackage{picture}
\usepackage[percent]{overpic}
\usepackage{siunitx}
\usepackage{units}
\usepackage{wrapfig}
\usepackage{subfloat}
\usepackage{rotating}
\usepackage{xcolor}
\newcommand{\ee}{e^{+} e^{-}}

\newcommand{\epm}{e^+e^-}

\def \ee {e^+e^-}

\newcommand{\pip}{\pi^{+}}

\newcommand{\piz}{\pi^{0}}

\def \Kl {\mathrm{K_L}}
\def \Ks {\mathrm{K_S}}
\newcommand{\sign}{dd \to ^{4}He\pi^0} 
\newcommand{\bg}{dd \to ^{3}Hen\pi^0 }
\def \ppi {$p\pi^0$}
\def \ppipi {$p\pi^0\pi^0$}
\def \peta {$p\eta$}
\def \pw {$p\omega$}
\def\ifm#1{\relax\ifmmode#1\else$#1$\fi} 
 \def\epm{\ifm{e^+e^-}}
\def \gammap{$\gamma p$}
\def \gammapa{\gammap$\rightarrow$}
\def \prodppi {\gammapa\ppi}
\def \prodpeta {\gammapa\peta}

\def \prodppipi{\gammapa\ppipi}
\newcommand{\epem}{e$^+$e$^-$}

\newcommand{\Del}{$\Delta$}

\def\gam{\ifm{\gamma}}
\def\mt{\ifm{M_{\rm trk}}}
 \def\pic{\ifm{\pi^+\pi^-}} 
 \def\po{\ifm{\pi^0}}
         \def\dif{\hbox{d}}

\usepackage{braket}
\newcommand{\kaon}{K^0}
\newcommand{\akaon}{\bar{K}^0}

\newcommand{\Km}{K_{-}}
\newcommand{\Ts}{$\mathcal{T}$}
\newcommand{\CPs}{$\mathcal{CP}$}

\newcommand{\etapiee}{\eta \rightarrow \pi^0 \mbox{ e}^+ \mbox{ e}^-}
\newcommand{\Etapiee}{\eta \rightarrow \pi^0 \mbox{ e}^+ \mbox{ e}^-}

\begin{document}

\thispagestyle{empty}

$\phantom{.}$

\begin{flushright}
{\sf  MITP/14-089 \\
MCTP-14-08
  } 
\end{flushright}

\hfill

\begin{center}
{\Large {\bf MesonNet 2014 International Workshop\\
Mini-proceedings} \\
\vspace{0.75cm}}

\vspace{1cm}

{\large 29$^{\mathrm{th}}$ September - 1$^{\mathrm{st}}$ October, Laboratori Nazionali di Frascati, Italy}

\vspace{2cm}

{\it Editors}\\
Simona Giovannella (Frascati), Andrzej Kupsc (Uppsala), and Pere Masjuan (Mainz)
\vspace{2.5cm}

ABSTRACT

\end{center}

\vspace{0.3cm}

\noindent
The MesonNet International Workshop was held in the Laboratori
Nazionali di Frascati from September the 29$^{\mathrm{th}}$ to October
the 1$^{\mathrm{st}}$, 2014, being the concluding meeting of the
MesonNet research network within EU HadronPhysics3 project. MesonNet
is a research network focused on light meson physics gathering
experimentalist and theoreticians from Europe and abroad.  An overview
of the research projects related to the scope of the network is
presented in these mini-proceedings.

\medskip\noindent
The web page of the conference:
\begin{center}
\url{https://agenda.infn.it/conferenceDisplay.py?confId=8209} 
\end{center}
\noindent
contains the presentations.
\vspace{0.5cm}

\noindent
We acknowledge the support of the EU HadronPhysics3 project and thank INFN - Laboratori Nazionali di Frascati for its hospitality.

\vspace{5mm}
\noindent
This work is a part of the activity of the MesonNet:
\begin{center}
[\url{https://sites.google.com/site/mesonnetwork/}]
\end{center}

\newpage

{$\phantom{=}$}

\vspace{0.5cm}

\tableofcontents

\newpage

\section{Introduction to the Workshop}

\addtocontents{toc}{\hspace{1cm}{\sl S.~Giovannella, A.~Kupsc, and P.~Masjuan}\par}

\vspace{5mm}

\noindent
{\sl S.~Giovannella$^1$, A.~Kupsc$^2$, and P.~Masjuan$^3$}

\vspace{5mm}

\noindent
$^1$Laboratori Nazionali di Frascati dell'INFN, Italy\\
$^2$Department of Physics and Astronomy, Uppsala University, Sweden\\
$^3$PRISMA Cluster of Excellence, Institut f\"ur Kernphysik, Johannes Gutenberg-Universt\"at, Mainz D-55099, Germany\\ 

\vspace{5mm}

MesonNet is a research network within EU HadronPhysics3 project (2012
-- 2014) and is a continuation and an extension of the PrimeNet
network which was active 2009 -- 2011
\cite{Hoistadt:2011iv,Adlarson:2012bi}. The main objective is the
coordination of light meson studies at different European accelerator
research facilities: COSY (J\"ulich), DA$\Phi$NE (Frascati), ELSA
(Bonn), GSI (Darmstadt) and MAMI (Mainz). The network includes also EU
researchers carrying out experiments at VEPP-2000 (BINP), CEBAF (JLAB)
and heavy flavor-factories (Babar, Belle II, BESIII experiments).  The
scope of the studies there are processes involving lightest neutral
mesons: $\pi^0$, $\eta$, $\omega$, $\eta'$, $\phi$ and the lightest
scalar resonances. The emphasis is on meson decay studies but we
include also meson production processes and meson baryon interactions.
The majority of the participants of the network are experimentalists
while close collaboration with theory groups is essential for the
planning of experiments and the interpretation of the data.

Three main workshops were organized by MesonNet: a workshop on Meson
Transition Form Factors in Cracow, Poland, May
2012~\cite{Czerwinski:2012ry1}, an international workshop halfway of
the project, in Prague, June 2013~\cite{Amaryan:2013eja1} and the
present workshop, in Frascati, end of September 2014.

We have organized this concluding meeting in order to review and summarize all the topics studied during the project course: light meson dynamics and decays, meson baryon interactions, and the studies related to fundamental particle physics problems such as rare decays, the anomalous magnetic moment of the muon, and searches for physics beyond the standard model.

The network has been successful in identifying research topics in the light meson physics  for which  a close
collaboration  between  experiment  and  theory is mandatory to achieve a significant
progress.   The  first  project  is  related to  the  studies of
$\eta$ and $\eta'$ hadronic decays which are sensitive tools for investigations of $\pi\pi$ and $\pi\eta$ interactions, symmetry breaking, and serve as a test of effective field theories.
Of particular interest and focus in the meeting were
the
isospin-violating $\eta$ meson decays into three pions occurring due to light  quark  mass  difference  $m_d-m_u$.   They could 
 provide  one of the best constraints  for the light quark
mass  ratios  and sensitive tests  of the three flavor chiral perturbation theory  convergence. The intermediate goal is  to resolve issues in the experimental
and theoretical description of  the decays. The meeting included a special session devoted to these topics. The session  was concluded with a discussion on different dispersive approaches and the use of the experimental Dalitz plot information including systematic uncertainties for the fits of the subtraction constant and the extraction of the light quark-mass difference. 

   The  second  joined   project aims  at  a  systematic determination of the transition form factors of the $\pi^0$ and $\eta$ mesons  with focus  on the  cases involving  two virtual  photons.  The
knowledge of the form factors is an important input for the calculations of the
Standard  Model contributions to the anomalous magnetic moment of the muon $(g-2)_{\mu}$  and to rare
$\pi^0$  and $\eta$ decays  into a  lepton-antilepton pair.   The muon
$g-2$  and  the  branching  ratio  for  $\pi^0\to  e^+e^-$  decay  are
currently among the few observables where hints of  a deviation  from the
Standard Model  predictions are reported. There were several presentations related to the meson 
transition form factors during the meeting concluded  by a  discussion session focused  
on role of the new precise experimental data and new theory approaches to the hadronic light-by-light
contribution to  $(g-2)_{\mu}$. This project is one of the main topics in the proposal for a new EU
network
{\it Hadron Precision Physics} -- a common initiative with {\it Working Group on Radiative
Corrections and MC Generators for Low Energies}\footnote{\url{http://www.lnf.infn.it/wg/sighad/}} \cite{Actis:2010gg1}.

The detailed program, which consisted of 31 talks and 14 posters, was arranged by a program committee having the members: 
Reinhard Beck, Johan Bijnens, Simon Eydelman, Ingo Froehlich, Simona Giovannella, Dieter Grzonka, Christoph Hanhart, Volker Hejny, Bo H\"oistad, Tord Johansson, Karol Kampf, Bernd Krusche, Bastian Kubis, Andrzej Kupsc, Stefan Leupold, Pere Masjuan, Pawel Moskal, Michael Ostrick, Teresa Pe\~na, Piotr Salabura, Susan Schadmand.

The workshop was held from the 29$^{\mathrm{th}}$ of September to the
1$^{\mathrm{st}}$ of October, at the Laboratori Nazionali di Frascati
dell'INFN, Italy and had 62 participants.

Webpage of the conference is 
\begin{center}
\url{https://agenda.infn.it/conferenceDisplay.py?confId=8209} 
\end{center}
\noindent
where detailed program and talks can be found.

Financial support is gratefully acknowledged from the European Union
Seventh Framework Capacities Programme FP7/2007-2013 HadronPhysics3 project grant
agreement n$^\circ$ 283286.

\begin{figure*}[h!]
\begin{center}
\begin{subfigure}[b]{0.6\textwidth}
\centering
\includegraphics[width=5cm]{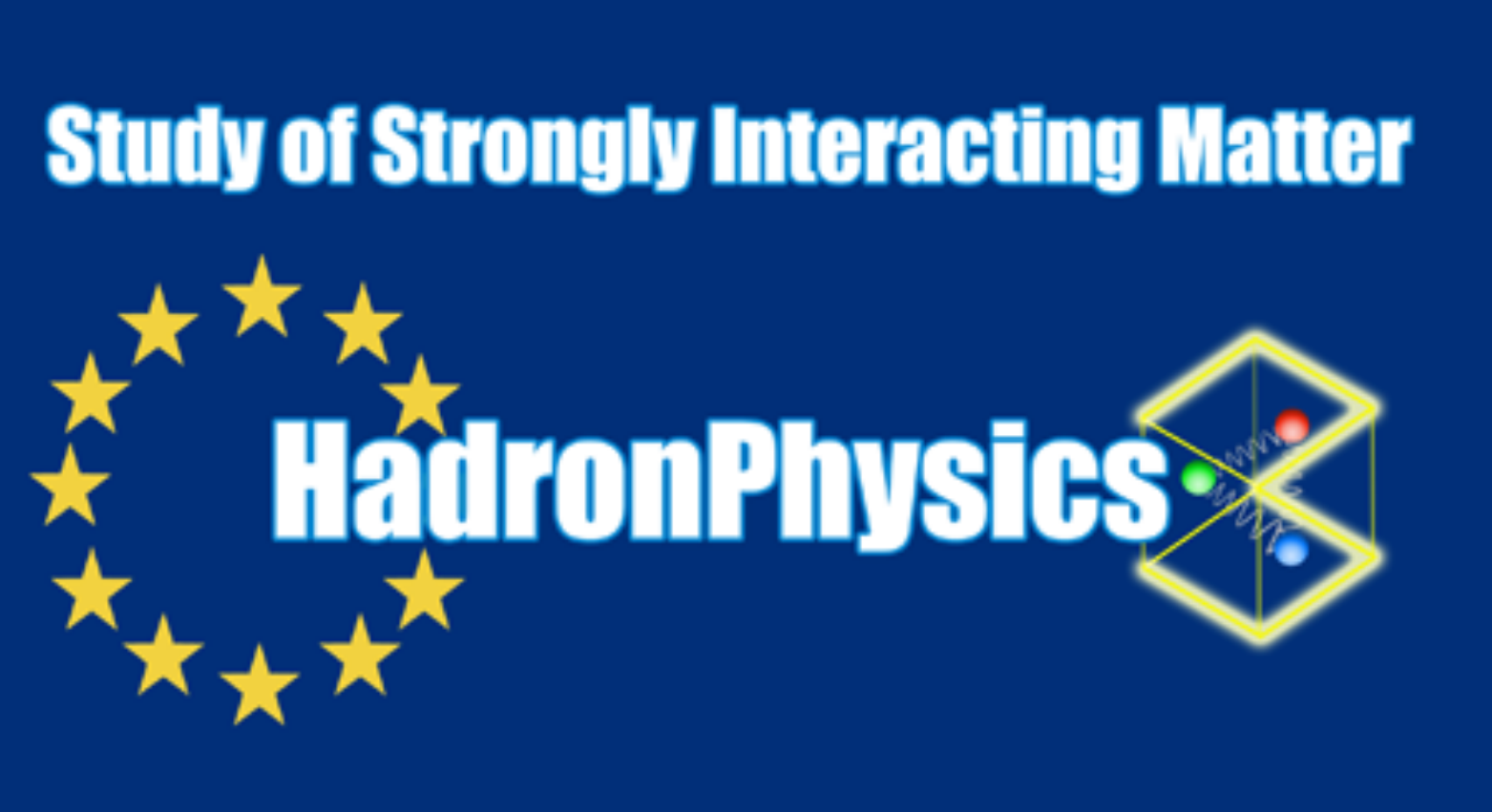}
\end{subfigure}
\begin{subfigure}[b]{0.5\textwidth}
\end{subfigure}
\label{default}
\end{center}
\end{figure*}

\newpage

\section{Summaries of the talks}

\subsection{Chiral Perturbation Theory and $\eta\to3\pi$: an Introduction}
\addtocontents{toc}{\hspace{2cm}{\sl J.~Bijnens}\par}

\setcounter{figure}{0}
\vspace{5mm}

{\sl J.~Bijnens}

\vspace{5mm}

\noindent
Department of Astronomy and Theoretical Physics, Lund University, Lund, Sweden\\
\vspace{5mm}

In the talk I first gave a short introduction to Chiral Perturbation
Theory (ChPT).
More extensive lectures can be found linked from the webpage \cite{chptwebpage}.
One of the problems in using effective field theories like ChPT is the rapidly
increasing number of parameters, usually called Low-Energy Constants (LECs),
when going to higher orders. The second part of the talk was discussing the
recent update and review of the LECs of three flavour ChPT
\cite{Bijnens:2014lea1}. The main problem is the number of LECs
versus the number of observables. The original paper by Gasser and Leutwyler
introducing three flavour ChPT \cite{Gasser:1984gg} needed to use large $N_c$
arguments to obtain all next-to-leading order (NLO) LECs, the $L_i^r$.
More observables were later included and in addition calculated to
NNLO, $p^6$ or two-loop order. This lead to first fit of the $L_i^r$
at NNLO \cite{Amoros:2000mc,Amoros:2001cp1} with a simple estimate of the
NNLO LECs, the $C_i^r$, and large $N_c$ input. When many more needed NNLO
calculations were performed, see the review \cite{Bijnens:2006zp},
a new study was done using the additional observables but the same estimates
of the $C_i^r$ \cite{Bijnens:2011tb}. Surprisingly, large $N_c$ relations were
not too well satisfied. On the other hand, ChPT was tested by finding relations
independent of the $C_i^r$ and found to work reasonably
well \cite{Bijnens:2009zd}. This puzzle made us try to include much more
information on the $C_i^r$ and redo the fit essentially with the same
observables as in \cite{Bijnens:2011tb}. The main conclusions of
\cite{Bijnens:2014lea} are that there exist reasonable choices of the
$C_i^r$ that give a good fit to all included observables and a reasonable
convergence. Imposing the large $N_c$ relation on $L_4^r$ causes
the large $N_c$ relations for $L_1^r,L_2^r,L_6^r$ to be obeyed as well.
The fit BE14 in \cite{Bijnens:2014lea1} is now the standard
fit for the $L_i^r$ from the continuum, but should be used together with the associated values of the $C_i^r$.

\begin{figure}[t!]
\begin{center}
\includegraphics[width=0.8\textwidth,bb= 92 692 436 766]{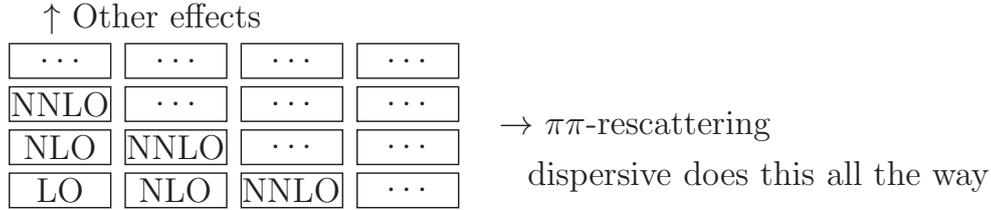}
\end{center}
\caption{A schematic view of the corrections to $\eta\to3\pi$ from
$\pi\pi$ rescattering and other effects. The labels indicate what is included at each order of ChPT.}
\label{fig1}
\end{figure}

The third part of the talk was an introduction to the ChPT results
for $\eta\to3\pi$ and an update of the results using the fit BE14 discussed
above \cite{Bijnens:2014lea1}. An introduction can be found in
\cite{Bijnens:2002qy}.
One main reason to study $\eta\to3\pi$ is that it proceeds via
isospin violation. Since the contributions from electromagnetism
are small \cite{Baur:1995gc1,Ditsche:2008cq1} this decay gives good access
to $m_u-m_d$ via $Q^2=(m_s^2-\hat m^2)/(m_d^2-m_u^2)$ or
$R=(m_s-\hat m)/(m_d-m_u)$. The LO calculation was done using current algebra,
the NLO using ChPT in \cite{Gasser:1984pr1} and NNLO
more recently in~\cite{Bijnens:2007pr1}. Note the twenty year timescale for
adding an order. One problem is that the two-loop calculation
of \cite{Bijnens:2007pr1} does not agree well with the observed
Dalitz plot parameters. The experimental situation is reviewed in the next talk
\cite{Li}. Using instead the new fit of the $L_i^r$ (BE14) as input,
the conflict remains. The new ChPT values are (preliminary!)
$a=-1.356$, $b=0.430$ and $d=0.063$.
 This is a problem since ChPT is needed to normalize the $\eta\to3\pi$
amplitude. There are two dispersive analyses of $\eta\to3\pi$
in progress. These were discussed by Knecht \cite{Knecht} and Passemar \cite{Passemar1}. The former is
partly published \cite{Kampf:2011wr1}. More details can be found in their talks.
The main reason is that, as was shown already in \cite{Gasser:1984pr1} and
references therein, $\pi\pi$- rescattering is important.
Schematically, we could split the amplitude corrections in $\pi\pi$ and others
as shown in Fig.~\ref{fig1}. Now, even though the total corrections
in ChPT are large, if we can remove the $\pi\pi$-rescattering from the NNLO
ChPT result and then put the rescattering effects back in to all orders
we should get both a better converging ChPT part and a better description
of the decay, thus leading to a much more precise value of $m_d-m_u$.
The separation is however nontrivial and should be done in cooperation with
the groups doing the dispersive analysis.
Another topic that deserves mention is the relation between the neutral
and charged Dalitz plot parameters derived in \cite{Bijnens:2007pr1} and
further studied in \cite{Schneider:2010hs1}.

\newpage

\subsection{Dispersive construction of the two-loop amplitude for $\eta \to \pi^+\pi^-\pi^0$}
\addtocontents{toc}{\hspace{2cm}{\sl M.~Knecht}\par}

\vspace{5mm}

{\sl M.~Knecht}

\vspace{5mm}

\noindent
Centre de Physique Th\'eorique, Marseille, France \\

\vspace{5mm}

The decays $\eta \to \pi^+\pi^-\pi^0$ and $\eta \to \pi^0\pi^0\pi^0$ are $\Delta I =1$
transitions, and thus require isospin breaking. The latter is provided by two sources,
electromagnetic interactions on the one hand, and the quark mass difference $m_d - m_u$
on the other hand. 
As far as the former are concerned, contributions of the order ${\cal O}(e^2 E^0)$
vanish \cite{Sutherland:1966zz,Bell:1996mi}, and corrections of the order
${\cal O}(e^2 m_q)$, $q=u,d,s$, were found to be quite small \cite{Baur:1995gc,Ditsche:2008cq}.
Thus, to a very good approximation, the amplitudes for the decays of the $\eta$ meson
into three pions are proportional to the isospin-breaking quark mass difference $m_d - m_u$,
e.g. $A^{\eta\to \pi^+\pi^-\pi^0} (s,t,u) = (\sqrt{3}/4 R) f (s,t,u)$, 
with $R\equiv (m_s - m_{ud})/(m_d - m_u)$.
Measuring the corresponding decay rates gives thus directly  
information on $m_d - m_u$,
provided one knows $f (s,t,u)$ sufficiently well.

The amplitude $f (s,t,u)$ has been computed in the chiral expansion, at
orders ${\cal O} (E^2)$ \cite{Cronin:1967jq,Bell:1996mi}, ${\cal O} (E^4)$ \cite{Gasser:1984pr}, 
and ${\cal O} (E^6)$ \cite{Bijnens:2007pr,Bijnens}. The convergence is however slow,
due to strong $\pi\pi$ rescattering effects. Furthermore, the two-loop expression
involves many unknown ${\cal O} (E^6)$ low-energy constants (LEC), and the very long complete analytic 
expression has not been published. It is available as a fortran code from the authors.
Other approaches have therefore been considered in order to improve the situation. For instance, 
the iterative resummation of the $\pi\pi$ rescattering effects can be handled numerically 
in a dispersive framework \cite{Kambor:1995yc,Anisovich:1996tx,Passemar}. Or a more compact
explicit representations of $f (s,t,u)$ at NNLO can also be worked out in a non-relativistic
framework \cite{Schneider:2010hs}.

In Ref. \cite{Kampf2011}, an analytic two-loop representation of $f (s,t,u)$ has
been constructed using general properties, like analyticity, unitarity, crossing, and
chiral counting. The method is a direct adaptation of the one discussed in \cite{Stern:1993rg}
and used to construct the two-loop amplitude of $\pi\pi$ scattering in \cite{Knecht:1995tr}.
This construction reproduces the structure of the two-loop amplitude \cite{Bijnens:2007pr},
and gives a valid description of $f (s,t,u)$ in the physical region
whenever contributions of orders ${\cal O} (E^8)$ and higher are small, and if the contributions 
from intermediate states other than two pions can be appropriately described by a polynomial.
This last requirement could in principle be dispensed with by including also other
two-meson intermediate states in the construction. The expression of the amplitude obtained this way 
involves, in particular, 6 parameters that can be related to
the ${\cal O}(E^6)$ LECs of the two-loop amplitude \cite{Bijnens:2007pr}.
They can also be determined by a fit to the experimental Dalitz plot distribution. 
Doing so with the data obtained by KLOE \cite{Ambrosino:2008ht}, we found
that the resonance saturation estimate of these LECs used in \cite{Bijnens:2007pr}
does not provide a good description of the data. The situation in this respect looks somewhat better
if instead one uses the data recently released by WASA at COSY \cite{Adlarson:2014aks}.

Finally, in order to extract $R$ from the comparison with the experimental decay rate,
one also needs to know the normalization of 
the amplitude. This missing piece of information can in principle be provided by 
the two-loop amplitude of \cite{Bijnens:2007pr}. In order to avoid the dependence with respect
to the poorly known ${\cal O} (E^6)$ LECs, one may use only the imaginary part of this amplitude.
Furthermore, in order to have a reasonable certitude that higher order effects are under control,
we have looked for a region in the Dalitz plane where the corrections to
the imaginary part are not too important when going from one loop to two loops.
Such a region has been identified \cite{Kampf2011} in the vicinity of the $t=u$ line, for values
of $s$ slightly below the physical region of the decay. A fit to the two-loop amplitude
of \cite{Bijnens:2007pr} in this region and to the KLOE results \cite{Ambrosino:2008ht} 
eventually gives the determination \cite{Kampf2011} $R=37.7(2.2)$.
This value compares well, for instance, with the results \cite{Aoki:2013ldr1} from lattice QCD.

It would be interesting to see how this analysis will be affected by the newer data of 
WASA at COSY \cite{Adlarson:2014aks,Caldeira} and of KLOE \cite{Caldeira}.
A substantial improvement in precision on the determination of $R$ is also to be expected
from the planed high-statistics experiment GlueX at JLab \cite{Somov}.

\newpage

\subsection{Constraints on QCD order parameters from $\eta\to 3\pi$ decays}
\addtocontents{toc}{\hspace{2cm}{\sl M.~Koles\'ar}\par}

\vspace{5mm}

{\sl M.~Koles\'ar, J.~Novotn\'y}

\vspace{5mm}

\noindent
Institute of Particle and Nuclear Physics, Charles University, Prague\\


The $\eta$$\,\to\,$$3\pi$ decays are a valuable source of information on low energy QCD. Yet they were not used for an extraction of the 3-flavor chiral symmetry breaking order parameters until now.  We use a Bayesian approach in the framework of resummed chiral perturbation theory \cite{DescotesGenon:2003cg} to obtain constraints on the quark condensate and pseudoscalar decay constant in the chiral limit, as well as the mass difference of light quarks.

\begin{table}[b] \small
\begin{center}
\begin{tabular}{|c|c|c|c|}
	\hline \rule[-0.2cm]{0cm}{0.5cm} phenomenology & $Z(3)$ & $X(3)$ \\
	\hline \rule[-0.2cm]{0cm}{0.5cm} NNLO $\chi$PT (main fit) \cite{Bijnens:2014lea} & 0.59 & 0.63\\
	\rule[-0.2cm]{0cm}{0.5cm} NNLO $\chi$PT (free fit) \cite{Bijnens:2014lea} & 0.51 & 0.48 \\
	\rule[-0.2cm]{0cm}{0.5cm} NNLO $\chi$PT ("fit 10") \cite{Amoros:2001cp} & 0.89 & 0.66 \\
	\hline \rule[-0.2cm]{0cm}{0.5cm} lattice QCD & $Z(3)$ & $X(3)$ \\
	\hline \rule[-0.2cm]{0cm}{0.5cm} RBC/UKQCD+Re$\chi$PT \cite{Bernard:2012ci} & 0.54$\pm$0.06 & 0.38$\pm$0.05\\
	\rule[-0.2cm]{0cm}{0.5cm} RBC/UKQCD+large $N_c$ \cite{Ecker:2013pba} & 0.91$\pm$0.08 & \\
	\rule[-0.2cm]{0cm}{0.5cm} MILC 09A \cite{Bazavov:2009fk} & 0.72$\pm$0.06 & 0.62$\pm$0.07 \\
	\hline
\end{tabular}
\end{center}
	\caption{Chosen results for the three flavor order parameters.}
	\label{tab2}
\end{table}\normalsize

Our calculation closely follows the general procedure outlined in \cite{Kolesar:2008jr}. Our experimental input are the $\eta$$\,\to\,$$3\pi$ decay widths \cite{Beringer:1900zz1} and the lowest order Dalitz parameter $a$ \cite{KLOE:2008ht}

\[
\Gamma_+ = 300 \pm 12\,\mathrm{eV},\quad 
\Gamma_0 = 428 \pm 17\,\mathrm{eV},\quad
a = -1.09 \pm 0.02.
\]

Leading order low energy constants (LECs) are expressed in terms of convenient free parameters

\[
Z = \frac{F_0^2}{F_{\pi}^2}\, ,\ \
X = -\frac{2\hat{m}\Sigma_0}{F_{\pi}^2M_{\pi}^2}\, ,\ \
r = \frac{m_s}{\hat{m}}\, ,\ \
R = \frac{(m_s-\hat{m})}{(m_d-m_u)},
\]
	
\noindent where $F_0$ is the chiral decay constant, $\Sigma_0$ the chiral condensate and $\hat{m}$=$(m_u+m_d)/2$. We fix $r\,$=\,27.5$\,\pm\,$0.4, a lattice averaging result \cite{Aoki:2013ldr}. For constraints on $X$ and $Z$ we use the value $R\,$=\,37.8$\,\pm\,$3.3 \cite{Kampf:2011wr}. At next-to-leading order, the LECs $L_4$-$L_8$ are algebraically reparametrized using chiral expansions of two point Green functions. For $L_1$-$L_3$ we use the estimate described in \cite{Kolesar:2011wn}. The $O(p^6)$ and higher order LECs, notorious for their abundance, are collected in a relatively smaller number of higher order remainders, treated as a source of statistical uncertainty in the Bayesian framework. For numerical integration, we use Monte Carlo sampling with $10^5$ samples per grid element, the total number of samples being 4$\cdot$$10^6$.

Our results show the $\eta$$\,\to\,$$3\pi$ decays to be sensitive to the values of three flavor chiral order parameters. As can be seen, when assuming $R$\,=\,37.8\,$\pm$\,3.3, there is some tension with available results. The $\eta$$\,\to\,$$3\pi$ data seem to prefer a larger value of the ratio of the chiral order parameters than recent $\chi$PT and lattice fits: $Y$\,=\,$X/Z$\,$\sim$\,1.5. The results also appear to rule out large values of $Z$.

As expected, it's hard to constrain $R$ without information on $X$ and $Z$. Even in this case a significant chunk of the parameter space can be excluded at 2$\sigma$ C.L. When integrating $Z$ out and approximating a normal distribution, we obtain $R$\,=\,37\,$\pm$\,9. A similar procedure when out-integrating $R$ yields $Z$\,=\,0.39$\,\pm\,$0.18.

\begin{figure}[t]
	\epsfig{figure=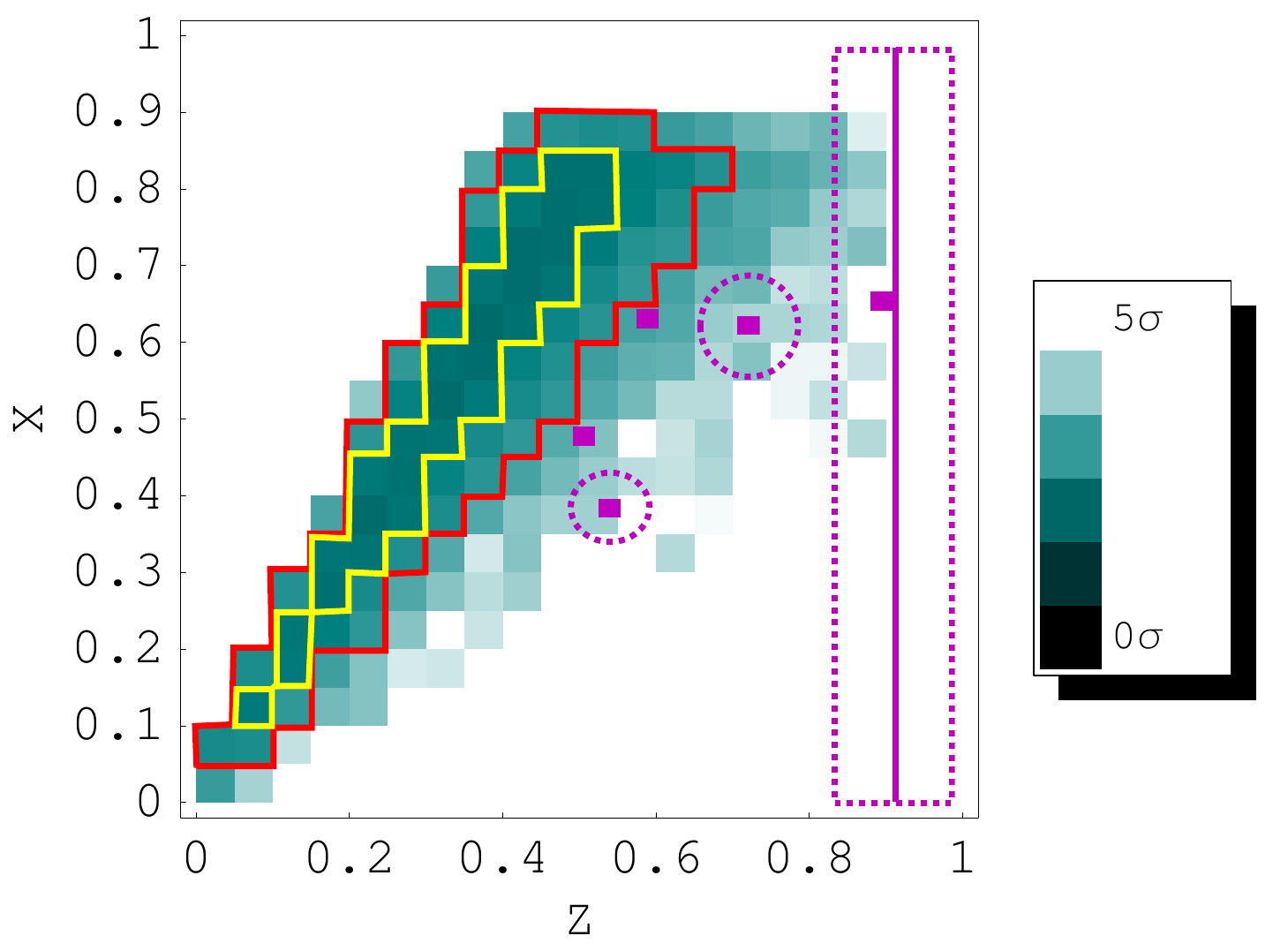,width=0.52\textwidth}
	\epsfig{figure=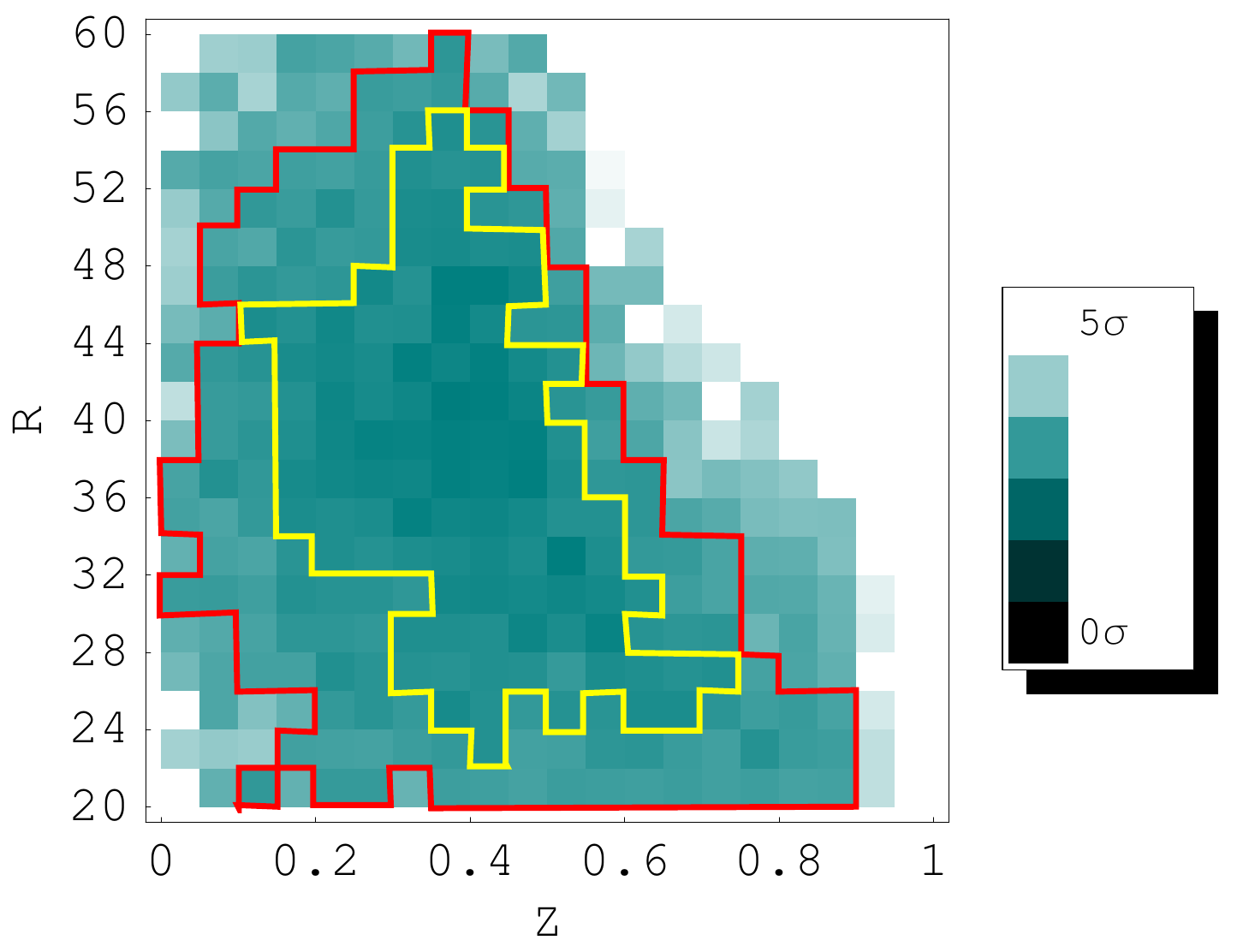,width=0.5\textwidth}	
	\caption{Probability density $P(X,Z|\mathrm{data})$ for $R$\,=\,37.8$\,\pm\,$3.3 (left) and $R$ free (right)\newline
					 highlighted: yellow: 1$\sigma$ C.L. contour, red: 2$\sigma$ C.L. contour, purple: results from table \ref{tab2}}
	\label{fig1:kp}
\end{figure}

\newpage

\subsection{Experimental overview on $\eta \to \pi^+ \pi^- \pi^0$}
\addtocontents{toc}{\hspace{2cm}{\sl L.~Caldeira Balkest\aa{}hl}\par}

\vspace{5mm}

 {\sl L.~Caldeira Balkest\aa{}hl} \\on behalf of the KLOE-2 Collaboration

\vspace{5mm}

\noindent
Department of Physcis and Astronomy, Uppsala University, Sweden\\


Since the $\eta \to 3 \pi$ ( $\eta \to 3 \pi^0$ and  $\eta \to \pi^+ \pi^- \pi^0$) decays break isospin symmetry, they can occur in the Standard Model through the electromagnetic interaction or by the $m_d - m_u$ term of the QCD Langrangian. The electromagnetic contribution to these decays has been shown to be small, by Sutherland and Bell in the context of current algebra \cite{sutherland66, sutherland_bell68} and more recently in the chiral perturbation theory ($\chi$PT) framework at next-to-leading order, including terms up to $\mathcal{O}(e^2)$ \cite{baur_kambor_wyler96} and first order isospin breaking  $\mathcal{O}(e^2 \cdot(m_d - m_u))$ \cite{ditsche_kubis_meissner2009}. So  $\eta \to 3 \pi$ is mainly a strong interaction phenomenum and can be calculated in $\chi$PT.

With the light quark ratio, $Q$, defined as:
\[
 Q^2 \equiv \frac{m_s^2 - \hat{m}^2}{m_d^2 - m_u^2} \qquad \hat{m}= \frac{1}{2}(m_d+m_u),
\]
the decay rate of $\eta \to \pi^+ \pi^- \pi^0$, in leading order of isospin breaking and up to NLO $\chi$PT, is proportional to $Q^{-4}$. A precise value of $Q$ provides a constraint on the light quark masses, \emph{e.g.} neglecting the small term $\hat{m}^2/m_s^2$, the definition of $Q$ forms an ellipse in the $\frac{m_s}{m_d}, \frac{m_u}{m_d}$ plane \cite{leutwyler96}. The strategy is to extract $Q$ by using both accurate theoretical calculations and experimental measurements.

The slow convergence of the $\chi$PT series for the decay rate of  $\eta \to \pi^+ \pi^- \pi^0$, $\Gamma_{LO} =66$ eV, $\Gamma_{NLO} = 160 \pm 50 $ eV \cite{gasser_leutwyler85_eta3pi} and $\Gamma_{NNLO} =  295 \pm 17 $ \cite{bijnens_ghorbani2007}, compared to the experimental value $\Gamma_{exp} = 295 \pm 16 $ eV \cite{PDG12}, shows the need to experimentally verify the theoretical calculations in more detail. The Dalitz plot distribution of the decay can be used for this porpuse.

\begin{table}[hb]
\caption{Theoretical calculations on the Dalitz plot parameters of $\eta \to \pi^+ \pi^- \pi^0$. \label{tab:dpparteor}}

\begin{tabular}{l@{}| l l l l l }
\hline
\textbf{Calculations} & $-a$ & $b$ & $d$ & $f$ & $g$ \\ \hline
$\chi$PT LO\cite{bijnens_ghorbani2007} & 1.039 & 0.27 & 0 & 0& - \\
$\chi$PT NLO\cite{bijnens_ghorbani2007} & 1.371 & 0.452 & 0.053 & 0.027 & -\\
$\chi$PT NNLO\cite{bijnens_ghorbani2007} & 1.271(75) & 0.394(102) & 0.055(57) & 0.025 (160) & -\\
NREFT\cite{schneider_kubis_ditsche2011} & 1.213(14) & 0.308(23) & 0.050(3) & 0.083(19) & -0.039(2)\\
BSE\cite{borasoy_nissler2005} & 1.054(25) & 0.185(15) & 0.079(26) & 0.064(12) & -
\end{tabular}
\end{table}

\begin{table}[htb]
\caption{Experimental results on the Dalitz plot parameters of $\eta \to \pi^+ \pi^- \pi^0$. \label{tab:dpparexp}}

\begin{tabular}{l@{}| l l l l l }
\hline
\textbf{Experiment} & $-a$ & $b$ & $d$ & $f$ \\
\hline
Gormley\cite{gormley70} & 1.17(2) & 0.21(3) & 0.06(4) & - \\
Layter \emph{et al}\cite{layter73} & 1.080(14) & 0.034(27) & 0.05(3) & - \\
CBarrel\cite{CBarrel98} & 1.22(7) & 0.22(11) & 0.06(fixed) & -\\
KLOE\cite{kloe2008} & $1.090(5)(^{+19}_{-8})$ & $0.124(6)(10)$ & $0.057(6)(^{+7}_{-16})$ & 0.14(1)(2)  \\
WASA-at-COSY.\cite{patrik2014} &1.144(18) & 0.219(19)(47)  & 0.086(18)(15) & 0.115(37) \\
KLOE prel.\cite{LiprocMENU2013, LiprocINPC2013} & $1.104(3)$ & $0.144(3)$ & $0.073(3)$ & $0.155(6)$ 
\end{tabular}
\end{table}

The usual Dalitz plot variables for the $\eta \to \pi^+ \pi^- \pi^0$ decay are
\[
X= \sqrt{3} \frac{T_{\pi^+} - T_{\pi^-}}{Q_\eta} \quad \text{ and }\quad Y= \frac{3T_{\pi^0}}{Q_\eta} -1\quad \text{ with }  \quad Q_\eta =T_{\pi^+} +  T_{\pi^-} + T_{\pi^0}
\]
where $T_{\pi^i}$ is the kinetic energy of $\pi^i$ in the $\eta$ rest frame. In these variables, the amplitude squared of the decay can be expanded as a polynomial,  $|A(X,Y)|^2 \simeq N(1+aY+bY^2+cX+dX^2+eXY + fY^3+ gX^2Y + hX Y^2 + lX^3)$ with the coefficients $a,b,\ldots$ referred to as the Dalitz plot parameters. Table \ref{tab:dpparteor} shows the status of theoretical calculations for these parameters, and Table \ref{tab:dpparexp} the experimental results. The two most recent experiments, WASA-at-COSY and KLOE, were presented in this talk. Two theoretical methods are not represented in Table \ref{tab:dpparteor}. These methods aim to improved the next-to-leading order $\chi$PT calculations by taking into account the final state interactions betweens $\pi$'s using dispersion relations \cite{colangelo_lanz_leutwyler_passemar2011, kampf_knecht_novotny_zdrahal2011}. These methods can take as input the experimental Dalitz plot distribution, together with a normalization to $\chi$PT, to extract a value for $Q$.

\newpage

\subsection{ Experimental Overview on $\eta^\prime$ Hadronic Decays }
\addtocontents{toc}{\hspace{2cm}{\sl S.~Fang}\par}

\vspace{5mm}

{\sl S.~Fang} \\for the BESIII collaboration\vspace{5mm}

\noindent
Institute of High Energy Physics, Beijing, China\\
\vspace{5mm}

A half century ago the $\eta^\prime$ was discovered in the bubble experiment.
Since then it has been attracted both theoretical and experimental
attentions due to its special role in understanding low energy Quantum
Chromodynamics (QCD). At present, the study of $\eta^\prime$ decays is still
listed in the physics programs of many experiments, including CLAS, Crystal
Ball, WASA-at-COSY, KLOE-2 and BESIII. In this talk, an experimental overview 
on $\eta^\prime$ hadronic decays is presented.

So far the Dalitz plot of $\eta^\prime\rightarrow \pi^0\pi^0\eta$ was only
performed by the GAMS-4$\pi$ experiment~\cite{gams-4pi-amblik} with a sample of $1.5\times 10^{4}$ events,
while the Dalitz plot parameters of $\eta^\prime\rightarrow \pi^+\pi^-\eta$
were determined by VES~\cite{ves}, CLEO~\cite{cleo} and
BESIII~\cite{bes3-etapppeta}, respectively, using both
generalized and linear representations. A detailed comparison between
experiments and theoretical predictions is given in
Ref.~\cite{bes3-etapppeta} which indicates
that the results are in general consistent with each other, but high precision measurements are also necessary to
test the theoretical predictions in the future.

Concerning the hadronic decays of $\eta^\prime\rightarrow 3\pi$, they are
believed to proceed via  $\eta^\prime\rightarrow\pi\pi\eta$ with the
assumption of $\pi^0-\eta$ mixing. Of great interest is the iso-spin
violation in these processes and the extraction of $m_d-m_u$ in terms of the ratio of the branching fractions of $\eta^\prime\rightarrow 3\pi$ and $\eta^\prime\rightarrow\pi\pi\eta$. The branching
fraction of $\eta^\prime\rightarrow 3\pi^0$ was measured to be 
$(1.6\pm 0.4)\times 10^{-3}$ by the GAMS experiment~\cite{gams-3pi0}, which is then updated to
be $(1.8\pm 0.4)\times 10^{-3}$~\cite{gams-4pi-3pi0}. Recently BESIII experiment reported the
result, $(3.56\pm0.22\pm0.34)\times 10^{-3}$~\cite{bes3-3pi0}, which is measured from $J/\psi$ radiative decays
and almost double of the previous measurements. The $\eta^\prime\rightarrow
\pi^+\pi^-\pi^0$ was first observed by CLEO and the branching fraction was
determined to be $(3.6^{+1.1}_{-0.9}\pm0.4)\times 10^{-3}$~\cite{cleo-3pi}. Subsequently
BESIII confirmed the observation and its branching fraction was
measured to be  $(3.83\pm 0.15\pm0.39)\times 10^{-3}$~\cite{bes3-3pi0}. Due to the low
statistics, only GAMS-4pi experiment reported a pretty rough measurement on
the Dalitz plot of $\eta^\prime \rightarrow 3\pi^0$. The investigation on
the dynamics of $\eta^\prime \rightarrow 3\pi$, in particular the P-wave
contribution in $\eta^\prime\rightarrow\pi^+\pi^-\pi^0$, is very important
to test the theoretical predictions.

Most recently the BESIII experiment reported the observation of
$\eta^\prime\rightarrow \pi^+\pi^-\pi^{+(0)}\pi^{-(0)}$ and the branching
fractions, $B(\eta^\prime\rightarrow\pi^+\pi^-\pi^+\pi^-)$=$(8.63\pm0.69\pm0.64)\times
10^{-5}$ and $B(\eta^\prime\rightarrow\pi^+\pi^-\pi^0\pi^0)$= 
$(1.82\pm 0.35\pm 0.18)\times10^{-4}$~\cite{bes3-4pi}, are consistent with the theoretical predictions
based on the combination of chiral perturbation theory (ChPT) and vector-meson dominance~\cite{guo}.
A search for the forbidden decay of $\eta^\prime\rightarrow 4\pi^0$ was
performed by the GAMS-4$\pi$ experiment and the upper limit at 90\%
confidence level is given to be $3.2\times 10^{-4}$~\cite{gams-4pi-4pi}. 

In addition the hadronic decays, BESIII also made progresses on
the search for $\eta/\eta^\prime$ rare or forbidden
decays~\cite{bes3-progress} via $J/\psi$ radiative or hadronic decays 
based on the sample of 225.3 million $J/\psi$ events. 
At present a sample of 1.3 billion $J/\psi$ events was accumulated at the BESIII detector, 
corresponding to about $6.8\times 10^6$ $\eta^\prime$ in accordance with 
the branching fraction of $J/\psi\rightarrow\gamma\eta^\prime$, which offers
a unique opportunity to update the study of $\eta^\prime$ decays. In summary, with much more data accumulated at the
WASA-at-COSY, Crystal Ball and BESIII detector, the $\eta^\prime$ hadronic
decays could be measured with higher precision, especially the Dalitz
decay parameters, allowing more stringent testing of the
predictions of ChPT.

\newpage

\subsection{$\eta$ Decay  Program at GlueX }
\addtocontents{toc}{\hspace{2cm}{\sl A.~Somov}\par}

\vspace{5mm}

{\sl A.~Somov} \\on behalf of the GlueX/JEF collaborations 

\vspace{5mm}

\noindent
Jefferson Lab, Virginia, USA\\

\vspace{5mm}

The new detector GlueX has been constructed in the experimental 
Hall-D at Jefferson Lab~\cite{gluex}, which will allow to carry out experiments 
using photon beams. Three experiments have already been approved 
in Hall-D: the GlueX experiment to search for gluonic excitations
in the spectra of light mesons, the PrimEx experiment to perform 
a precision measurement of the $\eta \to \gamma \gamma$ decay width 
via the Primakoff effect, and an experiment to measure the charged pion 
polarizability. 
The detector has a large and flat acceptance for both neutral and charged particles 
and allows for good identification of multi-particle final states.

The experimental program has been developed  to study various decays 
of eta mesons with the GlueX detector using a photon beam with energies 
of $E_\gamma = 9 - 12 {\rm GeV}$.  The data sample of $\eta$ 
decays collected with the approved experiments will be used to study the
Primakoff $\eta$ production and many non-rare $\eta$ decays. We propose to upgrade 
the inner part of the GlueX lead glass Forward Calorimeter with  high-granularity, 
high-resolution $PbWO_4$ crystals and perform a dedicated experiment, Jefferson Eta Factory (JEF)~\cite{jef}, 
to study $\eta$ rare decays. The experiment was conditionally approved by the Program 
Advisory Committee at Jefferson Lab in summer 2014. The main physics topics 
currently considered for the $\eta$ program are listed below: 

\begin{itemize}
\item Measurement of the quark mass ratio $Q = (m_s^2 - \hat{m})/(m^2_d - m^2_u)$, where
$\hat{m} = (m_u + m_d)/2$, using $\eta \to 3 \pi$ decays~\cite{Holstein:2001bt,Nefkens:2002sa}. 
The combination of JEF and GlueX running will acquire about 16.5 million reconstructed events for each 
$\eta \to \pi^+ \pi^- \pi^0$ and $\eta \to \pi^0 \pi^0 \pi^0$ decays, which is about a factor of 2.8 larger 
than existing  worlds datasets. Large statistics and the relatively flat acceptance of GlueX will allow 
significant reduction of the statistical error over the Dalitz distribution.
\item Perform measurements of the Dalitz distribution of the $\eta \to \pi^0 \gamma \gamma$ decays. 
The measurements will allow to better understand the contribution of scalar resonances in the calculation of $O(p^6)$ low-energy
constants (LEC) and to determine some LECs in the chiral Lagrangian~\cite{eta_chpt}.
 These low energy constants can be used to cross-check calculations for 
processes such as $K_L \to \pi^0\gamma\gamma \to \pi^0 l^+ l^-$ (CP - conserving background for
 C- and CP- violation searches in  $K_L \to \pi^0 l^+ l^-$)~\cite{Ng:1992yg}. Recently, $\eta \to \pi^0 \gamma \gamma$ 
decays have been measured by several low-energy experiments~\cite{Nefkens:2014zlt,Micco:2005rv,Prakhov:2008zz},
where $\eta$'s were produced with small boost. The large
background from $\eta \to 3 \pi^0$ and $\eta \to 2 \pi^0$ limits the precision on measurement of
$d \Gamma / d M_{\gamma\gamma}$, which is essential to distinguish among various production mechanisms.   
The highly boosted $\eta$'s produced in Hall-D is expected to have significantly smaller background, 
S/B ratio of 3:1. The expected number of reconstructed $\eta \to \pi^0 \gamma \gamma$ events at JEF is $1.4\cdot 10^3$.
\item A search of the leptophobic dark B boson , which couples predominantly to quarks and arises from a new 
$U(1)_B$ symmetry. The $\pi^0\gamma\gamma$ final 
state is ideal to search for the B-boson in the mass range $0.14\; {\rm GeV} \le m_B \le 0.62\; {\rm GeV}$.
The B-boson can be produced in $\eta \to B \gamma$ decay~\cite{Nelson:1989fx1}. The dominant decay channel of the B-boson 
in this energy range is $B \to \pi^0 \gamma$~\cite{Tulin:2014tya1}. The JEF sensitivity to the baryonic fine structure constant 
$\alpha_B$ corresponds to $10^{-7}$, which is about two orders of magnitude better than the existing 
bound. The measurements will indirectly constrain the existence of anomaly-cancelling fermions at the 
TeV-scale. In the $\eta^\prime$ mass range, the B-boson search will be performed using $\eta^\prime \to B \gamma$, 
$B \to \pi^+ \pi^- \pi^0$ decays.
\item A search for C violating and P conserving reactions such as $\eta \to 3\gamma$ (and $\eta \to 2\pi^0 \gamma$) 
decays. JEF is expected to reduce the branching ratio upper limits by 1-1.5 orders of magnitude.
\end{itemize}

\section*{Acknowledgement}
This work was supported by the U.S. Department of Energy.
Jeerson Science Associates, LLC, operates Jeerson Lab for
the U.S. DOE under U.S. DOE contract DE-AC05-060R23177.

\newpage

\subsection{Study of the $\Ks\Kl\to\pi\ell\nu\: 3\pi^0$ process for a direct test of $\mathcal{T}$ symmetry at KLOE-2}
\addtocontents{toc}{\hspace{2cm}{\sl A.~Gajos}\par}

\vspace{5mm}

{\sl A.~Gajos}\\for the KLOE-2 Collaboration

\vspace{5mm}

\noindent
Institute of Physics, Jagiellonian University, Cracow, Poland\\

\vspace{5mm}

A direct \Ts~symmetry test, understood as an observation of a probability asymmetry between two processes connected by an exchange of initial and final states, is only available in few systems which offer reversible transitions, such as neutral meson systems. 

A test is possible with entangled $\kaon$ and $B^0$ mesons by comparing rates of their transitions between flavour and CP-definite states and their time inverses\cite{Bernabeu:2012nu,Bernabeu:2012ab}. Whereas the BaBar experiment already provided a significant \Ts-violation measurement with $B^0$ mesons \cite{Lees:2012uka}, the KLOE-2 detector is uniquely capable of testing the time-reversal symmetry with neutral kaons using this idea.

Neutral kaons are provided to the KLOE detector by the DA$\Phi$NE accelerator in collisions of electrons and positrons at $\sqrt{s}\approx 1020\text{MeV}=m_{\phi}$ thus copiously producing $\phi$ mesons which decay into entangled $\kaon\akaon$ pairs with a branching ratio of 34\%. The KLOE detector records kaon decays with a large ($R=2m$) drift chamber and a sampling electromagnetic calorimeter covering most of the solid angle (98\%) around the $\phi$ decay point and offering good timing resolution. The detector has recently been upgraded to KLOE-2 \cite{Moricciani:2012zza}, equipped with new calorimeters improving acceptance at small angles around the beam \cite{DOMENICI:2014qma} and a cylindrical-GEM inner tracker closely surrounding the interaction region to improve tracking and vertexing \cite{Balla:2013gua}.

Strangeness-definite and \CPs-definite states of neutral kaons can be recognized at \mbox{KLOE-2} by observation of hadronic and semileptonic decays which identify the decaying kaon in a respective basis \cite{Bernabeu:2012nu}. Observation of full transitions between these states takes advantage of quantum entanglement between kaons which makes the first decaying kaon identify the state of its partner to be orthogonal before its decay. Through time-dependent rates of double kaon decays, KLOE-2 can measure the following ratios of transition probabilities:
\begin{equation*}
  {\frac{P[\kaon \to \Km;\Delta t]}{P[\Km \to \kaon;\Delta t]} } \: \sim \: \frac{\mathrm{I}(\pi^+\ell^-\bar{\nu},3\pi^0;\Delta t)}{\mathrm{I}(\pi\pi,\pi^-\ell^+\nu;\Delta t)} \quad \text{and} \quad
  {\frac{P[\akaon \to \Km;\Delta t]}{P[\Km \to \akaon;\Delta t]} } \: \sim \: \frac{\mathrm{I}(\pi^-\ell^+\nu,3\pi^0;\Delta t)}{\mathrm{I}(\pi\pi,\pi^+\ell^-\bar{\nu};\Delta t)},
\end{equation*}
whose deviation from unity in the asymptotic region of $\Delta t \gg \tau_S$ would be a signal of \Ts~symmetry violation. Simulations have shown that with the expected statistics of 10fb\textsuperscript{-1}, KLOE-2 is capable of performing a statistically significant test \cite{Bernabeu:2012nu}.

The time-reversal symmetry test at KLOE-2 requires reconstruction of the double kaon decay events $\phi\!\to\! \Ks\Kl\!\to\! \pi^\pm\ell^\mp\nu\;3\pi^0$ and $\phi\!\to\!\Ks\Kl \!\to\! 2\pi \; \pi^\mp\ell^\pm{\nu}$, of which the $\Kl\to 3\pi^0$ decay demands special treatment as it only contains neutral particles and thus is not recorded by tracking detectors. Therefore its reconstruction must be based solely in the calorimeter information provided by up to 6 recorded photons from 3$\pi^0$ decays.

A dedicated reconstruction procedure for the $\Kl\to 3\pi^0\to 6\gamma$ process was prepared with a view to provide information on the $\Kl$ decay time with a resolution $\mathcal{O}(1\tau_S)$ which is needed by the \Ts-test. The method is based on assigning a sphere constituted by possible $\gamma$ origin points to each of the $\gamma$ hits in the calorimeter and finding the decay vertex location as a common origin point of all photons, i.e. an intersection of the spheres. As their radii are dependent on the $\Kl$ decay time, it is directly found in the reconstruction without the need to be estimated from kaon path length and momentum. Further details on this reconstruction procedure can be found in Ref.~\cite{Gajos:2014pxa}.

Good timing resolution of the KLOE-2 calorimeter and a dedicated kinematic fit following the reconstruction allowed to obtain a resolution of the $\Kl$ decay proper time at a constant level of $\sim 2\tau_S$ independently of the kaon path length. This is a promising result for the future test of time-reversal symmetry at KLOE-2 and studies on the required processes will be continued towards this test.

This work was supported in part by the EU Integrated Infrastructure Initiative Hadron Physics Project under contract number RII3-CT- 2004-506078; by the European Commission under the 7th Framework Program through the `Research Infrastructures' action of the `Capacities' Program, Call: FP7-INFRASTRUCTURES-2008-1, Grant Agreement No. 227431; by the Polish National Science Centre through the Grants No. DEC-2011/03/N/ST2
/02641, 
2011/01/D/ST2/00748,
2011/03/N/ST2/02652,
2013/08/M/ST2/\linebreak[1]00323,

2013/11/B/ST2/04245,
and by the Foundation for Polish Science through the MPD Program and the project HOMING PLUS BIS/2011-4/3.

\newpage

\subsection{Status of $K_S \rightarrow \pi e \nu$ branching ratio and lepton charge asymmetry
measurements with the KLOE detector}
\addtocontents{toc}{\hspace{2cm}{\sl D.~Kami{\'n}ska}\par}

\vspace{5mm}

{\sl D.~Kami{\'n}ska} \\ for the KLOE-2 Collaboration

\vspace{5mm}

\noindent
Institute of Physics, Jagiellonian University, Cracow, Poland\\

\vspace{5mm}

 The physical $K_S$ and $K_L$ mesons are mixtures of states $K^0$ and $\bar{K^0}$		with parameters 
	$\epsilon_K$ and $\delta_K$ accounting for $\mathcal{CP}$ and $\mathcal{CPT}$ violation, respectively.
	This sensitivity of neutral kaon system to discrete symmetries made it one of the best
	candidates for the search of $\mathcal{CPT}$ violation. One of the possible tests is based on studies of 
	charge asymmetry in 	semileptonic	decay ($K\rightarrow \pi e \nu$). 
 The difference between semileptonic charge asymmetry in $K_S$ decays ($A_S$) and the analogous
	asymmetry in $K_L$ decays~($A_L$) is related only~to parameters describing $\mathcal{CPT}$
	violation~\cite{handbook_cp}. 
			
	The charge asymmetries for $K_S$ and $K_L$ kaons were determined by KTeV and KLOE experiments,
	respectively. 			
	A value of 
   $A_{L} = (3.322 \pm 0.058_{stat} \pm 0.047_{syst}) \times 10^{-3}$ 
	was obtained by examination of around	$1.9$ millions $K_{L} \rightarrow \pi e \nu$ decays produced in
	collisions of proton beam with a BeO target~\cite{ktev_kl_charge_asymm}.
	Analogous charge asymmetry for $K_S$ meson was determined with $0.41 \mbox{ fb}^{-1}$ total
 luminosity data sample:  $A_{S} = (1.5 \pm 9.6_{stat} \pm 2.9_{syst}) \times
	10^{-3}$~\cite{kloe_final_semileptonic}. Until now
	both values are the most accurate measurements and they are consistent within error limits which
	suggests
	conservation of $\mathcal{CPT}$ symmetry. However, accuracy on $A_L$ determination is more than two
 orders of magnitude bigger than this of the $A_S$ and the uncertainty on  $A_S$ is dominated by
	the data sample statistics which is three times larger than the systematic contribution.
	Therefore, further studies of $K_S \rightarrow \pi e \nu$ decay using larger statistical samples
	can improve the precision of $\mathcal{CPT}$ test.

	Especially suited for studies of rare semileptonic decays of $K_S$ is the KLOE experiment located at
	the DA$\Phi$NE $\phi$ factory. The KLOE detector consists of two main parts: a drift
	chamber~\cite{drift_chamber} and a	barrel shaped electromagnetic calorimeter~\cite{calorimeter},
	both inserted into electromagnetic field~(0.5~T). Around $60 \%$ of produced $K_L$ mesons reach
	electromagnetic calorimeter and can be identified by the deposited energy and characteristic value of velocity. Moreover, due
	to the pair production of neutral kaons, identification of $K_L$  meson on one side allows to tag a $K_S$
	meson on the other side of $\phi$ meson decay point. Further selection of $K_S \rightarrow \pi e
	\nu$ decays required a vertex with two oppositely charged particles near the Interaction
	Point.	These particles must reach the calorimeter and deposit their energy inside~it in
	order to apply the Time of Flight technique. This technique aims at rejecting background, which
	consists mainly of $K_S \rightarrow \pi^+ \pi^-$ decay, and at identifying
	the	final charged states ($\pi^+ e^- \bar{\nu}$ and $\pi^- e^+ \nu$).
	Based on an integrated luminosity of
	$1.7~\mbox{fb}^{-1}$ around $10^5$ of $K_S \rightarrow \pi e \nu$ events were reconstructed and
	will be used to determine branching ratio and charge asymmetry of $K_S$ semileptonic decays. The analysis is still in progress and preliminary result will be available soon.
 
	Further improvements of both statistical and systematical uncertainties are expected due to the
	installation of new sub-detectors in the KLOE-2 setup~\cite{prospects_kloe} and the luminosity upgrade
	of the DA$\Phi$NE	collider~\cite{dafne_upgrade}.

This work was supported in part by the EU Integrated Infrastructure Initiative Hadron Physics Project under contract number RII3-CT- 2004-506078; by the European Commission under the 7th Framework Program through the `Research Infrastructures' action of the `Capacities' Program, Call: FP7-INFRASTRUCTURES-2008-1, Grant Agreement No. 227431; by the Polish National Science Centre through the Grants No. 
 DEC-2011/03/N/ST2
 /02641, 
 2011/01/D/ST2/00748,
 2011/03/N/ST2/02652,
 2013/08/M/ST2/0 0323,
	DEC-2014/12/S/ST2/00459,
	2013/11/B/ST2/04245
 and by the Foundation for Polish Science through the MPD Program and the project HOMING PLUS BIS/2011-4/3.

\newpage

\subsection{Hadronic light-by-light scattering in the muon $g-2$: a dispersive approach}
\addtocontents{toc}{\hspace{2cm}{\sl M.~Hoferichter}\par}

\vspace{5mm}

{\sl M.~Hoferichter,$^{a,b,c}$ G.~Colangelo,$^c$ M.~Procura,$^c$ and P.~Stoffer$^c$}

\vspace{5mm}

\noindent
$^a$ Institut f\"ur Kernphysik, Technische Universit\"at
Darmstadt, Germany\\
$^b$ ExtreMe Matter Institute EMMI, GSI Helmholtzzentrum f\"ur
Schwerionenforschung GmbH, Germany\\
$^c$ Albert Einstein Center for Fundamental Physics and Institute for Theoretical Physics,
	    Universit\"at Bern, Switzerland\\

\vspace{5mm}

The main uncertainty in the Standard-Model prediction for the anomalous magnetic moment of the muon originates from strong interactions. While hadronic vacuum polarization is intimately related to $e^+e^-\to\text{hadrons}$ via a dispersion integral, a similarly data-driven approach has only recently been suggested for hadronic light-by-light scattering (HLbL) (see~\cite{Kurz:2014wya,Colangelo:2014qya} for even higher-order hadronic contributions). Our framework~\cite{Colangelo:2014dfa,Colangelo:2014pva,Mainz} exploits the analytic structure of the HLbL tensor, concentrating on pseudoscalar poles and two-meson intermediate states, which dominate at low energies.\footnote{A different approach, which aims at a dispersive description of the muon vertex function instead of the HLbL tensor, has
recently been presented in~\cite{Pauk:2014rfa}. An alternative strategy to reduce the model dependence in HLbL is based on lattice QCD~\cite{Blum:2014oka}.} 
Restricting the framework to pions,
the experimental input required for such a program concerns the doubly-virtual pion transition form factor~\cite{Hoferichter:2014vra}
and the partial waves for $\gamma^*\gamma^*\to\pi\pi$~\cite{GM,Hoferichter:2011wk,Moussallam13,Hoferichter:2013ama,Dai:2014zta}, which can again be reconstructed dispersively (see~\cite{Hanhart_eta} for a similar approach to the $\eta$, $\eta'$ transition form factor).

A crucial step in the derivation of our dispersive formalism~\cite{Colangelo:2014dfa} concerns the construction of a suitable basis for the HLbL tensor, in such away that the coefficient functions are free of kinematic singularities and thus amenable to a dispersive representation~\cite{Bardeen:1969aw,Tarrach:1975tu,Leo:1975fb}. In particular, the requirement of the absence of kinematic singularities mandates the presence of certain non-diagonal kernel functions, if the problem is formulated in terms of helicity amplitudes, a phenomenon that already occurs for $\gamma^*\gamma^*\to\pi\pi$ and can be checked there explicitly for the $1$-loop ChPT amplitudes. Moreover, the formalism can only be established rigorously for contributions that do not involve double-spectral regions, i.e.\ simultaneous cuts in two channels. For this reason, the sQED pion loop augmented with pion vector form factors (FsQED), as identified on the level of the Mandelstam representation, is separated and evaluated based on Feynman loop integrals.

In this talk, we presented a first numerical evaluation of $S$-wave $\pi\pi$ intermediate states. First, we find that just with $S$-waves and despite the double-spectral regions the FsQED contribution can be reproduced at the $(5\text{--}10)\%$ level. Second, we included $\pi\pi$ rescattering in the $\gamma^*\gamma^*\to\pi\pi$ partial waves in a simplified formalism involving a Born-term left-hand cut and a finite matching point below the $K\bar K$ threshold. In this setup, we find that the sum of $I=0$ and $I=2$ rescattering contributes $\sim-5\times 10^{-11}$ and, taken together with FsQED, $\sim-20\times 10^{-11}$ to HLbL scattering in the muon $g-2$.

\newpage

\subsection{Dalitz decays of $\pi^{0},\eta$ and $\eta^{\prime}$ mesons through Pad\'{e} approximants}
\addtocontents{toc}{\hspace{2cm}{\sl S.~Gonz\`{a}lez-Sol\'{i}s}\par}

\vspace{5mm}

{\sl S.~Gonz\`{a}lez-Sol\'{i}s}

\vspace{5mm}

\noindent
Institut de F\'{i}sica d'Altes Energies, Universitat Aut\`{o}noma de Barcelona, Catalonia\\
\vspace{5mm}

\begin{figure}[h!]
\begin{minipage}[c]{0.45\linewidth}
\includegraphics[scale=0.5]{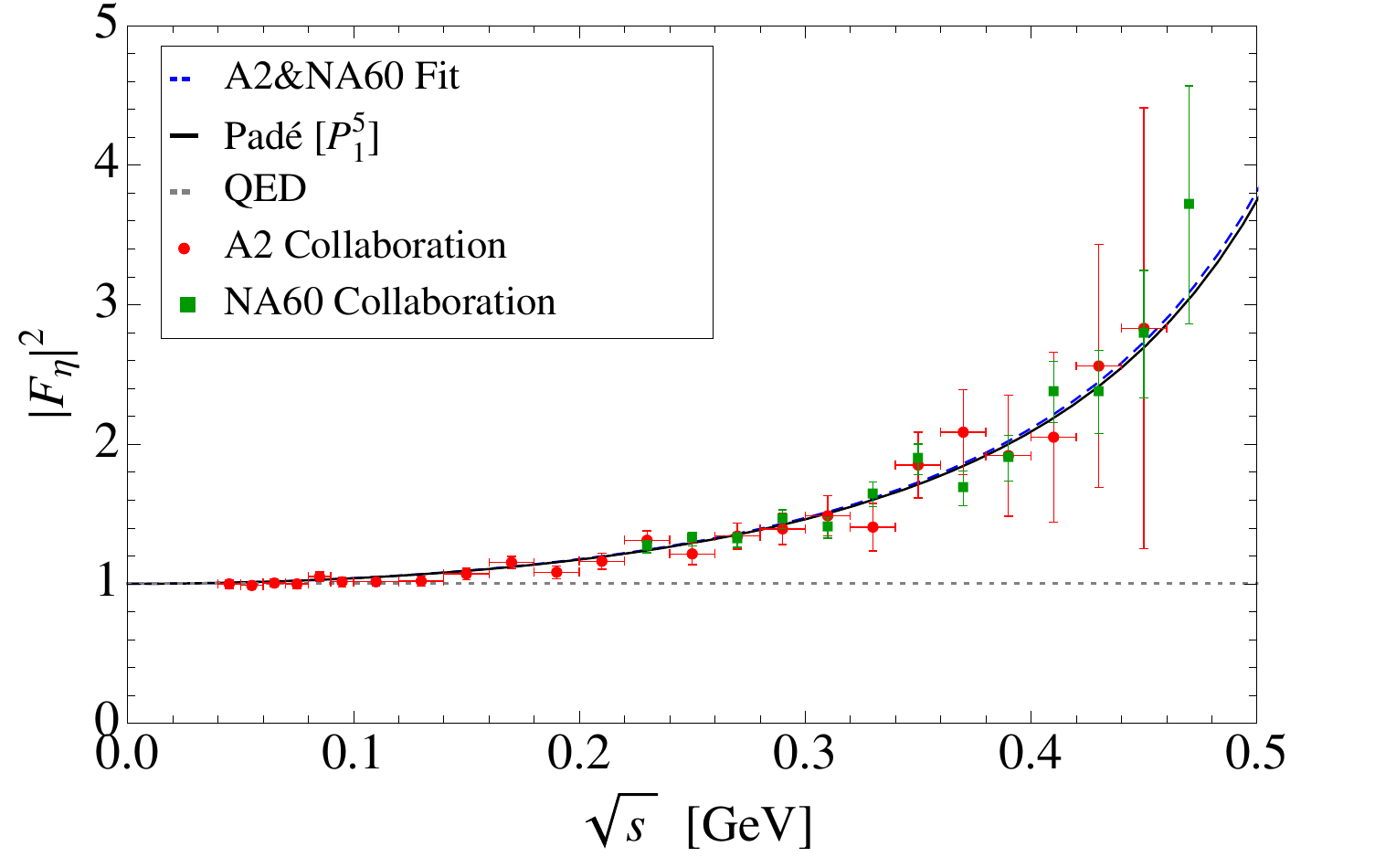}
\caption{{\bf{left}}: $P_{1}^{5}$ description (black curve) of the TFF $\eta\to\gamma^{*}\gamma$ vs. Ref.\cite{TL} and their VMD fits. {\bf{right}}: Our (preliminary) $\mathcal{BR}$ predictions. $P_{1}^{6}$ for $\eta^{\prime}$TFF \cite{Masjuan:2012wy}.}
\label{ffeta}
\end{minipage}
\begin{minipage}{\textwidth}
  \begin{tabular}{l|l|l}
Decay & This work & Experiment \\
    \hline 
$\pi^{0}\to e^{+}e^{-}\gamma$&$1.166(71)\%$&$1.174(35)\%$\\
$\eta\to e^{+}e^{-}\gamma$&$6.60(4)\cdot10^{-3}$&$6.90(40)\cdot10^{-3}$\\
$\eta\to\mu^{+}\mu^{-}\gamma$&$3.24(21)\cdot10^{-4}$&$3.1(4)\cdot10^{-4}$\\
$\eta^{\prime}\to e^{+}e^{-}\gamma$&$4.10\cdot10^{-4}$&$<9\cdot10^{-4}$\\
$\eta^{\prime}\to\mu^{+}\mu^{-}\gamma$&$0.54\cdot10^{-4}$&$1.08(27)\cdot10^{-4}$\\
$\pi^{0}\to e^{+}e^{-}e^{+}e^{-}$&$3.364(14)\cdot10^{-5}$&$3.34(16)\cdot10^{-5}$\\
$\eta\to e^{+}e^{-}e^{+}e^{-}$&$2.70(3)\cdot10^{-5}$&$2.4(2)(1)\cdot10^{-5}$\\
$\eta\to\mu^{+}\mu^{-}\mu^{+}\mu^{-}$&$3.89(30)\cdot10^{-9}$&$<3.6\cdot10^{-4}$\\
$\eta\to e^{+}e^{-}\mu^{+}\mu^{-}$&$2.34(15)\cdot10^{-6}$&$<1.6\cdot10^{-4}$\\
$\eta^{\prime}\to e^{+}e^{-}e^{+}e^{-}$&$1.840\cdot10^{-6}$&unobserved\\
$\eta^{\prime}\to\mu^{+}\mu^{-}\mu^{+}\mu^{-}$&$2.39\cdot10^{-9}$&unobserved\\
$\eta^{\prime}\to e^{+}e^{-}\mu^{+}\mu^{-}$&$3.94\cdot10^{-7}$&unobserved\\
\end{tabular}
\end{minipage}
\end{figure}

The transition form factor (TFF) encodes the effect of the strong dynamics of the anomalous $\mathcal{P}\gamma\gamma^{*}$ ($\mathcal{P}=\pi^{0},\eta^{(\prime)}$) vertices relevant for describing the single and double Dalitz decays $\mathcal{P}\rightarrow\ell^{+}\ell^{-}\gamma$ and $\mathcal{P}\rightarrow\ell^{+}\ell^{-}\ell^{+}\ell^{-}$ $(\ell=e$ or $\mu)$. Vector Meson Dominance (VMD) is the most common way of parameterizing this effect. It occurs after  the spectral (dispersive) representation of the form factor in $Q^{2}$ (with $q^{2}=-Q^{2}$ the momentum of the virtual photon) 
\begin{equation}
F(Q^{2}+i\epsilon)=\frac{1}{\pi}\int\limits_{0}^{\infty}ds\frac{\rho(s)}{s-Q^{2}}
\end{equation}
when one obviates the branch cut starting at the threshold $s=4m^{2}$ and consequently the spectral function reduces to $\rho(s)=\rm{{Im}}F(s)=M_{eff}^{2}\pi\delta(s-M_{eff}^{2})$ where ${\rm{M_{eff}}}^{2}$ is a kind of effective mass which just accounts for the position of the pole. This description, though simple, has successfully been employed for describing lots of phenomena but, sometimes, is not the best choice and something more elaborated is required. This may be the case of the space-like partner TFF $\gamma^{*}\gamma\to\mathcal{P}$ which, benefited from experimental data Ref.\cite{SL}, has been examined by the use of Pad\'{e} approximants (PA) \cite{Baker1}, $P_{N}^{M}(Q^{2})=\frac{\sum_{i=0}^{M}a_{i}(Q^{2})^{i}}{\sum_{j=0}^{N}b_{j}(Q^{2})^{j}}$, in Ref.\cite{Masjuan:2012wy1}. Such a mathematical approach appears to be a much appropriate tool for describing experimental data which in turn is codified within the coefficients of the polynomials. It is our proposal here to extend their findings in the space-like energy regime to describe the transition $\mathcal{P}\to\gamma^{*}\gamma$ in the time-like energy region with the final objective of making accurate branching ratio $(\mathcal{BR})$ predictions of the Dalitz decays we are interested in \cite{SergiRafel}. The use of Pad\'{e} approximants can be viewed as a purely mathematical method which inherently contains the physical information of the TFF but, in doing such extrapolation, one should bear in mind that the whole physical time-like TTF can not be reproduced because this region may contain isolated poles and branch cuts unable to be exactly described within this approach. This deficiency obviously occurs also within the VMD framework which in fact it can be seen as the first element, $P_{1}^{0}(Q^{2})$, of the general sequence of PA. To further improve on the VMD one typically uses a Breit-Wigner representation by resuming self energy insertions, $\Sigma(s)$, in the propagator and considering its imaginary part. In our case, the main contribution to the imaginary part would appear at the first (QCD) threshold $Q^{2}=4M_{\pi}^{2}$ as ${\rm{Im}}\Sigma(s)=M_{\rho^{0}}\Gamma_{\rho^{0}\to\pi^{+}\pi^{-}}(s)$. If one neglects this contribution, the VMD approach is recovered and the TFF is maintained analytic everywhere but on the poles. Strictly, one can not directly associate the mathematical poles of the PA to the physical resonance parameters which in turn can be deduced following the prescriptions of Ref.\cite{Masjuan:2013jha}. These poles lie beyond the available phase space for the Dalitz decays of the $\eta$ \cite{Masjuan:2012wy1}. In Fig.\ref{ffeta} we compare, with nice agreement, our time-like TFF description $\eta\to\gamma^{*}\gamma$ with the ones experimentally determined in $\eta\to e^{+}e^{-}\gamma$ (red points) and $\eta\to\mu^{+}\mu^{-}\gamma$ (green squares) Ref.\cite{TL}. This seems to indicate that the contribution coming from the unconsidered imaginary part is quite negligible. We, moreover, can corroborate this statement by looking the effect of introducing it into the denominator.

Contrarily the large mass of the $\eta^{\prime}$ increases the phase space and our approach would be ill-defined when crossing such singularities.\hspace{-0.1cm} We have considered, in this case, two options: to integrate the mass spectra up to some cut below the poles ($m_{\eta}$ in this work) or to regulate the denominator introducing an imaginary part, again through $\Gamma_{\rho^{0}\to\pi^{+}\pi^{-}}$(s), enabling to integrate up to $m_{\eta^{\prime}}$ \cite{SergiRafel}. For examining the double Dalitz decays, the double off-shell TFF, which depends on both photon virtualities, is required. Our ansatz, due to the lack of experimental information in this case, has been the standard factorization approach $F_{\mathcal{P}\gamma^{*}\gamma^{*}}(q_{\gamma_{1}}^{2},q_{\gamma_{2}}^{2})=F_{\mathcal{P}\gamma^{*}\gamma}(q_{\gamma_{1}}^{2},0)F_{\mathcal{P}\gamma\gamma^{*}}(0,q_{\gamma_{2}}^{2})$, where the right hand side is already known (the use of Chisholm approximants \cite{PabloSanchezPuertas} will also be considered in \cite{SergiRafel}). By looking at Fig.\ref{ffeta} we conclude that our $\mathcal{BR}$ predictions are supported by current measurements \cite{SergiRafel}.

\newpage

\subsection{On hadronic light-by-light contribution to the muon $g-2$ factor}
\addtocontents{toc}{\hspace{2cm}{\sl T.~Husek}\par}

\vspace{5mm}

{\sl T.~Husek, T.~Kadav\'y, K.~Kampf, J.~Novotn\'y}

\vspace{5mm}

\noindent
Institute of Particle and Nuclear Physics\\ Faculty of Mathematics and Physics, Charles University in Prague
\vspace{5mm}

Measurements of the muon anomalous magnetic moment belong to the most precise ones in particle physics. Quite recently, this quantity was measured with a remarkable accuracy by the E821 experiment at Brookhaven National Laboratory~\cite{Bennett:2006fi} with the current rescaled result $a_\mu=(g-2)/2=(116\,592\,089\pm63)\times10^{-11}\,.$
There has been found a significant and still persisting discrepancy between the theoretical prediction and the measured data. One of the greatest source of the theoretical uncertainty stems from the so called hadronic light-by-light (HLbL) contribution.

It is possible to study this problem from different points of view. We can try to focus on the related processes in low energy sector to test and improve existing models or try to calculate the appropriate Green functions.

Recently, the rare decay $\pi^0\to e^-e^+$ engrossed the attention of the theorists in connection with a new precise branching ratio measurement done by KTeV-E799-II experiment at Fermilab~\cite{Abouzaid:2006kk} which was determined to be
$
B(\pi^0\rightarrow e^+e^-(\gamma),\,x_\text{D}>0.95)=(6.44\pm0.33)\times10^{-8}\,.
$
Subsequent comparison with theoretical predictions of the SM were made~\cite{Dorokhov:2007bd1} and it has been found the 3.3\,$\sigma$ discrepancy between the theory and the experiment. Aside from the attempts to find the corresponding mechanism within the physics beyond the SM, also the possible revision of the SM predictions has been taken into account. For the theory overview see~\cite{Masjuan1}.

The full two-loop virtual radiative corrections (pure QED) and soft-photon bremsstrahul. were determined~\cite{Vasko:2011pi1} with the result 
$
\delta^\text{(2-loop)}(0.95)\equiv\delta^{\text{virt.}}+\delta_\text{soft}^\text{BS}(0.95)=(-5.8\pm0.2)\,\%\,,
$
which differs significantly from the previous approximative calculations done in~\cite{Bergstrom:1982wk1} or~\cite{Dorokhov:2008qn1}, where for $\delta^\text{(2-loop)}\left(0.95\right)$ they got -13.8\,\% and -13.3\,\%, respectively. The original disagreement was thus reduced to the level of 2\,$\sigma$.

Since the soft-photon approximation was used to compensate the IR divergences, the last step was to calculate the exact bremsstrahlung. Recently, it has been found, that for the KTeV kinematic cut, the soft-photon approximation is the relevant approach~\cite{Husek:20141}.

NLO radiative corrections in the QED sector did not solve the discrepancy using the contact interaction coupling finite part set to the value $\chi^\text{(r)}(M_\rho)=2.2\pm0.9$, which was theoretically modeled using the LMD approximation to the large-$N_{C}$ spectrum of vector meson resonances in~\cite{Knecht:1999gb}. Thus we can numerically fit the coupling to the result of the KTeV experiment using all available corrections to get the final model independent effective value (including higher order corrections and alternatively new physics) with the result
$
\chi^\text{(r)}(M_\rho)=4.5\pm1.0
$
\cite{Husek:20141}.
To check the stability of this result, the leading log estimation of the correction in the strong sector was done using Weinberg consistency relation to find
$
\Delta^\text{LL}\chi^\text{(r)}(M_\rho)
\doteq 0.081\,.
$
Moreover, $\chi^\text{(r)}$ is universal for $P\to l^+l^-$ processes up to $\mathcal{O}(m_l^2/\Lambda_{\chi \text{PT}}^2$). If the measurements of other related processes are not compatible with the result $\chi^\text{(r)}(M_\rho)=4.5\pm1.0$\,, something is not under control and it may be a sign for new physics. Such an interplay between precise measurement and theoretical calculations of
presented rare decay represents a valuable feedback for modeling the $F_{\pi^0\gamma^*\gamma^*}$ form-factor.


Another possibility to examine the HLbL contribution to the $g-2$ factor is via the perturbation theories of low-energy QCD: Chiral Perturbation Theory ($\chi$PT) and Resonance Chiral Theory (R$\chi$T). Using this approach we can include contributions of the resonance exchanges to the $g-2$. Both mentioned methods lead to the calculations of the chiral Green functions, the time ordered products of quantum fields, that, in general, enable to determine the amplitudes of physical processes using the LSZ reduction formula.

There are only five nontrivial Green functions in the odd-intrinsic parity sector of QCD: $\langle VVP \rangle$, $\langle VAS \rangle$, $\langle AAP \rangle$, $\langle VVA \rangle$ and $\langle AAA \rangle$, but only the first one contributes to the HLbL.

The $\langle VVP \rangle$ Green function is the most important example in the odd-intrinsic parity sector of QCD with a lot of phenomenological applications. Its contribution to the muon $g-2$ factor is due to a simplification of the $\langle VVVV \rangle$ Green function, representing the HLbL contribution to the $g-2$. The tensor structure of the $\langle VVP \rangle$ correlator can be written in the form
$$
\Pi_{\mu\nu}^{abc}(p,q)=\Pi(p^{2},q^{2},r^{2})\,d^{abc}p^{\alpha}q^{\beta}\varepsilon_{\mu\nu\alpha\beta}
$$
with the high-energy behavior within the OPE framework for high values of all independent momenta
$$
\Pi((\lambda p)^{2},(\lambda q)^{2},(\lambda r)^{2})=\frac{B_{0}}{\lambda^{4}}\frac{F^{2}}2\,\frac{p^{2}+q^{2}+r^{2}}{p^{2}q^{2}r^{2}}+\mathcal{O}\bigg(\frac{1}{\lambda^{6}}\bigg)\,.
$$

Substituting the coupling constraints obtained from OPE into $\Pi^{\mathrm{R\chi T}}(p^{2},q^{2},r^{2})$ allows us to extract the fully off-shell form-factor introduced in~\cite{Kampf:2011ty}
$$
\mathcal{F}_{\pi^{0}\rightarrow\gamma\gamma}^{\mathrm{R\chi T}}(p^{2},q^{2},r^{2})=\frac{2}{3F}\frac{r^{2}}{B_{0}}\Pi^{\mathrm{R\chi T}}(p^{2},q^{2},r^{2})\,.
$$
Using the form-factor we get the value for the muon anomalous magnetic moment
$
a_{\mu}^{\text{LbyL},\,\pi^{0}}=(65.8\pm 1.2)\cdot 10^{-11}
$
\cite{Kampf:2011ty}\,.
The updated result using Belle data \cite{Roig:2014uja} gives us the most recent value
$
a_{\mu}^{\text{LbyL},\,\pi^{0}}=(66.6\pm 2.1)\cdot 10^{-11}\,.
$

\newpage

\subsection{The role of experimental data on the hadronic light-by-light of the muon g-2}
\addtocontents{toc}{\hspace{2cm}{\sl  P.~Sanchez-Puertas}\par}

\vspace{5mm}
{\sl P.~Sanchez-Puertas}

\vspace{5mm}

\noindent
PRISMA Cluster of Excellence, Institut f\"ur Kernphysik, Johannes Gutenberg-Universt\"at, Mainz D-55099, Germany\\

\vspace{5mm}

The new planned experiments at Fermilab and J-PARC aiming to measure the anomalous magnetic moment of the muon $(g-2)_{\mu}$ to a precision around $0.14$ and $0.1$~ppm~\cite{Roberts:2010cj} demand an improvement on theoretical uncertainty estimations, at present of around $0.5$ppm~\cite{Jegerlehner:2009ry}. Such theoretical uncertainty comes entirely from 
hadronic contributions. Among them, the hadronic light-by-light contribution (HLBL) seems to be the hardest to reduce. The lack of ability to link it to experimental data, as has been done for the hadronic 
vacuum polarization through the use of dispersion relations (DR)~\cite{Jegerlehner:2009ry}, makes hard to reduce its theoretical error, which, eventually, will become the dominant one (for recent advances on DR, see~\cite{Hoferichter}). \\

This situation stimulated us to look for a different approach  which makes full use of data and gets systematic uncertainties under control. For the main HLBL piece, the 
pseudoscalar-exchange~\cite{Knecht:2001qf}, the hadronic element which drives the theoretical uncertainty is the  pseudoscalar transition form factor (TFF)  $F_{P\gamma^*\gamma^*}(Q_1^2,Q_2^2)$. 
In Refs.~\cite{Masjuan:2012wy,Escribano:2013kba,Escribano:2014}, it was proposed that such a data-driven model-independent approach is possible for the single-virtual  $F_{P\gamma^*\gamma}(Q^2,0)$ TFF through the use of Pad\'e approximants (PA)~\cite{Baker}. We stress that Pad\'e theory was applied here, on one hand, for reconstructing the TFF~\cite{Masjuan:2007ay}, and on the other, for extracting the required parameters~\cite{Masjuan:2008fv}. To recover the double-virtual TFF, for which no data is available yet, we assumed factorization~\cite{Escribano:2013kba}: $F_{P\gamma^*\gamma^*}(Q_1^2,Q_2^2) = F_{P\gamma^*\gamma}(Q_1^2,0)\times F_{P\gamma\gamma^*}(0,Q_2^2)$.\\


The aim of this work~\cite{DoubleVirt}, is to extend such approach to the double-virtual case which is required in our calculation without assuming factorization ideas which violate the OPE behavior~\cite{Knecht:2001xc}. This is done by extending PAs to the bivariate case as proposed by 
Chisholm~\cite{Chisholm},
\begin{equation}
P^N_M(Q^2) = \frac{\sum_i^N a_{i}Q^{2i}}{\sum_j^M b_jQ^{2j}} \ \ \rightarrow P^N_M(Q_1^2,Q_2^2) = \frac{\sum_{i,j}^N a_{i,j}Q_1^{2i} Q_2^{2j}}{\sum_{k,l}^M b_{k,l}Q_1^{2k} Q_2^{2l}}; \ \ a_{i,j}=a_{j,i}, \ \ a_{i,0}(b_{i,0})=a_i(b_i).
\end{equation}
The strong point of this approach comes from its convergence properties, which makes 
it very accurate in the space-like~(SL) low-energy region (the one entering the HLBL), with its ability to implement the high-energy behavior at once. 
This means that we obtain a full SL representation, and, for the first time, model-independent approach, using this method.\\


In analogy with the single-virtual case~\cite{Masjuan:2012wy}, such coefficients, related to the TFF series expansion, may be determined from experimental data alone. However, the lack of data for the 
double virtual TFF makes this procedure impossible at the moment. Therefore, we use the lowest-order bivariate approximant we can obtain,
\begin{equation}
\label{eq:Chisholm}
P^0_1(Q_1^2,Q_2^2) = \frac{a_{0,0}}{1+b_{1,0}(Q^2_1,Q^2_2) + b_{1,1}Q_1^2 Q_2^2}.
\end{equation}
$a_{0,0}$ and $b_{1,0}$ are known from the $P\rightarrow\gamma\gamma$ decay and the TFF slope respectively~\cite{Masjuan:2012wy,Escribano:2013kba,Escribano:2014}, while  $b_{1,1}$, related to double-virtual effects, would be extracted once double-virtual experimental data from BES-III~\cite{Redmer} becomes available.  Meanwhile, as a generous estimate, we consider ranging $b_{1,1}$ from its OPE value, $b_{1,1}=0$, to  $b_{1,1} = 2b_{1,0}^2$, well beyond the factorization result $b_{1,1} = b_{1,0}^2$. We show some representative values in Tab.~\ref{tab:HLBL}.
\begin{table}[h!]
\center
  \begin{tabular}[t]{lllll} \hline
             & $\pi^0$ & $\eta$ & $\eta'$  &  Total \\ \hline 
   {\small{$b_{1,1} = 0$}} \hspace{1cm}   & $6.64(33)$ & $1.69(6)$ & $1.61(21)$  &  $9.94(40)_{stat}(50)_{sys}$ \\
   {\small{$b_{1,1} = b_{0,1}^2$}}    & $5.53(27)$ & $1.30(5)$ & $1.21(12)$  &  $8.04(30)_{stat}(40)_{sys}$ \\
   {\small{$b_{1,1} = 2b_{0,1}^2$}}   & $5.10(23)$ & $1.16(7)$ & $1.07(15)$  &  $7.33(28)_{stat}(37)_{sys}$ \\ \hline
  \end{tabular}
  \caption{Our $a_{\mu}^{HLbL;P}$ results for different $b_{1,1}$ values within the considered range. }
  \label{tab:HLBL}
\end{table} 
The statistical uncertainty comes from the slope determination, and forthcoming low-energy data from BES-III~\cite{Redmer} will help diminishing these quantities, except for the $\eta$, where existing 
low-energy time-like data~\cite{Aguar-Bartolome:2013vpw} allows for the most precise slope determination ever~\cite{Escribano:2014}. The systematic error in Tab.~\ref{tab:HLBL} ($5\%$) comes from our PA 
reconstruction~\cite{Escribano:2013kba} and may be improved by going to higher order (i.e. $P^1_2(Q_1^2,Q_2^2)$) approximants. From our considered $b_{1,1}$ range we quote
\begin{equation}  
a_{\mu}^{HLbL;P} = ( 9.94(40)(50) \div 7.33(28)(37) )\times10^{-10}.
\end{equation}
Such band yields the biggest uncertainty, around $0.2$~ppm, and has been ignored in current estimations~\cite{Moverview}. In principle, this could be constrained using the $\pi^0\rightarrow e^+e^-$ decay, which  depends on the low-energy double-virtual TFF as well. However, the experimental result implies unexpected high $b_{1,1}$ values which would translate into $a_{\mu}^{HLbL;\pi^0} \simeq 2\times10^{-10}$~\cite{Masjuan}. The implication on double Dalitz decays~\cite{Sergi} still has to be checked.\\

\newpage

\subsection{ Measurement of$\,$ hadronic cross section at KLOE/KLOE-2}
\addtocontents{toc}{\hspace{2cm}{\sl V.~De Leo}\par}

\vspace{5mm}

{\sl V.~De Leo} \\on behalf of the KLOE/KLOE-2 Collaborations

\vspace{5mm}
\noindent
Universit\`{a} degli Studi di Messina\\

\vspace{5mm}

The anomalous magnetic moment of the muon parametrized as $a_\mu \equiv \frac{g_\mu-2}{2}$,
can be accurately measured and, within the Standard Model (SM) framework, precisely predicted\cite{Fred}.
\,The experimental value of $a_\mu$ ($(11 659 208.9 \pm 6.3) \times 10^{-10}$)
measured at the Brookhaven Laboratory differs from SM estimates by 3.2 - 3.6 $\sigma$\cite{Fred}. 
A large part of the uncertainty on the theoretical estimates comes from the leading order contribution 
$a_\mu^{had,LO}$, which at low energies is not calculable by perturbative QCD, but has to be evaluated with a dispersion integral using measured hadronic cross sections. Therefore,
improved precision in the $\pi\pi$ cross section would result in a reduction of the uncertainty on the
LO hadronic contribution to $a_\mu$ , and in turn to the SM prediction for $a_\mu$. 
The measurement of the $\sigma(e^+e^-\rightarrow \pi^+\pi^-)$ cross section allows to determine the pion form factor $\vert F_\pi  \vert^2$ and the two pion contribution to the muon anomaly $a_\mu$. 
In the 2008 and 2010 two measurements of the $\sigma(e^+e^- \rightarrow \pi^+\pi^-\gamma)$ have been 
performed at $DA\Phi NE$ with the KLOE detector\cite{Ambrosino,Ambrosino10}. The last KLOE measurement of the $e^+e^- \rightarrow \pi^+\pi^-$ cross section (KLOE12) has been 
obtained from the ratio between the pion and muon ISR differential cross section\cite{Babusci}.   
The data sample is the same of the KLOE08 analysis and corresponds to an integrated luminosity of
$239.2\, pb^{-1}$ collected in 2002 with the small angle photon selection. The selection between the $\pi\pi\gamma$ and $\mu\mu\gamma$ events is obtained 
assuming the final state of two charged particles
with equal mass $M_{TRK}$ and one photon: the $M_{TRK} < 115\,MeV$ for the muons and $M_{TRK} > 130 \,MeV$ for the pions. \\
The $\pi/\mu$ separation has been crosschecked with different methods as 
a kinematic fit or tighter cuts on the quality of the charged tracks. 
Trigger, particle identification and tracking efficiencies have been checked from data
control samples. The $\mu\mu\gamma$ cross section measurement is compared with
the one obtained by PHOKHARA MC, and a good agreement has been found.
Then the pion
form factor has been extracted using the following equation \cite{Babusci} 
\begin{equation}
{\vert F_\pi(s') \vert}^2 = \frac{3}{\pi} \frac{s'}{\alpha^2 \beta_\pi^3} \sigma^0_{\pi\pi(\gamma)}(s')
(1+\delta_{VP}) (1-\eta_\pi(s'))
\label{F_pi}
\end{equation}
where $\sigma^0_{\pi\pi(\gamma)}$ is the bare cross section, $\delta_{VP}$ is the VP correction
and $\eta_\pi$ accounts for FSR radiation assuming point-like pions\cite{Babusci}. 
The result for the  $a_\mu^{\pi\pi}$ has been compared with the previous ones from KLOE08 and KLOE10 showing good agreement. The preliminary combination of the last three KLOE results (KLOE08, KLOE10, KLOE12) has been also performed using the Best Linear Unbiased Estimate (\textit{BLUE}) method \cite{Valassi}.

\newpage

\subsection{Dispersive analysis of the $\pi^0$ transition form factor}
\addtocontents{toc}{\hspace{2cm}{\sl B.~Kubis}\par}

\vspace{5mm}

{\sl B.~Kubis}

\vspace{5mm}

\noindent
Helmholtz-Institut f\"ur Strahlen- und Kernphysik (Theorie) and\\ Bethe Center for Theoretical Physics,
Universit\"at Bonn, Germany

\vspace{5mm}

A major challenge to an improved theoretical prediction of the anomalous magnetic moment of the muon $(g-2)_\mu$ lies 
in the determination of the contribution from hadronic light-by-light scattering, including a reliable estimate
of the associated uncertainty.  Recently, an effort has been started to analyze the dominant individual
contributions using dispersion relations: the one- and two-pion contributions~\cite{Amaryan:2013eja,dispHlbl,roadmap,Martin}, 
as well as the $\eta$ and $\eta'$ pole terms~\cite{Stollenwerk,etaTFF,Andreas}.

The strength of the $\pi^0$ pole is determined by the singly- and doubly-virtual
$\pi^0$ transition form factor, which can be analyzed in dispersion theory~\cite{pi0TFF}.
In the most important energy range (roughly up to 1\,GeV), the isovector and isoscalar part of 
$\gamma^* \to \pi^0 \gamma^{(*)}$ are dominated by two- and three-pion intermediate states, respectively.
While the dispersion relation for two pions requires the (charged) pion vector form factor and 
the anomalous amplitude $\gamma^{(*)}\pi \to \pi\pi$ as input, three pions can be simplified due to the 
dominance of the narrow isoscalar resonances $\omega$ and $\phi$ and demand an understanding of the 
vector-meson transition form factors $\omega,\,\phi\to\pi^0\gamma^*$.  All of these components can in turn be 
reconstructed dispersively.

The process $\gamma\pi\to\pi\pi$ at zero energy and pion masses is determined---as is the decay of the 
$\pi^0$ into two real photons---by the Wess--Zumino--Witten anomaly.  A dispersive representation~\cite{g3pi} 
can be used to extract the anomaly from data in the full elastic region.  A similar analysis provides
decay amplitudes for $\omega,\,\phi\to3\pi$~\cite{V3pi}, which have been shown to reproduce high-statistics
data for the $\phi\to3\pi$ Dalitz plot~\cite{KLOE:phi} with excellent accuracy.  
A comparably precise experimental determination of the $\omega\to3\pi$ Dalitz plot would be highly desirable~\cite{Schott,Heijkenskjold}.
The corresponding partial waves,
again combined with the pion vector form factor, yield a dispersive representation of the 
vector-meson transition form factors~\cite{omegaTFF}.  Sum rules for the decays $\omega,\,\phi\to\pi^0\gamma$
work rather well, although the description of data on $\omega\to\pi^0\mu^+\mu^-$~\cite{NA60-1,NA60-2} remains problematic.
Recent studies analyzing more general constraints on the $\omega\pi$ transition form factor derived from analyticity
and unitarity suggest that at least some of the data points are likely problematic~\cite{ACK}.

As a final step, a parametrization of the cross section data for $e^+e^-\to3\pi$ allows for
a full dispersive reconstruction of $\pi^0\to\gamma^*\gamma^*$. 
So far, the singly-virtual form factor has been analyzed and compared to data explicitly~\cite{pi0TFF}:
without adjusting any further parameters, the existing (time-like) data for 
$e^+e^-\to\pi^0\gamma$~\cite{eepi0g_SND_1,eepi0g_SND_2,eepi0g_CMD2}
are described very well.
Given the imaginary part in the time-like region, we can reconstruct the transition form factor in the space-like
region by another dispersion relation.  In particular in the low-energy region $Q^2\lesssim (1.1\,\text{GeV})^2$,
we give a very accurate prediction (with uncertainties of less than 5\%) for the upcoming high-precision BESIII
measurements~\cite{Redmer1}.
Expanding the singly-virtual form factor around zero, 
$F_{\pi^0\gamma^*\gamma}(q^2,0) = F_{\pi\gamma\gamma} [1+a_\pi\times q^2/M_{\pi^0}^2+\mathcal{O}(q^4)]$, 
we can calculate the slope from a sum rule to excellent precision, $a_\pi=(30.7\pm0.6)\times 10^{-3}$.
Even curvature terms can be determined similarly.  These expansion coefficients essentially determine
the decay rate and spectrum of $\pi^0\to e^+e^-\gamma$.  
Work on the generalization to the doubly-virtual $\pi^0$ transition form factor and 
the determination of its contribution to $(g-2)_\mu$ is in progress.

\newpage

\subsection{Dispersive Approach to the $\eta$ Transition Form Factor}
\addtocontents{toc}{\hspace{2cm}{\sl A.~Wirzba}\par}

\vspace{5mm}

{\sl A.~Wirzba}

\vspace{5mm}

\noindent
Institut f\"ur Kernphysik and Institute for Advanced Simulation,\\  Forschungszentrum J\"ulich, D-52425 J\"ulich, Germany\\
\vspace{5mm}

In this talk I review the dispersive approach to the $\eta$ transition form factor, in part already published in  Refs.~\cite{aw:us1,aw:us2}, of a collaboration with
C.\,Hanhart, A.\,Kup\'s\'c, \mbox{U.-G.\,Mei{\ss}ner}, F.\,Stollenwerk, and recently G.\,Tukhashvili and T.\,Dato.  

The ultimate goal of our investigation is the construction of the doubly virtual $\eta\to \gamma^\ast\gamma^\ast$ (and
$\eta'\to \gamma^\ast\gamma^\ast$) transition form factor
which then can lead to some constraint on the hadronic light-by-light contribution to the theoretical  $g{-}2$ value of the muon.
Here we present---as interim result---the nearly model-independent construction of the 
singly virtual transition form factor for $\eta\to \gamma^\ast\gamma$. For that purpose, we start with the model-independent  	
fit\,\cite{aw:us1}  of the arbitrarily normalized experimental spectra of the radiative two-pion decays $\eta\to \pi^+\pi^-\gamma$~\cite{aw:WASA,aw:KLOE} and 
$\eta'\to \pi^+\pi^-\gamma$~\cite{aw:CrystalBarrel} by the
isovector form factor of the pion (see Ref.\,\cite{aw:Hanhart_FF} for
a recent  parametrization of this quantity which accounts for the pion-pion final-state interaction), solely
multiplied by a linear polynomial $P(s_{\pi\pi})= 1 +\alpha s_{\pi\pi}$. The reaction-specifics
of the production vertex  are incorporated by 
the coefficient $\alpha$ (and the branching ratios of the cross sections~\cite{aw:PDG}). 
In contrast to the $\gamma \pi\to \pi\pi$ case the  radiative $\eta$ and $\eta'$  two-pion decays 
have the advantage that the  left-hand cuts are chirally and kinematically suppressed because of the $p$-wave nature of the
$\eta^{(\prime)}\pi$ scattering~\cite{aw:Bernard_left,aw:Kubis_left} while the right-hand cut is the same
as for the pion  form factor (and  pertinent Omn\`es function).

From this input a once-subtracted dispersion integral is derived that predicts the isovector part of
the singly virtual $\eta\to\gamma^\ast\gamma$ transition form factor~\cite{aw:us2}. The isoscalar contribution follows
from saturating the corresponding part of the $e^+e^- \to \eta \gamma$ cross section with the  tabulated~\cite{aw:PDG} masses and widths of 
$\omega$ and $\phi$  bosons~\cite{aw:us2}.

Our prediction of  the slope at the origin (and the momentum dependence) of the $\eta$ transition form factor 
is consistent with all recent data\,\cite{aw:CELLO,aw:NA60,aw:Berghauser,aw:A2}
and the parametrization of Ref.\,\cite{aw:Escribano},
 but differs from some previous theoretical analyses, see Ref.~\cite{aw:Ametller}. 

In summary, both processes, $\eta\to \pi^+\pi^-\gamma$ and $\eta\to \gamma^\ast\gamma$ are characterized
by two scales:
a universal one, proportional to the mass of the $\rho$ meson from the $\pi\pi$ final-state interaction, and, respectively, a reaction-specific
one from the production vertices (and left-hand cuts). 

We plan to incorporate the  $\rho'$ and
$\rho''$ resonances and further non-two-pion channels into the $\eta\to \pi^+\pi^-\gamma$ amplitude that enters the
dispersion integral. Especially, we are looking forward to up-coming data on the $e^+ e^- \to \eta\pi^+\pi^-$  cross sections and 
spectra---the latter as function of $s_{ee}$ {\em and}   $s_{\pi\pi}$\,---from the SND and CMD-3 collaborations at VEPP-2000~\cite{aw:Eidelman} and,
maybe, from elsewhere. As a matter of fact, the BaBar collaboration
had already published  $e^+ e^- \to \eta\pi^+\pi^-$ cross section and spectral data~\cite{aw:BaBar} 
which indicate both $\rho'$ dominance in the $e^+e^-$ leg and
$\rho$ dominance in the $\pi\pi$ one. 
Unfortunately, the  spectral data are only available in    $s_{ee}$-averaged form, such that we are forced to unfold
them in a model-dependent way.  Our preliminary results indicate the presence of an
additional $s_{ee}$ dependence in the linear polynomial, {\it i.e.} $P(s_{ee},s_{\pi\pi}) = 1 + \alpha(s_{ee}) s_{\pi\pi}$, and, moreover,
non-factorization---already below  $2\,\mathrm{GeV}$---of the two photon legs of doubly virtual $\eta\to \gamma^\ast\gamma^\ast$
form factor.

\newpage

\subsection{Photoproduction and Decay of Light Mesons in CLAS}
\addtocontents{toc}{\hspace{2cm}{\sl D.~Schott}\par}

\vspace{5mm}

{\sl D.~Schott} \\on behalf of the CLAS Collaboration

\vspace{5mm}

\noindent
The George Washington University, Washington, DC, USA

\vspace{5mm}

We presented the current results of light meson decays from CLAS (CEBAF Large Acceptance Spectrometer) at Jefferson Lab in Newport News, VA, USA. The experimental data included in this summary are produced by tagged photons incident on a hydrogen target. The main particles of interest are pseudoscalar - $\pi^0$, $\eta$, $\eta'$ and $\rho$, $\omega$ and $\phi$ vector mesons. Physics topics covered include Dalitz Decays of $\pi^0$, $\eta$, $\eta'\rightarrow e^+ e^- \gamma$ and $\omega\rightarrow e^+ e^- \pi^0$; Radiative Decays of $\eta,\eta'\rightarrow \pi^+ \pi^- \gamma$; and Hadonic Decays of $\eta$, $\eta'$, $\omega\rightarrow \pi^+ \pi^- \pi^0$, $\pi^+ \pi^- \eta$. Using this data we are able to significantly expand world data base of photo production mesons in the range of 
$E_{\gamma}=1.3-5.45~GeV$. This is important in order to to allow the addition of Regge analysis to the standard PWA. The preliminary analysis of the hadronic decay channels $\eta$, $\omega\rightarrow \pi^+ \pi^- \pi^0$ include the ongoing amplitude analysis in partnership with the Joint Physics Analysis Center (JPAC) to get an insight in low energy QCD. The abundant statistics available will allow analyses using CLAS data to gain access to transition form factors in time-like domain, box anomaly term, quark mass ratio, test fundamental C and CP symmetries, search for dark photon, and invisible decays. 

\newpage

\subsection{Studies of the $\omega$ meson at KLOE}
\addtocontents{toc}{\hspace{2cm}{\sl  W. I. Andersson}\par}

\vspace{5mm}
{\sl B. Cao, L. Heijkenskj\"old, W. I. Andersson}\\on behalf of the KLOE-2 Collaboration
\vspace{5mm}

\noindent
Uppsala University\\

\vspace{5mm}

The study of light mesons is an active research field of the strong interactions. At low energies perturbative quantum chromodynamics breaks down due to the growing coupling constant. Effective field theories and dispersion calculations provide an alternative framework to study the light mesons. Below, three ongoing studies concerning $\omega$ mesons in data collected by KLOE experiment are presented.

The $\phi \to \omega \gamma$ is a $C$-parity violating decay and its branching ratio is expected to be heavily suppressed. This decay was last searched for in the Lawrence Radiation Laboratory in 1966 \cite{Lindsey:1966zz} and the current upper bound of the branching ratio is less than 5\% \cite{PDG-2012}. The production of $\omega$ meson together with an initial state radiation photon has the same final state particles and can be used in the search for the $\phi \to \omega \gamma$ decay by measuring the $\omega \gamma$ final state cross section around the $\phi$ meson mass region. A data sample of $\mathcal{L} = 1.6$ fb$^{-1}$ integrated luminosity recorded at $\sqrt{s} = 1019$ MeV/c$^2$ center of mass energy has been used to study the $e^+e^- \to \omega \gamma_{\mathrm{ISR}}$ production channel. A preliminary analysis scheme provides an 54\% efficiency for signal events, while the sum of all events has an efficiency of 3\%.

The transition form factors of light mesons could play a significant role in the prediction of the magnetic moment of the muon \cite{Jegerlehner:2009ry1}. The Vector Meson Dominance (VMD) model is often used to predict such form factors but fails to describe the $\omega \to \mu^+ \mu^- \pi^0$ decay measured by the NA60 experiment \cite{Arnaldi}. Alternative theoretical models attempts to describe the form factor but a second high statistics measurement is desired \cite{Terschlusen:2011pm}. The same KLOE data sample has been used as a benchmark for the feasibility study of the $\omega \to l^+ l^- \pi^0$ decay. It is found that $(32 \pm 2) \cdot 10^2$ $\omega \to e^+e^- \pi^0$ decays and $(25 \pm 7) \cdot 10$ $\mu^+\mu^+ \pi^0$ decays are expected. The branching ratio errors from PDG and the detector efficiency errors are taken into account when evaluating the error in the number of expected events. The error in the detector efficiency is evaluated from the mean of ten repeated simulations.

The KLOE collaboration has, in a previous investigation~\cite{Kloewpi0}, measured $1.3\cdot10^6\ \omega \to \pi^+ \pi^- \pi^0$ decays in the $e^+e^- \to \omega \pi^0$ production reaction. The study presented here has considered the possibility of using this data set for producing a high statistics Dalitz plot. Such distribution would provide an ample testing ground for the predictions made by two different theoretical approaches~\cite{Niecknig,Terschlusen}. However, when modelling the full interaction using a VMD matrix element~\cite{Akhmetshin}, the interference between the two final state $\pi^0$ is shown to have a non-negligible effect ($\sim 10 \%$) on the Dalitz plot distribution. A possibility to extract the Dalitz plot distribution is to use minimally disturbed parts of the Dalitz plot and utilize the symmetry of the distribution to predict the full shape.

\newpage

\subsection{Search for $C$-violation and Physics beyond SM in the Decay $\eta \rightarrow \pi^0 \mbox{ e}^+ \mbox{e}^-$ with WASA-at-COSY}
\addtocontents{toc}{\hspace{2cm}{\sl K.~Demmich}\par}

\vspace{5mm}

{\sl K.~Demmich} \\for the WASA-at-COSY Collaboration

\vspace{5mm}
\noindent
Institut f\"{u}r Kernphysik, Westf\"{a}lische Wilhelms-Universit\"{a}t M\"{u}nster\\
\vspace{5mm}

The strong and the electromagnetic interaction are invariant with respect to the $C$-parity, i.e. to the exchange of a particle with its anti-particle. The observation of a $C$-violating reaction would, therefore, give hints to physics beyond the standard model.

\begin{figure}[b]
	\includegraphics[width=.25\columnwidth]{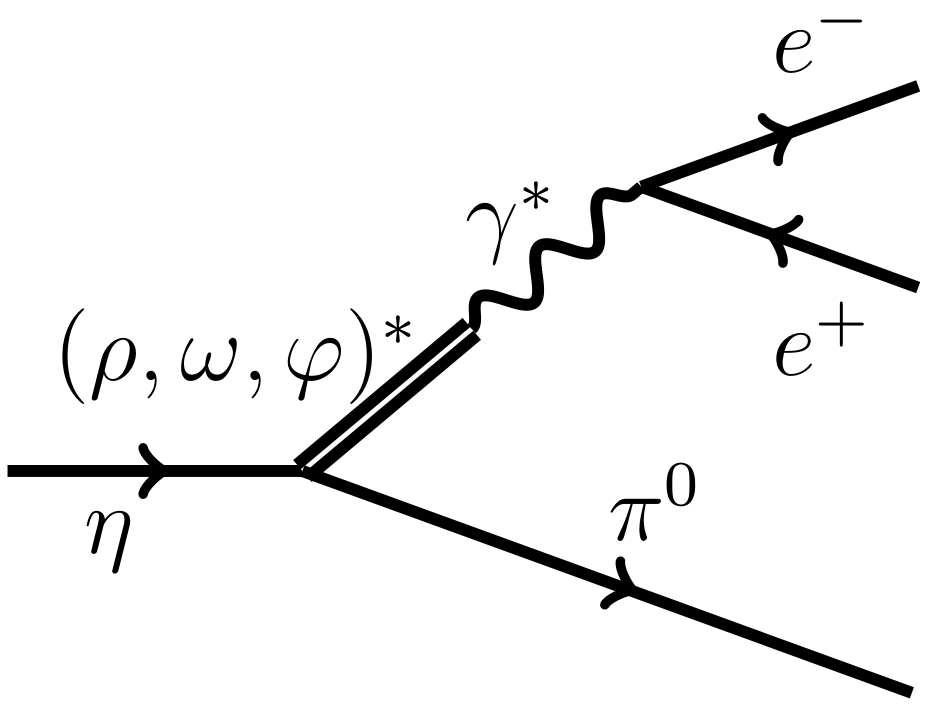} \hfill
  \includegraphics[width=.25\columnwidth]{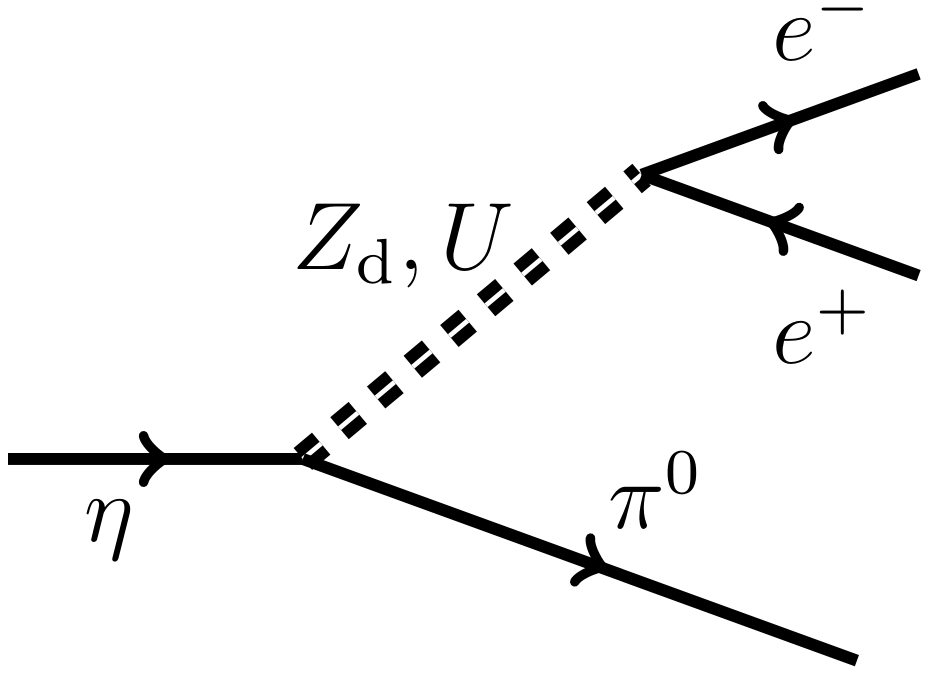} \hfill
  \begin{picture}(0.31\columnwidth,3cm)
	\put(0,0){\includegraphics[width=.3\columnwidth]{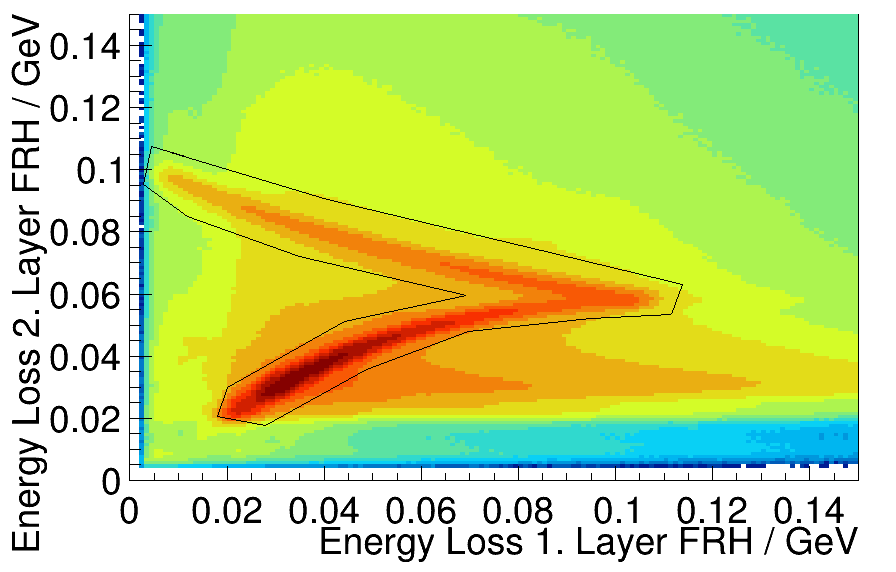}} 
	\put(2.9cm, 2.8cm){\tiny{pp data sample}}
	\put(4.2cm, 0.7cm){\rotatebox{90}{\small{preliminary}}}
  \end{picture}
	\caption{Left and middle: possible reactions for $\etapiee$ assuming a vector dominance model or a Z-like/photon-like dark U boson. Right: energy loss plot for proton identification.}
	\label{fig:1}
\end{figure}

An appropriate probe for the search of a $C$-violating process are $\eta$-meson decays. The decay channel $\etapiee$ can concurrently be possible by the coupling of the two leptons to one virtual photon -- assuming the vector dominance model -- or to two virtual photons. The second process is of higher order and by this strongly suppressed and allows, therefore, for the search of the first $C$-violating reaction. Moreover, the decay could also be realized via a dark photon (U) or a Z-like dark U boson ($\text{Z}_\text{d}$) as an intermediate state (see Fig.~\ref{fig:1}).
The decay $\etapiee$ has not been observed yet and only an upper limit for the branching ratio assuming a phase space distribution of the decay products is quoted by the PDG to be $\textrm{BR}(\Etapiee) < 4 \times 10^{-5}$ \cite{PDG}. 
The high statistics on $\eta$-mesons collected at WASA-at-COSY allows for detailed studies on this decay. These data enables a high sensitive measurement of the relative branching ratio of this $C$-violating decay mode and for testing the vector dominance model hypothesis. Moreover, the analysis will give access to the coupling constants of the dark bosons and will set constraints to the mass region of the Z$_\text{d}$.

A detector setup uniquely suitable for investigating $\eta$-decays is the WASA installation located at the COSY accelerator facility at the Forschungszentrum Jülich, Germany.
WASA-at-COSY is dedicated to investigate the physics of light mesons produced in the collision of protons and/or deuterons. The forward detector is designed to reconstruct the forward scattered particles with high precision and the central detector provides a nearly $4\pi$ acceptance for charged and neutral decay particles.
In order to analyze rare $\eta$-decays with high accuracy two different data samples were taken with the WASA detector.

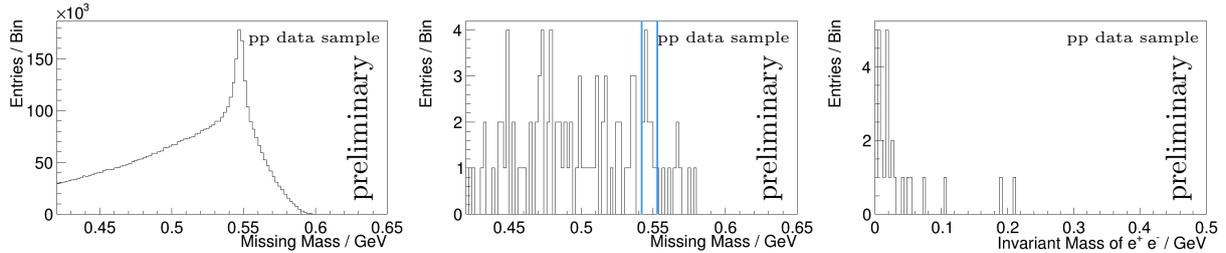
\begin{figure}[h]
  \begin{picture}(0.32\columnwidth,3cm)
	\put(0,0){\includegraphics[width=.32\columnwidth]{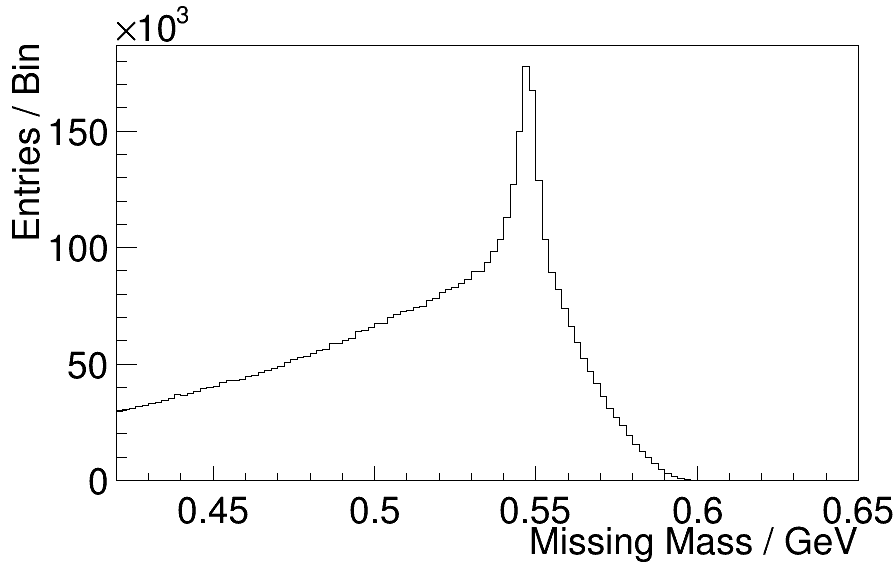}} 
	\put(3.2cm, 2.8cm){\tiny{pp data sample}}
	\put(4.5cm, 0.7cm){\rotatebox{90}{\small{preliminary}}}
  \end{picture}
  \begin{picture}(0.32\columnwidth,3cm)
	\put(0,0){\includegraphics[width=.32\columnwidth]{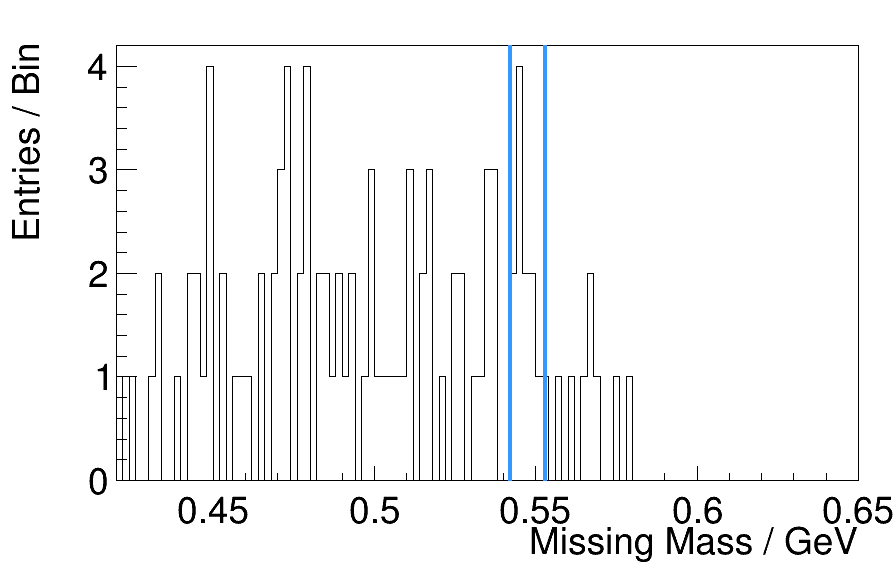}} 
	\put(3.2cm, 2.8cm){\tiny{pp data sample}}
	\put(4.5cm, 0.7cm){\rotatebox{90}{\small{preliminary}}}
  \end{picture}
  \begin{picture}(0.32\columnwidth,3cm)
	\put(0,0){\includegraphics[width=.32\columnwidth]{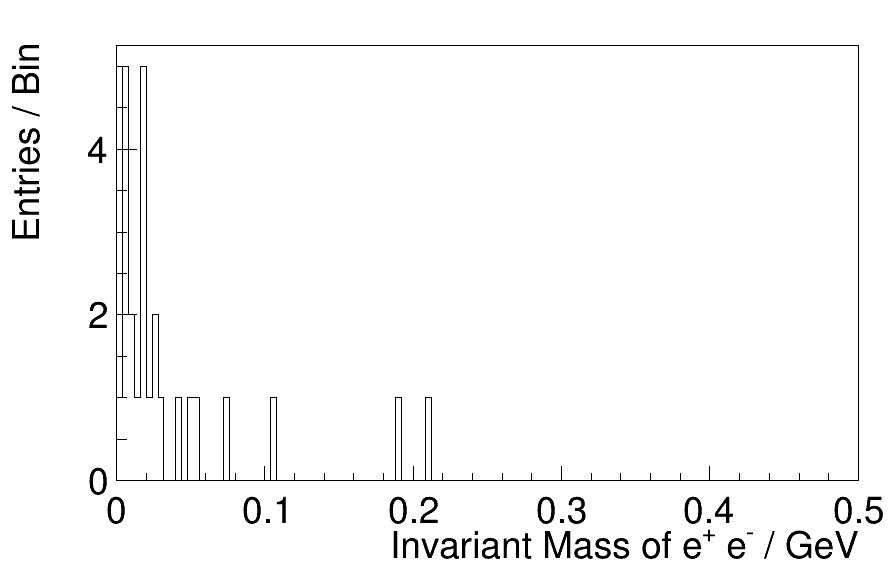}} 
	\put(3.2cm, 2.8cm){\tiny{pp data sample}}
	\put(4.5cm, 0.7cm){\rotatebox{90}{\small{preliminary}}}
  \end{picture}
	\caption{Left/middle: missing mass after proton selection and before final missing mass cut (blue lines). Right: invariant mass of $\mbox{ e}^+\mbox{ e}^-$ after cut on missing mass.}
	\label{fig:2}
\end{figure}

In 2008 and 2009 a data sample with $\approx 30 \times 10^6$ $\eta$-events was taken by the means of proton-deuteron scattering $\text{p} + \text{ d} \rightarrow {}^3\text{He} + \eta$. The clear  ${}^3\text{He}$ signal allows for a very effective separation of the $\eta$-signal from background originating from direct (multi-)pion production. A second data sample was taken in 2008, 2010, and 2012 by proton-proton scattering $\text{p} + \text{ p} \rightarrow \text{ p} + \text{ p} + \eta$. The higher cross section results in a very high statistic with $\approx 500 \times 10^6$ fully reconstructed $\eta$-events but provides a greater challenge to handle the larger hadronic background.
In a first analysis step, the reaction $\text{ p} + \text{ d} \rightarrow {}^3\text{He} + \text{ X}$ and $\text{ p} + \text{ p} \rightarrow \text{ p} + \text{ p} + \text{ X}$, respectively, has to be identified. Energy loss plots of the subsequent scintillator layers of the forward detector allow for an effective particle identification (see Fig. \ref{fig:1}). Despite the different production reactions, both data samples can be analyzed with a similar procedure. Firstly, a selection on the decay signature (two oppositely charged tracks, corresponding to the lepton pair, and two neutral tracks, corresponding to the photon pair originating from the decayed pion) is performed. Further cuts are used to reduce the background from conversion, bremsstrahlung and split-offs. A discrimination of electrons and pions is done by a neural network based on the energy loss and momentum correlation. Cuts on the invariant mass of the photon pair close to the pion mass as well as on the invariant mass of all decay particles close to the $\eta$ mass are done. A kinematic fit testing 
$\text{ p} \, \text{ d} \rightarrow {}^3\text{He}  \, \text{ e}^+  \, \text{ e}^-  \, \gamma  \, \gamma$ 
and $\text{ p} \, \text{ p} \rightarrow \text{ p} \, \text{ p} \, \text{ e}^+ \, \text{ e}^- \, \gamma \, \gamma$, respectively, reduces the background significantly. After a last cut on the missing mass close to the $\eta$-mass the number of remaining events can be determined depending on the $\text{ e}^+ \text{ e}^-$-invariant mass as can be seen in Fig.~\ref{fig:2} for a data sample.

The sample taken in 2008 by the means of proton-deuteron collisions is already analyzed by A. Winnem\"oller \cite{winne}. 
For the combined analysis of the full pd-data set cut conditions based on Monte Carlo simulations are currently optimized by F. Bergmann \cite{FB}. The analysis of the large data set obtained in proton-proton scattering is in progress and already shows a very good background suppression.\\\\
The support from COSY-FFE grants is kindly acknowledged.

\newpage

\subsection{Search for the dark photon in $\pi^0$ decays by NA48/2 at CERN }
\addtocontents{toc}{\hspace{2cm}{\sl  E.~Goudzovski}\par}

\vspace{5mm}

{\sl E.~Goudzovski}\\on behalf of the NA48/2 collaboration\footnote{Cambridge, CERN, Chicago, Dubna, Edinburgh, Ferrara, Firenze, Mainz, Northwestern, Perugia, Pisa, Saclay, Siegen, Torino, Wien. Email: eg@hep.ph.bham.ac.uk.}

\vspace{5mm}

\noindent
School of Physics and Astronomy, University of Birmingham, United Kingdom\\
\vspace{5mm}

Kaons represent sources of tagged neutral pion decays, mainly via the $K^\pm\to\pi^\pm\pi^0$ and $K_L\to3\pi^0$ decays. Therefore high intensity kaon experiments are suitable for studies of rare $\pi^0$ decays (e.g. $\pi^0\to e^+e^-$) and searches for new physics in the $\pi^0$ decay. One of these experiments is NA48/2 at CERN, which took data in 2003--2004 with narrow momentum band 60~GeV/$c$ $K^\pm$ beams and was exposed to $\sim2\times 10^{11}$ $K^\pm$ decays in its fiducial decay volume contained in a 114~m long vacuum vessel~\cite{ba07}.

The NA48/2 data sample is used to search for the dark photon (DP, denoted $A'$)  production in the $\pi^0$ decay, the DP being a hypothetical gauge boson appearing in hidden sector new physics models with an extra $U(1)$ gauge symmetry~\cite{ho86}. The DP is characterized by the (unknown) mass $m_{A'}$ and mixing parameter $\varepsilon$. The search at NA48/2 is performed via the decay chain $K^\pm\to\pi^\pm\pi^0$, $\pi^0\to A'\gamma$, $A'\to e^+e^-$, assuming that $A'$ decays into SM fermions only. Under this assumption, the expected DP decay width for $m_e\ll m_{A'}<2m_\mu$ is $\Gamma_{A'}\approx\alpha\varepsilon^2m_{A'}/3$~\cite{batell09}, and the distance between DP production and decay points can be neglected in the NA48/2 conditions for $m_{A'}>10$~MeV/$c^2$ and $\varepsilon^2>10^{-7}$. This has two consequences: 1) the main NA48/2 ``three-track vertex'' trigger is highly efficient for the considered decay chain; 2) the DP signature is identical to that of the Dalitz decay $\pi^0_D\to e^+e^-\gamma$, which therefore represents an irreducible background and determines the sensitivity.

\begin{figure}[tb]
\begin{center}
\resizebox{0.48\textwidth}{!}{\includegraphics{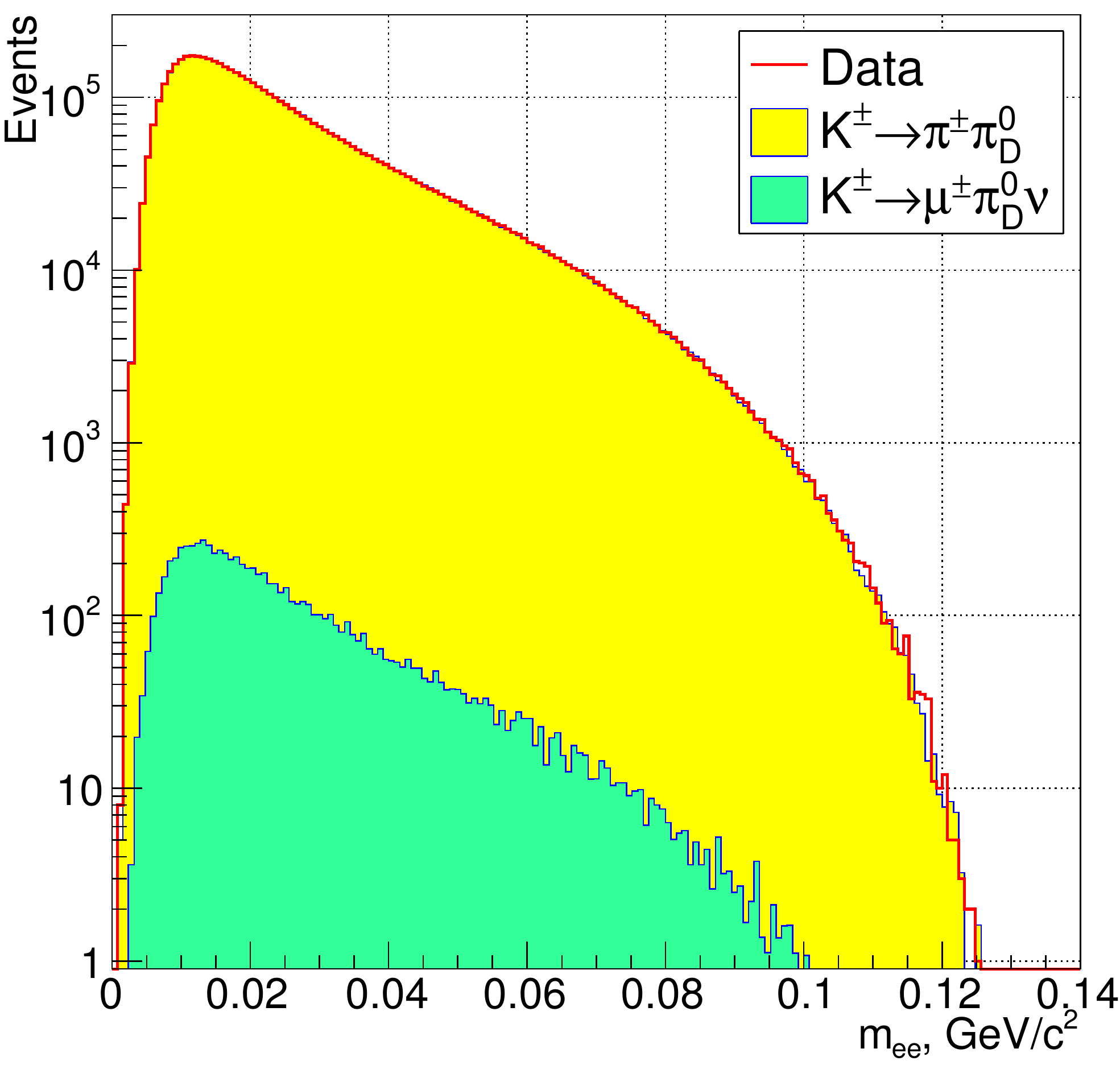}}%
\resizebox{0.48\textwidth}{!}{\includegraphics{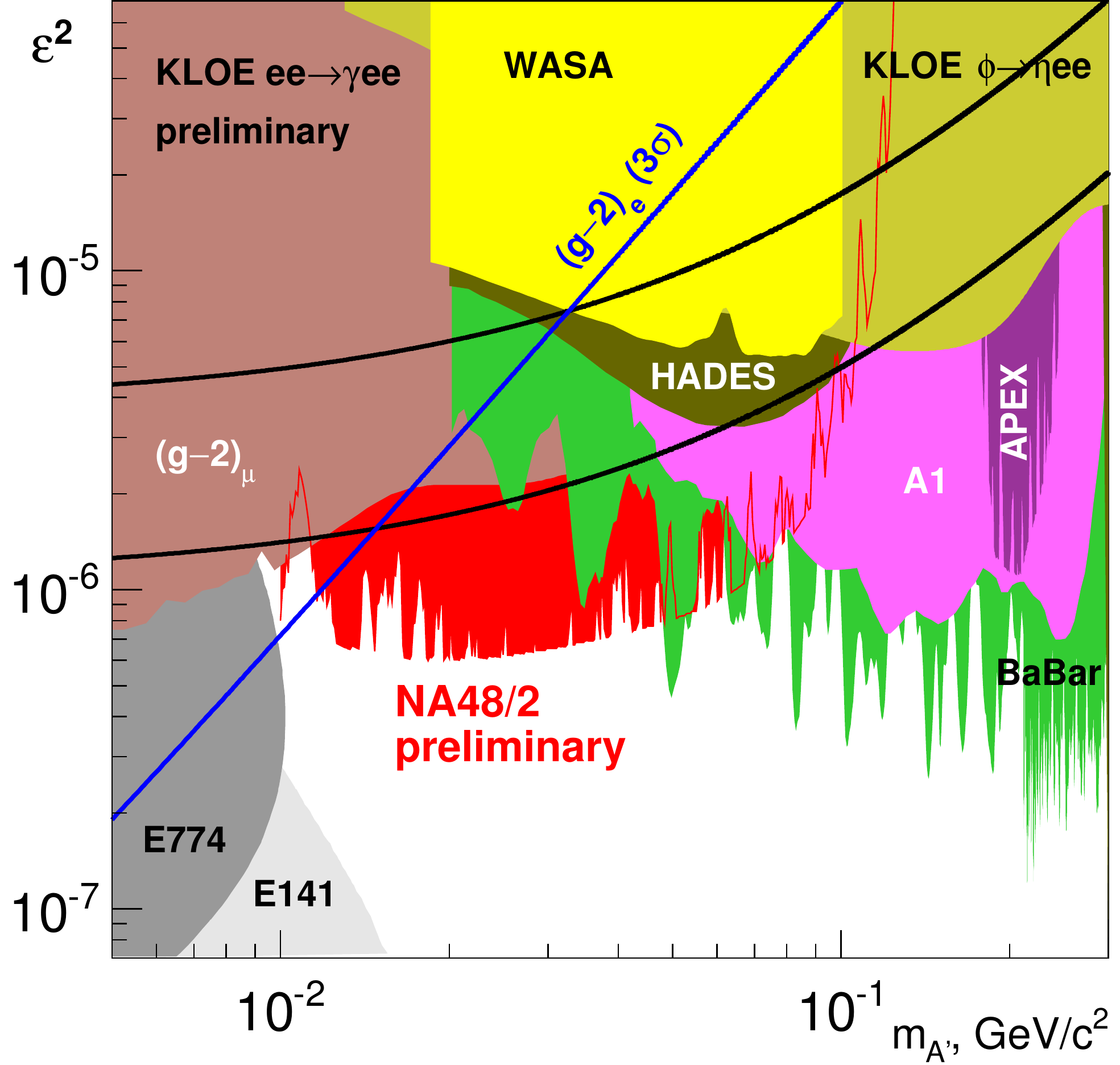}}
\end{center}
\vspace{-9mm} \caption{(a) Di-electron invariant mass spectrum of the reconstructed $\pi^0_D$ decay candidates: data and MC expectation. A DP would produce a narrow spike in the spectrum. (b) Experimental DP exclusion limits in terms of mass and mixing parameter. The band corresponding to the explanation of the muon $g-2$ puzzle is also displayed.}
\vspace{-2mm}
\label{fig:na48-darkphoton}
\end{figure}

In total, $4.687\times 10^6$ fully reconstructed $\pi^0_D$ decay candidates in the $e^+e^-$ invariant mass range $m_{ee}>10~{\rm MeV}/c^2$ with negligible background have been selected (mainly originating from $K^\pm\to\pi^\pm\pi^0$ decays, with 0.15\% from $K^\pm\to\pi^0\mu^\pm\nu$ decays). Their $m_{ee}$ spectrum is shown in Fig.~\ref{fig:na48-darkphoton} (left). The $\pi^0_D$ decay is simulated using the lowest-order differential decay rate and radiative corrections~\cite{mi72} recently revised using no approximations to improve the numerical precision~\cite{husek}. The PDG average value of the slope $a$ of the $\pi^0$ transition form factor (TFF), $F(x)=1+ax$, $x=(m_{ee}/m_{\pi^0})^2$, determined mainly by a measurement in the space-like region from $e^+e^-\to e^+e^-\pi^0$~\cite{cello}, is modified to obtain a satisfactory fit to $m_{ee}$ spectrum over the whole range $m_{ee}>10~{\rm MeV}/c^2$. A TFF slope measurement is in progress.

A DP mass scan is performed in the mass range $10~{\rm MeV}/c^2 < m_{ee} < 125~{\rm MeV}/c^2$ with a varying mass step equal to half of the resolution on $m_{ee}$. The resolution can be approximately parameterized as $\sigma_{m_{ee}}\approx 0.012\,m_{ee}$. In total, 398 DP mass hypotheses are tested. For each hypothesis, the signal region is defined as $\pm1.5\sigma_{m_{ee}}$, leading to the optimal sensitivity representing a trade-off between $\pi^0_D$ background fluctuation and acceptance. Confidence intervals for the number of $A'$ decay candidates in each mass hypothesis are computed from the number of observed data events in the signal region, the number of background events expected from simulation and their uncertainties using the Rolke--Lopez method~\cite{rolke}. The involved numbers of observed and expected events exceed $10^5$ for $m_{A'}<30~{\rm MeV}/c^2$. No statistically significant DP signal is observed, and upper limits at 90\% CL are set.

The limits for the mixing parameter $\varepsilon^2$ are obtained using ${\cal B}(\pi^0\to A'\gamma)/{\cal B}(\pi^0\to\gamma\gamma)=2\varepsilon^2(1-(m_{A'}/m_{\pi^0})^2)^3$ and ${\cal B}(A'\to e^+e^-)=1$ (the latter holds for $m_A'<2m_\mu$ assuming $A'$ decays to SM fermions only)~\cite{batell09}. The upper limits for $\varepsilon^2$ are most stringent at $m_{A'}\approx 20~{\rm MeV}/c^2$, where the acceptance for the considered decay chain is highest reaching 2.5\%, while the kinematic suppression of ${\cal B}(\pi^0\to A'\gamma)$ is small. The obtained preliminary DP exclusion limit, along  with constraints from other experiments, is shown in Fig.~\ref{fig:na48-darkphoton} (right). It represents an improvement over earlier limits in the mass range 10--60~MeV/$c^2$. A combination of this result with the others displayed in Fig.~\ref{fig:na48-darkphoton} rules out the DP as an explanation to anomalous muon $g-2$, assuming DP couples to SM fermions and photons only.

The search for the prompt DP decay is limited by the irreducible $\pi^0_D$ background, with the sensitivity to $\varepsilon^2$ improving as the inverse square root of the integrated beam flux. The NA62 experiment at CERN aiming to surpass NA48/2 by a factor of $\sim 50$ in terms of integrated flux in 2015--2017 can bring only a modest improvement. The improved $m_{ee}$ resolution and the expected downscaling of the $\pi^0$ trigger chain at NA62 should be noted.

\newpage

\subsection{Search for light vector boson production in $e^+e^- \rightarrow \mu^+ \mu^- \gamma$ interactions with the KLOE experiment}
\addtocontents{toc}{\hspace{2cm}{\sl  F.~Curciarello}\par}

\vspace{5mm}
{\sl  F.~Curciarello} \\on behalf of the KLOE-2 Collaboration

\vspace{5mm}

\noindent
Universit\`{a} degli Studi di Messina and INFN Sezione Catania\\

\vspace{5mm}

Many extensions of the Standard Model (SM)~\cite{Holdom,U_th1, Fayet,U_th2,U_th6} predict that dark matter (DM) is made up of non-barionic particles charged under a new kind of interaction usually called "dark force". 
This new interaction should be mediated by a new gauge vector boson, the $U$ boson (also referred to as dark photon or A'), that should be produced during dark matter annihilation processes and have a leptophilic decay channel. Such dark photon is associated to an abelian gauge symmetry that can communicate with the ordinary SM through a kinetic mixing portal providing therefore a small coupling between the $U$  and SM particles~\cite{Holdom,U_th1,U_th2,Fayet,U_th6}.
The coupling strength is expressed by a single factor, $\varepsilon$, equal to the ratio of dark
and standard model electromagnetic couplings~\cite{Holdom}. 
A $U$ boson with mass of $\mathcal{O}$(1GeV) and $\varepsilon$ in the range $10^{-2}--10^{-7}$; could
explain all puzzling effects observed in recent astrophysics experiments~\cite{Pamela,AMS,Integral,Atic,Hess,Fermi,Dama/Libra} and account also for the muon magnetic moment anomaly. For this reason, many efforts have been made in the last years  to find evidence of its existence, with unfortunately null result for the moment~\cite{Mami1,Apex,KLOE_UL1,KLOE_UL2,WASA,HADES,BaBar,a_mu}.

We used a data sample collected in 2002 at DA$\Phi$NE \, \epm\  collider with an integrated luminosity of 239.3 pb$^{-1}$ to investigate the radiative $\ee \to U\gam$, $U\to \mu^+ \mu^-$ process. This channel is considered a very clean and simple channel independent of details of the dark sector and has a reach in sensitivity of $10^{-6}--10^{-4}$, for $U$-boson masses, $M_{\rm U}$,  up to a few GeV~\cite{U_th1,Fayet,U_th2,U_th6}. The $U$-boson peak would appear in the dimuon mass spectrum. 
The $\mu^+ \mu^- \gam$ event selection requires two tracks of opposite charge with $50^\circ\!<\!\theta\!<\!130^\circ$ and an undetected photon whose momentum
points at small polar angle ($\theta\!<\!15^\circ,\ \!>\!165^\circ$)~\cite{KLOE_U,KLOE_pi_FF}.  Pions and muons are identified by means of the variable $M_{\rm trk}$ defined as the mass of particles $x^+,\ x^-$ in the $\epm\to x^+ x^- \gam$ process.  
The $M_{\rm trk}$ values between  80--115 identify muons while $M_{\rm trk}$ values $>$130 MeV identify pions. To improve $\pi/\mu$ separation a cut based on the quality of fitted tracks has been implemented resulting in a suppression of the left tail of $\pi\pi\gam$ \mt\ distribution up to 40\%~\cite{KLOE_U}.
At the end of the analysis  chain, the residual background is obtained by fitting the  $M_{\rm trk}$ data distribution with MC simulations describing signal, \pic\gam \,and \pic\po\ backgrounds plus a distribution obtained from data for the \epm\gam\ ~\cite{KLOE_U,KLOE_pi_FF}.
Finally, we derived the differential cross section $\dif\sigma_{\mu \mu \gamma}/\dif M_{\mu\mu}$ achieving an excellent agreement between the measurement and the PHOKHARA~\cite{PHOKHARA} simulation. No structures are visible in the $M_{\mu\mu}$ spectrum.
To exclude small  $U$-boson signals  we extracted the limit on the number of $U$-boson candidates
through the CLS technique~\cite{CLS_Technique} by comparing expected and observed $\mu^+\mu^- \gam$ yield and a MC generation of the $U$-boson signal which takes into account the $M_{\mu\mu}$ invariant mass resolution (1.5~MeV to 1.8~MeV as $M_{\mu\mu}$ increases). The limit on number of $U$-boson events has been converted in terms of the kinetic mixing parameter $\varepsilon^2$ and results to be
of 1.6~$\times 10^{-5}$ and 8.6 $\times 10^{-7}$ in the 520-980 MeV energy range.
This limit represents the first one derived by a direct study of the $\mu^+\mu^-\gamma$ channel.

\newpage

\subsection{Transition Form Factors - Experimental Overview}
\addtocontents{toc}{\hspace{2cm}{\sl  M.~Mascolo}\par}

\vspace{5mm}
{\sl M.~Mascolo}

\vspace{5mm}

\noindent
Laboratori Nazionali di Frascati dell'INFN, Frascati, Italy\\

\vspace{5mm}

The study of meson Transition Form Factors (TTFs) is motivated by a number of different reasons, such as the calculation of the hadronic Light-by-Light contribution to the Standard Model value of the anomalous magnetic moment of the muon or the search for quark-gluon plasma in heavy-ion collisions \cite{hion}. Moreover, TFFs represent a strong ``benchmark'' for theoretical modeling of different processes, being a field in which high precision measurements are possible. In particular, the most important theoretical advances in modeling the conversion decay of a light vector resonance (V) into a light pseudoscalar meson (P) and a lepton pair ($l^+l^-$), $V \to P \,\, \gamma^{*} \to P \,\, l^+l^-$, were mostly driven by the $\sim\!\!10\,\sigma$ discrepancy between the experimental data of NA60 \cite{na601} and Lepton G \cite{lepg}, and the Vector Meson Dominance (VMD) ansatz \cite{VMD} prediction for the $\omega \to \pi^0 \gamma^{*}$ transition for factor.

In recent years, several theoretical models have been developed to justify this discrepancy \cite{mod1, ivs, mod3}. In this scenario, a measurement of the pion TFF in Dalitz decays of $\phi$ is extremely valuable. In particular, the study of $\phi \to \pi^0 \gamma^* \to \pi^0 e^+e^-$, which was never measured so far, would allow to expand the range of explored $q^2$ (the squared 4-momentum of the virtual photon) to the $\rho$ resonance mass region.

The KLOE experiment \cite{kloe} is involved in the measurement of TFF in Dalitz decays of vector mesons through the study of $\phi \to \eta e^+e^-$ and $\phi \to \pi^0 e^+e^-$ processes. Analyzing the $\phi \to \eta e^+e^-$ decay channel (with $\eta \to \pi^0\pi^0\pi^0$) in the data sample collected at DA$\Phi$NE collider ($\sqrt{s} = m_{\phi}$), during the 2004-2005 data taking campaign, a precise measurement of both the BR($\phi \to \eta e^+e^-$) and the transition form factor slope, $b_{\phi\eta}$, was obtained. 

Measured values are respectively: BR($\phi \to \eta e^+e^-$) = (1.075 $\pm$ 0.007 $\pm$ 0.038) $\times 10^{-4}$, and $b_{\phi\eta}$ = (1.17 $\pm$ 0.10 $^{+0.07}_{-0.11}$) GeV$^{-2}$, both in agreement with VMD predictions. The Branching Ratio is in agreement with SND and CMD-2 results, and the TFF is a factor five better than previous SND measurement.

The analysis of the $\phi \to \pi^0 e^+e^-$ (performed on a data sample of the $\eta$ case) allowed the selection of about 9000 signal candidates, with a good agreement between data and Monte Carlo in all kinematical variables. A deviation of data from the Monte Carlo simulation (including a constant TFF parametrization) is observed at higher values of the $e^+e^-$ mass spectrum. This can be interpreted as the effect of a non-constant form factor playing a role in the decay.

Thanks to the statistics available for data, an improvement of a factor $\sim\!\!10$, with respect to the previous measurement of SND \cite{achasov} and CMD-2 \cite{akhm} experiments, is expected in the statistical error of the Branching Ratio measurement. The $F_{\phi \, \pi^0 \, \gamma^*}$ will be measured for the first time in this kinematical region; this will provide an strong consistency check of all theoretical model describing the TFF of the $\pi^0$ meson.

\newpage

\subsection{Recent experimental Results of Transition Form Factors from Meson Decays}
\addtocontents{toc}{\hspace{2cm}{\sl P.~Adlarson}\par}

\vspace{5mm}

{\sl P.~Adlarson}

\vspace{5mm}

\noindent
Institut f\"ur Kernphysik, Johannes Gutenberg-Universt\"at Mainz, Germany\\

\vspace{5mm}

The transition form factors (TFF) for pseudoscalar mesons describe the electromagnetic structure. The form factor is a scalar function, $F_{P}(q_1^2,q_2^2)$, where $q_{1,2}$ are the four momentum transfers of the virtual photons (for a recent review, see \cite{MesonNet:2012}). Experimentally TFFs can be studied in space- and time-like processes. The time-like TFF can be accessed by studying the single Dalitz decay $P \to l^+l^- \gamma$ which covers the  kinematical region $(2m_l)^2<q^2<m_P^2$. An observable which characterizes the energy dependence of the TFF is the slope parameter, defined as  $ b_{P} = \frac{d ln F(q^2,0)}{dq^2} |_{q^2 = 0 }$, where $q^2$ for the Dalitz decays is equal to the invariant mass squared of the lepton-antilepton pair, $m_{l^{+}l^{-}}^2$, after internal conversion of the virtual photons.

The decay $\eta\rightarrow e^+e^- \gamma$ has recently been re-measured by the A2 collaboration \cite{1Aguar-Bartolome:2013vpw}. The statistics of 2.2$\cdot10^4$ decay events is more than an order of magnitude compared to the previous measurement \cite{Berghauser:2011zz}. The result,  $ b_{\eta} $ = (1.95 $\pm$ 0.15$_{stat}$  $\pm$ 0.10$_{syst}$) GeV$^{-2}$, is in good agreement with the NA60 result for the $\eta$ Dalitz decay ($l = \mu$) \cite{Arnaldi:2009aa}. 

\begin{table}[ht] 
\begin{center}
\begin{tabular}{|l| l|}
\hline
Decay & BR WASA (PRELIMINARY) \\
	\hline \hline
$\eta\rightarrow \pi^+\pi^-\gamma$ & (4.68 $\pm$ 0.07$_{stat}$ $\pm$ 0.19$_{sys}$)$\cdot 10^{-2}$ \\
$\eta\rightarrow e^+e^-\gamma$ & (6.75 $\pm$ 0.06$_{stat}$ $\pm$ 0.29$_{sys}$)$\cdot 10^{-3}$ \\
$\eta\rightarrow \pi^+\pi^- e^+ e^- $ & (2.7 $\pm$ 0.2$_{stat}$ $\pm$ 0.1$_{sys}$)$\cdot 10^{-4}$ \\
$\eta\rightarrow e^+ e^- e^+ e^- $ & (3.2 $\pm$ 0.9$_{stat}$ $\pm$ 0.4$_{sys}$)$\cdot 10^{-5}$ \\
\hline \hline
Decay & BR BESIII \cite{Ablikim:2014nro} \\
	\hline \hline
$J/\Psi \rightarrow \eta^{\prime} e^+e^-$ & (5.81 $\pm$ 0.16$_{stat}$ $\pm$ 0.31$_{sys}$)$\cdot 10^{-5}$ \\
$J/\Psi \rightarrow \eta e^+e^-$ & (1.16 $\pm$ 0.07$_{stat}$ $\pm$ 0.06$_{sys}$)$\cdot 10^{-5}$ \\
$J/\Psi \rightarrow \pi^0e^+e^-$ & (7.56 $\pm$ 1.32$_{stat}$ $\pm$ 0.50$_{sys}$)$\cdot 10^{-7}$ \\
\hline
\end{tabular}
\end{center}
\label{table:etadecays}
\caption{\textit{Preliminary results of branching ratios from WASA-at-COSY (upper box) and final results from BESIII (lower box).}}
\end{table}

Branching ratios of several anomalous $\eta$ decays have been measured by the WASA-at-COSY collaboration. The analyses are based on $3.0\cdot10^{7}$ tagged $\eta $s produced in the reaction $pd \rightarrow  \mbox{}^{3}\mbox{He} \eta$. The branching ratios have been measured relative to $\eta\rightarrow \pi^+\pi^-\pi^0$. Preliminary values of the absolute branching ratios, normalised to the PDG BR($\eta\rightarrow \pi^+\pi^-\pi^0$) \cite{PDG1}, are presented in table 1. The $\eta$ Dalitz decay slope parameter value and branching ratios of the anomalous decays can be compared to recent theoretical calculations based on Pad\'{e} approximants \cite{Escribano:2013kba, GonsalezSoliz:2014} and to a dispersive approach \cite{Hanhart:2013vba}.

Recently, branching ratios for the rare charmonium decays $J/\Psi \rightarrow P e^+e^-$ ($P = \pi^0, \eta, \eta^\prime$) have been measured by the BESIII collaboration for the first time \cite{Ablikim:2014nro}. 

\newpage

\subsection{$\gamma\gamma$ Physics -- Experimental Overview}
\addtocontents{toc}{\hspace{2cm}{\sl C.~F.~Redmer}\par}

\vspace{5mm}

{\sl C.~F.~Redmer}

\vspace{5mm}

\noindent
Institut f\"ur Kernphysik, Johannes Gutenberg-Universt\"at Mainz, Germany\\

\vspace{5mm}

Two-photon physics at $e^+e^-$ colliders refers to measurements of scattering processes, in which photons are emitted 
from each of the leptons. In the fusion of the two photons, states of the quantum numbers $0,2^{\pm+}$ are produced, 
which are inaccessible in the dominating annihilation process. The mass of the produced states is, however, much 
smaller than the available center-of-mass energy.

The production cross section of spin zero particles is depending on the radiative width 
$\Gamma_{X\rightarrow\gamma\gamma}$ and the space like electromagnetic transition form factor (TFF) $F(Q_1^2, Q_2^2)$, 
with $Q_i^2 = -q_i^2$ being the virtualities of the exchanged photons. Thus, measurements of meson production in 
$\gamma\gamma$ reactions provide valuable information on their structure.

The kinematics in two-photon reactions favor small scattering angles of leptons. At experimental facilities, which do 
not feature special tagging detectors to register the scattered leptons, three different classes of measurements are 
feasible.

In so called untagged measurements two quasi-real photons are exchanged in the scattering process. The small 
virtualities correspond to small scattering angles, so that the scattered leptons remain undetected. The 
produced hadronic system is characterized by a vanishing transverse momentum. Recently, the KLOE-2 collaboration has 
used this technique to determine radiative width of the $\eta$ meson~\cite{EtaRW}. The cross section of $\eta$ 
production in $\gamma\gamma$ collisions has been measured from the complementary decay channels 
$\eta\rightarrow\pi^0\pi^0\pi^0$ and $\eta\rightarrow\pi^+\pi^-\pi^0$, making use of both, the drift chamber and 
calorimeter of the KLOE detector~\cite{KLOEH}. The width $\Gamma_{\eta\rightarrow\gamma\gamma}$ extracted from 
the cross section represents the most precise measurement to date.

In the second class of experiments, so called single-tag measurements, one of the scattered leptons is registered in 
the detector. The signature corresponds to processes with one large virtuality and one quasi-real photon. This 
technique allows to study TFFs as a function of a single virtuality, $F(Q^2,0)$. Information on the $Q^2$ dependence of 
TFFs is an important input to theory calculations of the contribution of the hadronic Light-by-Light scattering to the 
anomalous magnetic moment of the muon $(g-2)_\mu$. Available measurements of the $\pi^0$ TFF suffer from low statistics 
or cover regions of large momentum transfer~\cite{pinot}, while regions of low momentum transfer are most relevant for 
$(g-2)_\mu$\cite{PM,LMDV}.Recently, the BESIII collaboration started measurements of TFFs of the pseudoscalar mesons 
$\pi^0, \eta$ and $\eta^\prime$. First results for the $\pi^0$ in the region of momentum transfer $0.3\leq Q^2 
[\textrm{GeV}^2] \leq3.1$ are expected soon~\cite{CFRsfb}. A measurement of the $\pi^+\pi^-$ TFF was started 
recently~\cite{GUOsfb}. It gains additional motivation from the recently proposed dispersive treatment of hadronic 
Light-by-Light scattering~\cite{disp}. In contrast to previous measurement of $\pi^+\pi^-$ production in two-photon 
collisions~\cite{twopi}, BESIII will provide the first measurement in the invariant mass region of $2m_{\pi}\leq 
M_{\pi\pi}[\textrm{GeV}]\leq2.0$ and the momentum transfer region of $0.2\leq Q^2 [\textrm{GeV}^2] \leq2.0$.

Finally, in so called double-tagged measurements both of the scattered leptons are registered in the detector. 
Currently, experimental information on TFFs depending on two virtualities $F(Q_1^2,Q_1^2)$ is not available, due to 
vanishing cross sections for the corresponding events kinematics. First Monte Carlo studies have been started at BESIII. 
The aim is to exploit the large data sets, originally collected for charm physics and charmonium-like spectroscopy, for 
first double-tagged measurements of the $\pi^0$ TFF. The effort is driven by a recent comparison of the TFF in the VMD 
and LMD+V models by Nyffeler, which indicates a difference as large as 25\% for $F(1\textrm{GeV}^2,1\textrm{GeV}^2)$, 
due to a different momentum transfer dependence~\cite{LMDV}.

Dedicated tagging detectors allow for double-tagged measurements in various event kinematics. These kind of detectors 
are to be installed at the BESIII detector and are already available at the KLOE-2 experiment. Here, the taggers 
have been optimized to measure events with two quasi-real photons for a precise determination of 
$\Gamma_{\pi^0\rightarrow\gamma\gamma}$ and events with only one quasi real to determine the $\pi^0$ TFF in the range of 
$Q^2 < 0.1$~GeV~\cite{KLOEtag}. The latter will allow to reduce the model dependence of TFFs for the hadronic 
Light-by-Light scattering even further.

\newpage

\subsection{Theory overview on $P \to e^+e^-$ decays}
\addtocontents{toc}{\hspace{2cm}{\sl  P.~Masjuan}\par}

\vspace{5mm}
{\sl P.~Masjuan}\footnote{Supported by the Deutsche Forschungsgemeinschaft DFG through the Collaborative Research Center ``The Low-Energy Frontier of the Standard Model" (SFB 1044)} {\sl and P. Sanchez-Puertas}

\vspace{5mm}

\noindent
PRISMA Cluster of Excellence, Institut f\"ur Kernphysik, Johannes Gutenberg-Universt\"at, Mainz D-55099, Germany\\

\begin{wrapfigure}{l}{5.5cm}
\includegraphics[width=0.35\textwidth]{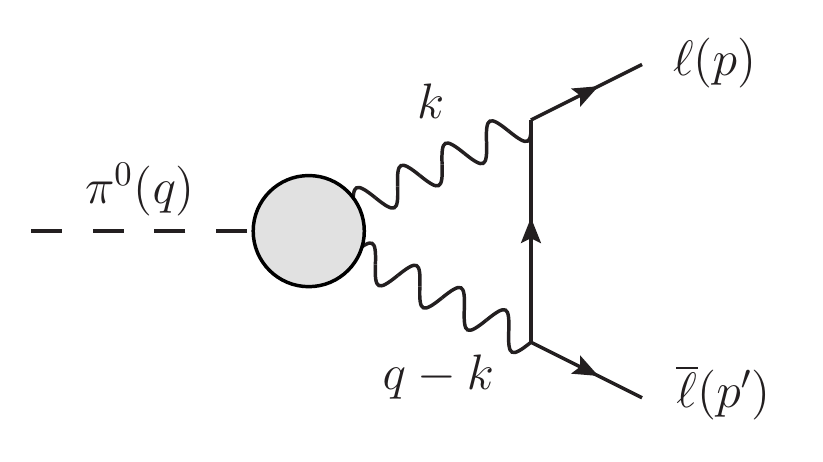}
\caption{\small{Feynman Diagram for $\pi^0 \to e^+e^-$ process.}}
\label{Pi0ee}
\end{wrapfigure}

Pseudoscalar decays into lepton pairs provide a unique environment for testing our knowledge of QCD. As such decays are driven by a loop process, encode, at once, low and high energies. For the $\pi^0$ decay, the process (neglecting electroweak corrections) proceeds in two steps as shown in Fig.~\ref{Pi0ee}. 
The loop does not diverge due to the presence of the pseudoscalar transition form factor on the $\pi^0 \to \gamma^* \gamma^*$ anomalous vertex~\cite{Adler:1969gk}, the $F_{P\gamma^*\gamma^*}(k^2,(q-k)^2)$ with $k^2,(q-k)^2$ space-like photon virtualities.

The most accurate measurement of the $\pi^0 \to e^+e^-$ was done by the KTeV collaboration~\cite{Abouzaid:2006kk2} and yields $\mathrm{BR}(P\to e^+e^-)=(7.48 \pm 0.29 \pm 0.25)\times 10^{-8}$, after radiative corrections~\cite{Bergstrom:1982wk}. Cutcosky rules provide the well-known unitary bound discussed by Drell~\cite{Drell}, $\mathrm{BR}(P\to e^+e^-) \leq \mathrm{BR}^{\mathrm{unitary}}(P\to e^+e^-)=4.69\times 10^{-8}$, which is a model-independent result. 

Due to the presence of the photon propagators, the kernel of the integral is peaked at very low energies of around the electron mass. Then, one can expand that kernel in terms of $m_e/m_P$ but also $m_e/\Lambda$, being $\Lambda$ the cut off of the loop integral, or the hadronic scale driven by the TFF. Recently, the Dubna group~\cite{Dorokhov:2007bd} resummed such power corrections using the Mellin-Barnes technique and found them negligible~\cite{Dorokhov:2009xs}. Then, using a Vector Meson Dominance for the TFF they 
found $\mathrm{BR}(\pi^0\to e^+e^-)=(6.23 \pm 0.09)\times 10^{-8}$~\cite{Dorokhov:2009xs}, $3\sigma$ off the KTeV result.

Prague group~\cite{Vasko:2011pi} reconsidered the radiative corrections used by the KTeV, based on~\cite{Bergstrom:1982wk}, and found that subleading diagrams were important.
Later one, they also studied the role of the soft-photon approximation by Bergstrom, finding it accurate enough~\cite{Husek:2014tna}. 
With such considerations, the new KTeV result is $\mathrm{BR}_{\mathrm{"KTeV"}}(\pi^0\to e^+e^-)=(6.87 \pm 0.36)\times 10^{-8}$.

In Mainz, we are investigating the role of the TFF on such decay~\cite{inprep}. Beyond considering the eventual effects of different New Physics scenarios (which are negligible), we noticed that the factorization approximation for the TFF, i.e., $F(Q_1^2,Q_2^2)=F(Q_1^2,0)\times F(Q_2^2,0)$, con induce large effects. We consider the reconstruction of the TFF of doubly virtuality by a data driven approach. That method, model independent, is based on the theory of Pad\'e approximants~\cite{Baker2} extended to the doubly virtual case. Having experimental data in some energy region, even without high precision ($30\%$ or even $50\%$ statistical error), one can attempt the reconstruction in a systematical way by a sequence of doubly virtual approximants fitted to such data. Such Chisholm approximants~\cite{CM}, can accommodate the high-energy constraints from QCD as well:

\begin{equation}\label{Chisholm}
P^0_1(Q_1^2,Q_2^2)=\frac{a_0}{1+a_1(Q_1^2+Q_2^2)+(2a_1^2-a_{1,1})Q_1^2Q_2^2}\, ,
\end{equation}
\noindent
where Bose symmetry is already implemented. Knowing the Taylor expansion of the $F(Q_1^2,Q_2^2)$, Eq.~(\ref{Chisholm}) would be unique: $a_0$ is determined from the $\Gamma (\pi^0 \to \gamma\gamma)$, $a_1$ is the slope of the single virtual TFF, and $a_{1,1}$ corresponds to the doubly-virtual slope.  

Experimental data for $F(Q_1^2,Q_2^2)$ is not available yet and we cannot extract $a_{1,1}$ from them. The OPE tells as that $\lim_{Q^2 \to \infty} F(Q^2,Q^2) \sim Q^{-2}$ and implies $a_{1,1}=2a_1^2$. 
We consider for the numerical analysis that $0 \leq a_{1,1} \leq 2 a_1^2$, and obtain the new Standard Model value:
\begin{equation}
BR_{\mathrm{SM}}(\pi^0\to e^+e^-)=(6.22 - 6.41) (4)\times 10^{-8}\, ,
\end{equation}
\noindent
where the error comes from $a_0$ and $a_1$ together with the evaluation of the systematic error from our approximation~\cite{inprep}, and the two main numbers from the ranging of $a_{1,1}$. 
To shrink the window here provided, experimental data would then be very welcome. This final number represents still a deviation of the measured BR between $2.6 - 1.4 \sigma$.

Forcing our approximant~(\ref{Chisholm}) to reproduce the KTeV result and then used for the $\pi^0$ contribution to the HLBL~\cite{g2}, we obtain $a_{\mu}^{HLBL;\pi^0}=2.9 \times 10^{-10}$, a deviation of $2- 3 \times 10^{-10}$ with respect to the standard result (see P. Sanchez-Puertas in this proceedings). Taking into account that the global theoretical SM error for the muon $(g-2)$ is $6 \times 10^{-10}$~\cite{g2}, the role of the $\pi^0 \to e^+e^-$ is certainly remarkable, and never been considered so far. Similar effect is also found for the $\eta \to \mu^+ \mu^-$ decay, indicating that the current precision of the SM error on the muon $(g-2)$ is underestimated.

\newpage

\subsection{Search for $\eta^\prime \to e^+e^-$ Decay at CMD-3 
 and Review of $P \to e^+e^-$ Measurements}
\addtocontents{toc}{\hspace{2cm}{\sl S.~Eidelman}\par}

\vspace{5mm}

{\sl S.~Eidelman}

\vspace{5mm}

\noindent
Budker Institute of Nuclear Physics SB RAS and Novosibirsk State University, 
Novosibirsk, Russia \\

\vspace{5mm}

Decays of  pseudoscalar mesons to lepton pairs are of interest 
for various reasons. First, as a relatively simple system they are a
good object to be considered in model calculations and to compare
theoretical predictions with experiment. Second, as it became clear
recently, pseudoscalar transition form factors (TFF) that are measured in 
such decays can be related to the calculations of the hadronic
light-by-light contribution to the muon anomalous magnetic 
moment~\cite{Colangelo:2014dfa1,Colangelo:2014pva1}.

Decays of pseudoscalars, $P \to l^+l^-$, provide one of the 
important ingredients for such calculations since they proceed via two
photons and in the unitarity limit (both photons are real) the branchings
are related~\cite{Landsberg:1986fd}: 
$$
{\cal B}_{P \to l^+l^-} = {\cal B}_{P \to \gamma\gamma}\frac{\alpha^2}{2\beta}
\left(\frac{m_e}{m_P}\right)^2\left[\ln(\frac{1+\beta}{1-\beta})\right]^2,
$$
where $\alpha$ is the fine structure constant, $m_e$ and $m_P$
are masses of electron and meson $P$, respectively,
and $\beta =\sqrt{1-4(\frac{m_e}{m_P})^2}$. This relation is
the lower bound only that can be significantly enhanced, by a factor of 5-10, 
by photon virtuality and the transition form factor~\cite{Landsberg:1986fd}.  
In practice, decays to an electron pair are strongly suppressed with respect
to those to a muon pair because of helicity suppression:
$$
{\cal B}_{P \to e^+e^-}/{\cal B}_{P \to \mu^+\mu^-} \propto 
m^2_e\Phi_{e^+e^-}/m^2_\mu\Phi_{\mu^+\mu^-}=
2.3 \cdot 10^{-5} \Phi_{e^+e^-}/\Phi_{\mu^+\mu^-},
$$
where $m_\mu$ is muon mass and $\Phi_{l^+l^-}$ is the phase space for the 
corresponding decay.

An interesting way to study decays  of any $C$-even
resonance to an electron pair and through them the corresponding 
transition form factors was suggested in Ref.~\cite{Vorobev:1997wb},
where a search for a number of  $C$-even resonances ($\eta^\prime(958)$,
$f_0(980),~a_0(980),~f_2(1270),~a_2(1320)$ and $f_0(1370)$) was performed
with the ND detector in an inverse reaction $e^+e^- \to R$ with a 
subsequent decay of the resonance $R$ into one of the modes convenient 
for detection. No signal was found in any of the final states studied and 
upper limits significantly exceeding the unitarity bounds were placed. 
Later they were improved with the SND detector for the    
$f_2(1270)$ and $a_2(1320)$ mesons~\cite{Achasov:2000zx}.

Recently the CMD-3 Collaboration performed a new search for the process  
$e^+e^- \to \eta^\prime(958) \to \eta\pi^+\pi^-,~\eta \to 2\gamma$  
using an integrated luminosity of 2.69 pb$^{-1}$ collected with the  
CMD-3 detector at the VEPP-2000 $e^+e^-$ collider at the center-of-mass (c.m.) 
energy $\approx m_{\eta^\prime}=957.78\pm0.06 $ 
MeV/$c^2$~\cite{Akhmetshin:2014hxv1}. 
From the absence of the signal they obtain
$$
\Gamma_{\eta^\prime\to e^+e^-}{\cal {B}}_{\eta^\prime\to\pi\pi\eta}
{\cal {B}}_{\eta\to\gamma\gamma}< 0.00041~{\rm eV~at~90\%~C.L.}
$$
and with ${\cal {B}}_{\eta^\prime\to\pi\pi\eta}$ and ${\cal {B}}_{\eta\to\gamma\gamma}$
from PDG:
$$
\Gamma_{\eta^\prime\to e^+e^-}<0.0024~{\rm eV},
$$  
a factor of 25 more stringent than that of ND, but 
still 300 times higher than the unitarity bound.

The table below lists the current status of the searches for
$P \to l^+l^-$ decays. Comparison of the branching fractions measured 
to the corresponding unitarity bounds shows that there might be  
a 5$\sigma$ inconsistency for the $\pi^0 \to e^+e^-$. 

{\small
\begin{table}[h]
{Status of $P \to l^+l^-$ Decay Searches}
\centering
{\small
\begin{tabular}{|l|c|c|c|c|c|}
\hline
Decay mode & ${\cal {B}}_{\rm exp}$ & Events & Group & 
${\cal {B}}_{\rm unit. bound}$ & Ref. \\ 
\hline
$\pi^0 \to e^+e^-$ & $(6.46 \pm 0.33) \cdot 10^{-8}$ & 794  & KTEV, 2007 & 
$4.8 \cdot 10^{-8}$ & ~\cite{Abouzaid:2006kk1}   \\
\hline
$\eta \to e^+e^-$ & $< 2.3 \cdot 10^{-6}$  & -- & HADES, 2014 & 
$1.8 \cdot 10^{-9}$ & \cite{Agakishiev:2013fwl}\\
\hline
$\eta \to \mu^+\mu^-$ & $(5.7 \pm 0.9) \cdot 10^{-6}$  &
 114 & SATURNEII, 1994 & $4.3 \cdot 10^{-6}$ & \cite{Abegg:1994wx}\\
\hline
$\eta^\prime \to e^+e^-$ & $< 1.2 \cdot 10^{-8}$ & -- & CMD-3, 2014 & 
$3.75 \cdot 10^{-11}$ & \cite{Akhmetshin:2014hxv1}\\
\hline
$K^0_L \to e^+e^-$ & $(9^{+6}_{-4}) \cdot 10^{-12}$ & 4 & B871, 1998 & 
$3.0 \cdot 10^{-12}$ & \cite{Ambrose:1998cc}\\
\hline
$K^0_L \to \mu^+\mu^-$ & $(6.84 \pm 0.11) \cdot 10^{-9}$  & 6210 & B871, 2000 & 
$6.8 \cdot 10^{-9}$ & \cite{Ambrose:2000gj}\\
\hline
\end{tabular}}
\end{table}

Further progress with $\pi^0 \to e^+e^-$ decays will most probably be related
to the NA62 experiment at CERN ($K^\pm \to \pi^\pm\pi^0$ decays with the 
expected data sample a few times higher than at KTeV).  
Promising numbers of $\pi^0,~\eta,~\eta^\prime$ can also come from  
hadronic collisions (Crystal Ball at MAMI, Crystal Barrel at ELSA, 
GLUEX and CLAS at JLAB, HADES at GSI). Huge samples of the $\eta$ and
$\eta^\prime$ can be collected in the radiative decays $\phi \to \eta\gamma$
at KLOE2 and $J/\psi \to \eta^\prime\gamma$ at BES-III, respectively. Finally,
the Super-tau-charm factory in Novosibirsk~\cite{Brambilla:2010cs} with a 
luminosity $\sim 10^{34}$~cm$^{-2}$s$^{-1}$ at the c.m. energy of 1 GeV can 
make feasible observation of the process 
$e^+e^- \to \eta^\prime \to \eta\pi^+\pi^-(\eta\pi^0\pi^0)$.

\newpage

\subsection{New weakly-coupled forces hidden in low-energy QCD}
\addtocontents{toc}{\hspace{2cm}{\sl S.~Tulin}\par}

\vspace{5mm}

{\sl S.~Tulin}

\vspace{5mm}

\noindent
York University, Toronto, ON, Canada
\vspace{5mm}

The discovery of a new gauge forces beyond the Standard Model (SM) would revolutionize our understanding of fundamental symmetries and interactions, and may provide a portal into the physics of dark matter.  Recently, there have been numerous experimental searches for new gauge bosons in the MeV -- GeV mass range (see \cite{Essig:2013lka} and references therein).  However, these searches have largely focused on {\it leptonic} signals from gauge boson decays to resonant $e^+ e^-$ or $\mu^+ \mu^-$ pairs.  

What if a new force couples predominantly to quarks over leptons?  Is it possible to detect a new weakly-coupled force within the nonperturbative regime of QCD?  While the idea of ``leptophobic'' forces is not a new one~\cite{Rajpoot:1989jb,Foot:1989ts,Nelson:1989fx,He:1989mi,Carone:1994aa}, this case had been regarded as a particularly challenging blind-spot for experiment, especially in the mass range 100 MeV -- 1 GeV that is the domain of low-energy QCD.  I showed in Ref.~\cite{Tulin:2014tya} that the situation is not as hopeless as it may seem {\it a priori}. In fact, there exist striking signatures that can be searched for studies of rare decays in high luminosity light meson factories.

The model considered herein is a leptophobic gauge boson $B$ that couples to baryon number with the Lagrangian
\begin{equation} \label{eq:Lint}
\mathrm{L} =  \tfrac{1}{3} g_B  \bar q \gamma^\mu q  B_\mu  \, ,
\end{equation}
where $g_B$ is the new gauge coupling and $\alpha_B = \tfrac{g_B^2}{4\pi}$ is the associated baryonic fine structure constant.  This interaction preserves the low-energy symmetries of QCD, namely invariance under charge conjugation ($C$), parity ($P$), and $SU(3)$ flavor symmetry.  In fact, the $B$ boson can be assigned the same quantum numbers, $I^G(J^{PC}) = 0^-(1^{++})$, as the $\omega$ meson.  The $B$ boson does not decay predominantly to $\pi^+ \pi^-$, which is forbidden by $G$-parity, and therefore would not be hidden by a large background for $\rho \to \pi \pi$.  Rather, for $m_\pi \lesssim m_B \lesssim$ 1 GeV, the $B$ boson decays as $B \to \pi^0 \gamma$ or $B \to \pi^+ \pi^- \pi^0$ (when allowed), similar to the $\omega$ meson.

The $B \to \pi^0 \gamma$ decay channel provides a new avenue for discovery.  The $B$ boson can induce to the following decays, for example,
\begin{equation}
\eta \to B \gamma \to \pi^0 \gamma \gamma \, , \qquad \phi \to \eta B \to \eta \pi^0 \gamma \, .
\end{equation}
The branching ratios for these processes to occur are
\begin{subequations} \label{pred}
\begin{align}
\mathcal{B}(\eta \to B \gamma \to \pi^0 \gamma \gamma) &\sim \tfrac{\alpha_B}{\alpha_{\rm em}} \times \mathcal{B}(\eta \to \gamma \gamma)  \, \mathcal{B}(B \to \pi^0 \gamma) \sim 39.4\% \times \tfrac{\alpha_B}{\alpha_{\rm em}} \\
\mathcal{B}(\phi \to \eta B \to \eta \pi^0 \gamma) &\sim \tfrac{\alpha_B}{\alpha_{\rm em}} \times \mathcal{B}(\phi \to \eta \gamma)  \, \mathcal{B}(B \to \pi^0 \gamma) \sim 1.3\% \times \tfrac{\alpha_B}{\alpha_{\rm em}} 
\end{align}\end{subequations}
where $\alpha_{\rm em}$ is the electromagnetic fine structure constant.  These estimates assume that $\mathcal{B}(B \to \pi^0 \gamma) \approx 1$, which is the case for $m_\pi \lesssim m_B \lesssim 620$ MeV~\cite{Tulin:2014tya}.  These channels mimic rare SM processes, which have the following observed branching ratios~\cite{Beringer:1900zz}
\begin{subequations} \label{obs}
\begin{align}
\mathcal{B}(\eta \to \pi^0 \gamma\gamma ) &= (2.7 \pm 0.5) \times 10^{-4} \\
\mathcal{B}(\phi \to \eta \pi^0 \gamma) &= (7.27 \pm 0.30) \times 10^{-5} \, .
\end{align}
\end{subequations}
Requiring that the $B$ boson contributions \eqref{pred} do not saturate the total observed rates \eqref{obs} for these channels, we must have $\alpha_B \lesssim 2 \times 10^{-5}$ and $\alpha_B \lesssim 5 \times 10^{-5}$, respectively.  It is remarkable that a new force for quarks in nonperturbative QCD regime is constrained to be $\sim 10^3$ times weaker than electromagnetism and $\sim 10^5$ times weaker than the strong force!

A much greater sensitivity can be obtained by searching in these channels for a $\pi^0 \gamma$ resonance, which would reconstruct $m_B$.  This is a new type of signal that has never been searched for previously.  But it can be easily included among the research goals for future meson facilities that are targeting $\eta \to \pi^0 \gamma\gamma$ and $\phi \to \eta \pi^0 \gamma$ for QCD-related studies~\cite{AmelinoCamelia:2010me,JEFpro}.  Preliminary simulations as part of the proposed Jefferson Eta Factory indicate that the $B$ boson sensitivity may be boosted by $\sim 2$ orders of magnitude through a $\pi^0 \gamma$ resonance search in $\eta \to \pi^0 \gamma \gamma$~\cite{JEFpro}.  

Lastly, it is important to consider the prospects for distinguishing the $B$ boson from the dark photon $A^\prime$ in the event of a discovery.  For $m_B \gtrsim m_\pi$, the smoking gun signature of the $B$ boson is observing the $\pi^0\gamma$ resonance.  However, for $m_B \lesssim m_\pi$, the leading decay is expected to be $B \to e^+ e^-$, which is identical to $A^\prime$.  Precision studies of the $\omega$ meson can provide a key diagnostic test.  The decay $\omega \to \pi^0 A^\prime$ can occur, while $\omega \to \pi^0 B$ is highly suppressed, being forbidden by isospin.  Therefore, the observation of an $e^+ e^-$ resonance in $\omega \to \pi^0 e^+ e^-$ would favor the $A^\prime$, while the nonobservation of this signal would favor the $B$ boson.


I wish to thank the organizers of the MesonNet 2014 meeting for their interest and hospitality.  This work was supported from the DOE under contract de-sc0007859 and NASA Astrophysics Theory Grant NNX11AI17G.

\newpage

\subsection{Hadron Properties from the Dyson-Schwinger Approach}
\addtocontents{toc}{\hspace{2cm}{\sl G.~Eichmann}\par}

\vspace{5mm}

{\sl  G.~Eichmann}

\vspace{5mm}

\noindent
Institut f\"ur Theoretische Physik, Justus-Liebig--Universit\"at Giessen, Germany\\

\vspace{5mm}
This contribution aims to highlight progress in the Dyson-Schwinger approach, where hadron masses, form factors, scattering amplitudes etc.
are calculated directly from their quark-gluon substructure. Dyson-Schwinger equations (DSEs) are the quantum equations of motion for QCD's $n$-point functions~\cite{DSEs}. In practice they are subject to truncations, which amounts to choosing the minimal model input for a given $n$-point function (e.g., the quark-gluon vertex) that allows one to calculate all remaining building blocks of hadronic matrix elements consistently: the dressed quark propagator and quark-photon vertex, the Bethe-Salpeter amplitudes of mesons, and the Faddeev amplitudes of baryons~\cite{Eichmann:2013afa}.

Maintaining this consistency is important for several reasons. Chiral symmetry and its dynamical breaking are preserved model-independently, i.e., the pion is both a $q\bar{q}$ bound state and the massless Goldstone boson in the chiral limit~\cite{Maris:1997hd}. Similarly, the requirements of electromagnetic gauge invariance are satisfied at the vertex level, which produces hadronic current matrix elements that are conserved automatically~\cite{gauging-of-eqs}. Another consequence of dynamical chiral symmetry breaking is the generation of a large 'constituent'-quark mass at low momenta. This is a natural outcome of the quark DSE if the gluonic ingredients in the equation exceed a certain strength, but the effect is also well-studied in lattice QCD~\cite{Bowman:2005vx}. While the quark mass function is not observable by itself, it transpires to the hadron level where it contributes, for example, the bulk of the proton mass. Finally, it is the very idea of Bethe-Salpeter and Faddeev equations to extract hadron properties from the pole structure of QCD's Green functions. These poles will appear wherever higher $n$-point functions are present; for example, vector-meson poles are retrieved in the dynamical solution for the quark-photon vertex~\cite{Maris:1999bh}, in the time-like structure of electromagnetic form factors, etc. The underlying origin of 'vector-meson dominance' is therefore quite transparent~\cite{Eichmann:2014qva}.
   
The idea is then to test the response of hadron properties to different truncations and interactions, for example: is it two-body or three-body forces that dominate the baryon excitation spectrum? What is the impact of pion-cloud effects in the structure of form factors and scattering amplitudes? Can one understand the various contributions to nucleon Compton scattering (handbag diagrams at the quark level, meson poles and nucleon resonances at the hadron level) from the underlying microscopic point of view? 
  
Much progress has been made in the meson sector over the last decade in computing meson masses, decay constants, form factors, and valence-quark distributions~\cite{Mesons}. A calculation of the $\pi\gamma\gamma$ transition form factor is also available~\cite{Maris:2002mz}. Results for baryons have come from quark-diquark models~\cite{qdq} but more recently also from the covariant three-quark Faddeev equation~\cite{Faddeev}. Nucleon and $\Delta$ electromagnetic~\cite{emffs}, $N\to\Delta\gamma$ transition~\cite{Eichmann:2011aa}, nucleon axial and pseudoscalar form factors~\cite{Eichmann:2011pv} and more have been calculated in these setups. Except for missing pion-cloud effects at low $Q^2$ and low pion masses, the space-like behavior of form factors is usually described quite well by these calculations. Among others, the present efforts aim to improve truncations~\cite{bRL}, and to study baryon excitations and pion cloud effects~\cite{exc}, tetraquarks~\cite{Heupel:2012ua}, nucleon Compton scattering~\cite{compton}, and the muon $g-2$~\cite{Goecke:2012qm}.

\newpage

\subsection{Photoproduction of Mesons of Quasifree Nucleons - selected results}
\addtocontents{toc}{\hspace{2cm}{\sl  B.~Krusche}\par}

\vspace{5mm}

{\sl B.~Krusche}

\vspace{5mm}

\noindent
Departement f\"ur Physik, Universit\"at Basel\\

\vspace{2mm}

Photoproduction of mesons off the proton has developed 
to the most important tool for the study of the excitation spectrum of the nucleon.
A huge effort has been made at several electron accelerators
(ELSA, ESRF, Jlab, MAMI) to study not only cross sections but also single
and double polarization observables for many different final states. Although 
the largest fraction of these data are still under analysis, first impact becomes
visible. The 2012 Review of Particle Physics (PDG) \cite{PDG_12} included for the first 
time nucleon resonances for which the main experimental evidence came from photon 
induced reactions.    

However, due to the isospin dependence of the electromagnetic interaction, one must study
these reactions also with a neutron target. This is the only access 
to the isospin decomposition of the electromagnetic excitation amplitudes. Furthermore,
due to selection rules (from SU(3) symmetry) the excitation of certain states
is strongly suppressed for the proton but not for the neutron. 
The data base for this type of reactions is still much less complete. Statistical as well
as systematic quality of the few existing data are much inferior to the corresponding
reactions off free protons. The reason is of course that only neutrons bound in light
nuclei (mostly the deuteron) can serve as targets. This complicates the experiments
since it requires coincident detection of the recoil neutrons, which typically results
at least in a reduction of counting statistics by a factor of $\approx$3 due to the 
neutron detection efficiency. For the interpretation of the data one must in addition 
take into account the Fermi motion of the bound nucleons and possible final state interaction 
(FSI) effects. 

However, progress during the last few years on this subject was large 
(see \cite{Krusche_11} for a summary). Currently, significant efforts are under way at
the MAMI and ELSA accelerators to study photoproduction of mesons off quasi-free
nucleons. In these modern measurements effects from Fermi motion are controlled by
a complete kinematical reconstruction of the final state and FSI effects are 
systematically studied by a comparison of the results for bound, quasi-free protons 
to the same reactions off free protons. The experiments set up at ELSA 
(Crystal Barrel/TAPS \cite{Aker_92,Gabler_94}) and MAMI (Crystal Ball/TAPS 
\cite{Starostin_01,Gabler_94}) are ideally suited for this program since the
almost $4\pi$ covering electromagnetic calorimeters allow the coincident detection
of photons, recoil protons and neutrons, and partly also charged pions. This means 
that even final states like for example $\pi^0\pi^0 n$, $\pi^0\eta n$ are accessible. 

In the following we will shortly summarize the recently published results and ongoing
data analysis. Total cross sections and angular distributions for single $\pi^0$ production 
off the deuteron throughout the second and third nucleon resonance region were measured 
at MAMI \cite{Dieterle_14}. The results clearly deviated with {\it all} existing 
predictions for the partial waves of this reaction (based on the analysis of the other 
three isospin channels). Large effects were for example found for the $P_{11}$ wave.  

Photoproduction off $\eta$ mesons off neutrons was studied with several experiments
at ELSA and at MAMI using deuterium \cite{Jaegle_08,Jaegle_11,Werthmueller_13,Werthmueller_14}
and $^3$He \cite{Werthmueller_13,Witthauer_13} targets. Special interest in this reaction
arose because the excitation function for the neutron shows a pronounced, narrow bump
around $W$=1680 MeV, which is absent for the proton. The very precise angular distributions
measured at MAMI \cite{Werthmueller_14} can be best reproduced by intricate interference
effects in the $S_{11}$ partial wave, which require a change of the sign of the 
$A_{1/2}^n$ helicity coupling with respect to the value currently listed by PDG.
Data for the polarization observables $E$, $T$, and $F$ from the MAMI facility are
under analysis (for $E$ also from ELSA). First, very preliminary results for $E$
seem to indicate that the bump in the neutron excitation function is indeed related to
the $J=1/2$ partial wave. Total cross sections and angular distributions have been also
measured for $\eta '$ mesons (ELSA) \cite{Jaegle_11a}. Furthermore, in connection with 
the search for $\eta$-mesic nuclei, coherent $\eta$ production was studied for $^3$He \cite{Pheron_12} 
and $^7$Li nuclei \cite{Maghrbi_13}. A detailed summary of all experimental results for
$\eta$ and $\eta '$ production (also for the proton and heavy target nuclei) is given in
\cite{Krusche_14}. 

For the photoproduction of pion pairs \cite{Oberle_13,Oberle_14} 
($\pi^0\pi^0 p$, $\pi^0\pi^0 n$, $\pi^0\pi^+ n$, $\pi^0\pi^- p$ final states) data for the 
beam-helicity asymmetry (circularly polarized beams, unpolarized targets) 
have revealed surprisingly large discrepancies with model predictions. An investigation
of all four isospin channels for the production of $\pi\eta$-pairs has confirmed a strong
dominance of the $\Delta\eta$ decay of the $D_{33}$(1700) resonance in the threshold region.
Analysis of the $E$, $T$, $F$ observables for all these double meson production channels
is still under way.

\newpage

\subsection{Recent Photoproduction Results from the CBELSA/TAPS Experiment off the Proton}
\addtocontents{toc}{\hspace{2cm}{\sl A.~Wilson}\par}

\vspace{5mm}

{\sl A.~Wilson} \\for the CBELSA/TAPS Collaboration

\vspace{5mm}

\noindent
Helmholtz-Institut f\"{u}r Strahlen- und Kernphysik,\\ Universt\"{a}t Bonn, Bonn, Germany\\
\vspace{5mm}

Essential information about the non-perturbative regime of Quantum Chromodynamics~(QCD) and its bound states can be obtained by studying the pattern of states in the excited baryon spectrum. However when comparing to predictions such as~\cite{Loring:2001kx,Edwards:2011jj}, many more higher mass states are predicted than are currently listed in the PDG~\cite{Agashe:2014kda}.  Because these states tend to be challenging to identify, experiments must expand the possibilities for detection and very carefully and unambiguously extract these states and their properties from the data. The current strategy to increase detection is to use the electromagnetic interaction and photon beams to excite the nucleon into its excited states~(photoproduction) and measure a multitude of final states. Almost all of the experimentally known states were found in pion-production experiments which feature an exclusive dependence on resonances coupling to $\pi N$. By photoproducing these states, resonances which have a small coupling to $\pi N$ have a chance to be produced and identified. However because these strongly decaying resonances are broad and overlapping, analyzing the photoproduction data requires the use of decomposition analysis methods such as a partial wave analysis~(PWA). Therefore to unambiguously extract the partial waves and consequently resonance quantum numbers from the data, polarization observables must be measured~\cite{Chiang:1996em}. To extract the most precise information about the resonance spectrum, cross sections and polarization observables from all the available reactions must be simultaneously analyzed. 

At the CBELSA/TAPS experiment, the measurement of single and double polarization observables in photoproduction for a variety of final states are continuing to advance our knowledge of the baryon spectrum.  The experimental setup features polarized beam photons and polarized target protons both of which can be polarized in several orientations. Linear or circular beam polarization is obtained by scattering high energy electrons, up to 3.2~GeV, from the Electron Accelerator ELSA in Bonn, Germany~\cite{Hillert:2006yb} off of a bremsstrahlung radiator. Target protons are longitudinally or transversely polarized using the Bonn Frozen Spin Target~\cite{Dutz:2004zz}. The heart of the detector systems are the electromagnetic calorimeters which have been optimized to detect mesons decaying to photons. For more information on the CBELSA/TAPS experiment, see \cite{Thiel:2012yj,Gottschall:2013uha,Hartmann:2014mya}. 

\begin{table}[ht]
\centering
\begin{small}
\begin{tabular}{|c|c|ccc|ccc|cccc|}

\hline 
Photon & &\multicolumn{3}{c|}{Target } & \multicolumn{3}{c|}{Recoil Nucleon }  & \multicolumn{4}{c|}{Target and Recoil } \\ 
Polarization& &\multicolumn{3}{c|}{Polarization } & \multicolumn{3}{c|}{Polarization }  &\multicolumn{4}{c|}{Polarization }  \\ [1mm]
 &  &  X & Y & Z &  &  &  & 	X & 	X & 	Z & Z\\ 
  &  &   &  &  & X' & Y' & Z' & X' & Z' & X' & Z'\\
\hline 
unpolarized & $\sigma$ & - & $T$ & - & - &$P$& - & 	$T_x$ & $T_z$  & $L_x$  &  $L_z$ \\ [1mm]

linear & $\Sigma$ & $H$ & (-$P$) & $G$ & $O_x$ & (-T) & $O_z$ & (-$L_z$) & (-$L_x$)&(-$T_z$) & (-$T_x$)  \\ [1mm]

circular & - & $F$& - & $E$ & $C_x$& - & $C_z$ & - & - & -  &- \\ 
\hline 
\end{tabular} 
\end{small}
\caption{Polarization observables for the photoproduction of one pseudoscalar meson, e.g. \prodppi}
\label{table:polobs}
\end{table}

For the reaction \prodppi, the data measured for polarization observables $G$, $E$, $T$, $P$, and $H$~\cite{Thiel:2012yj,Gottschall:2013uha,Hartmann:2014mya} (defined in Table~\ref{table:polobs}) show how the inclusion of double polarization observables can enhance the understanding of even resonances which are known from pion-production experiments. These data are driving the \prodppi~PWA solutions to become more unique and at the same time increasing our knowledge of the properties of these resonances.  

If a resonance couples weakly to $\pi N$ but strongly to $\eta N$, this resonance should be detected for the first time in the reaction \prodpeta. The polarization observables $E$, $G$, $T$, $P$, and $H$ have been measured and have been included in the Bonn-Gatchina Partial Wave Analysis, which simultaneously analyzes the data from a variety of final states~\cite{Anisovich:2011fc1}. These observables are providing the basis for the most precise information on resonance $p\eta$ decays.

Other photoproduction final states such as \ppipi~and \pw~are also being studied. The number of polarization observables required for unambiguous decomposition of these final states increases. However the power of considering this information can be shown in the comparison of PWA predictions to preliminary data on the polarization observable $T$ for \prodppipi; the comparison shows the significant effect polarization observables will have on PWA solutions. The \pw~final state differential cross sections, double polarization observable $E$, and spin-density matrix elements~(defined for the radiative decay in \cite{Zhao:2005vh}) have been measured and included in the Bonn-Gatchina PWA.  A preliminary description indicates the resonant behavior near threshold is due to the production of the $N(1700)3/2^+$ resonance and at higher energies is dominated by pomeron-exchange along with spin-parity $1/2^-$, $3/2^-$, and $5/2^+$ resonances being produced. 

The future for the CBELSA/TAPS experiment involves the continued measurement and analysis of polarization observables for many photoproduction final states. Currently, analyses are underway for the \ppi, \peta, $p\eta\prime$, \ppipi, and \pw~final states. The detector systems are currently being upgraded to increase the triggering rate to measure even more final states such as off the neutron. These new data will continue to enhance our understanding of the baryon spectrum and in turn the nonperturbative regime of QCD.

Supported by the {\it Deutsche Forschungsemeinschaft} (SFB/TR16) and {\it Schweizerischer Nationalfonds}.

\newpage

\subsection{Measurement of the beam asymmetry $\Sigma_{B}$ and the double polarization observable G in the reaction $\gamma p \rightarrow p \pi^0 \pi^0$}
\addtocontents{toc}{\hspace{2cm}{\sl K.~Spieker}\par}

\vspace{5mm}

{\sl K.~Spieker}\\ for the CBELSA/TAPS collaboration

\vspace{5mm}

\noindent
Helmholtz-Institut f\"ur Strahlen- und Kernphysik, University of Bonn, Germany
\vspace{5mm}

The excitation spectrum of baryons consists of many resonances which contribute selectively to distinct decay channels. To obtain information about the contributing resonances, Partial Wave Analyses (PWA) are performed to identify the resonances and characterize their properties. For an unambiguous partial wave analysis solution, the measurement of several well chosen single and double polarization observables is needed in different decay channels \cite{test}.\\ Compared to charged multi-pion-photoproduction, the double~$\pi^0$-photoproduction is highly sensitive to nucleon resonances since background processes like direct $\Delta \pi$  production, pion-exchange in the t-channel and low energy resonance born terms are strongly suppressed. Furthermore, the background contains no $\rho$(770) contribution due to isospin conservation. Consequently, the double~$\pi^0$ final state is perfectly suited for the study of nucleon resonances via polarization observables. Two of these observables are the beam asymmetry $\Sigma$ and the double polarization observable G which can be determined by using linearly polarized photons in combination with a longitudinally polarized target. Analogously to the single pseudoscalar meson photoproduction and if a pure polarized proton target would be available, the differential cross section in a quasi two-body approach is given by \cite{barker}:
\vspace*{-0.2cm}
\begin{small}
\begin{equation}
\frac{d\sigma}{d\Omega}=\left.\frac{d\sigma}{d\Omega}\right|_0\cdot \bigg(1-\delta_l \Sigma \cos 2(\varphi)+\delta_l \Lambda_z G \sin 2(\varphi)\bigg)
\label{diffcross1}
\end{equation}
\end{small}
\vspace*{-0.2cm}\\
\noindent where $ \left.\frac{d\sigma}{d\Omega}\right|_0$ indicates the unpolarized differential cross section,  $\delta_l$ ($\Lambda_z$) the polarization degree of the beam photons (the longitudinally polarized target) and $\varphi$ the angle between the beam photon plane and the reaction plane. \\
The polarization observables are measured at the~CBELSA/TAPS experiment at ELSA in Bonn. The linearly polarized photons  ($\delta_l \approx 60$\%) are created via bremsstrahlung of unpolarized electrons on a diamond crystal. The bremsstrahlung's photons collide with a longitudinally polarized frozen-spin butanol target  (C$_4$H$_9$OH) with a mean polarization of about 70\%. The setup covers nearly the full 4$\pi$ of the solid angle and has a high detection efficiency for final states with neutral mesons.\\
The reaction $\gamma p \to p \pi^0 \pi^0 \to p 4 \gamma$  was reconstructed by applying several cuts. Treating the proton as a missing particle, the missing mass can be calculated and has to be in the range (938~$\pm$~67)~MeV. Since the beam photons travel in z-direction towards the target, the transverse component of the total momentum has to be zero. Therefore, the difference between the azimuthal angle of the proton and the 2$\pi^0$-system was asked to be compatible with~(180~$\pm$~9.5)$^\circ$. As a next step, the difference between the reconstructed and calculated proton polar angle should be located in~(0~$\pm$~10)$^\circ$. Lastly the invariant mass of each $2\gamma$ system has to be compatible with~(135~$\pm$~20)~MeV. With all introduced cuts, 5.45 $\times$ 10$^5$ 2$\pi^0$ events have been selected and are used for the determination of the two observables ($\Sigma_B$,G). However, the selected data comprises events on bound and quasi-free protons. Since the double polarization observable G requires only the reaction initiated by quasi-free polarized protons, the dilution factor D has to be determined.

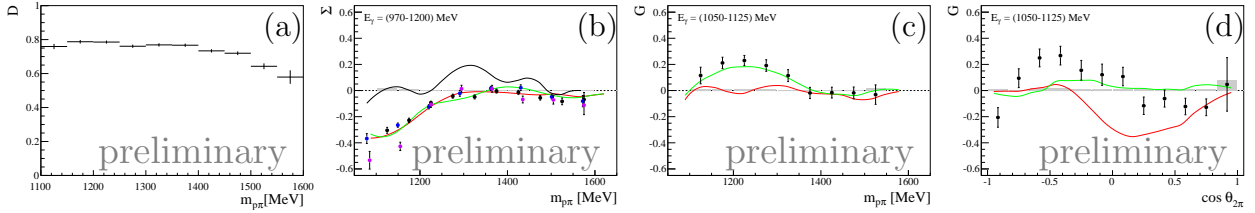
\begin{figure}[htb]
\hspace*{-0.5cm}
      \begin{overpic}[width=0.27\textwidth]{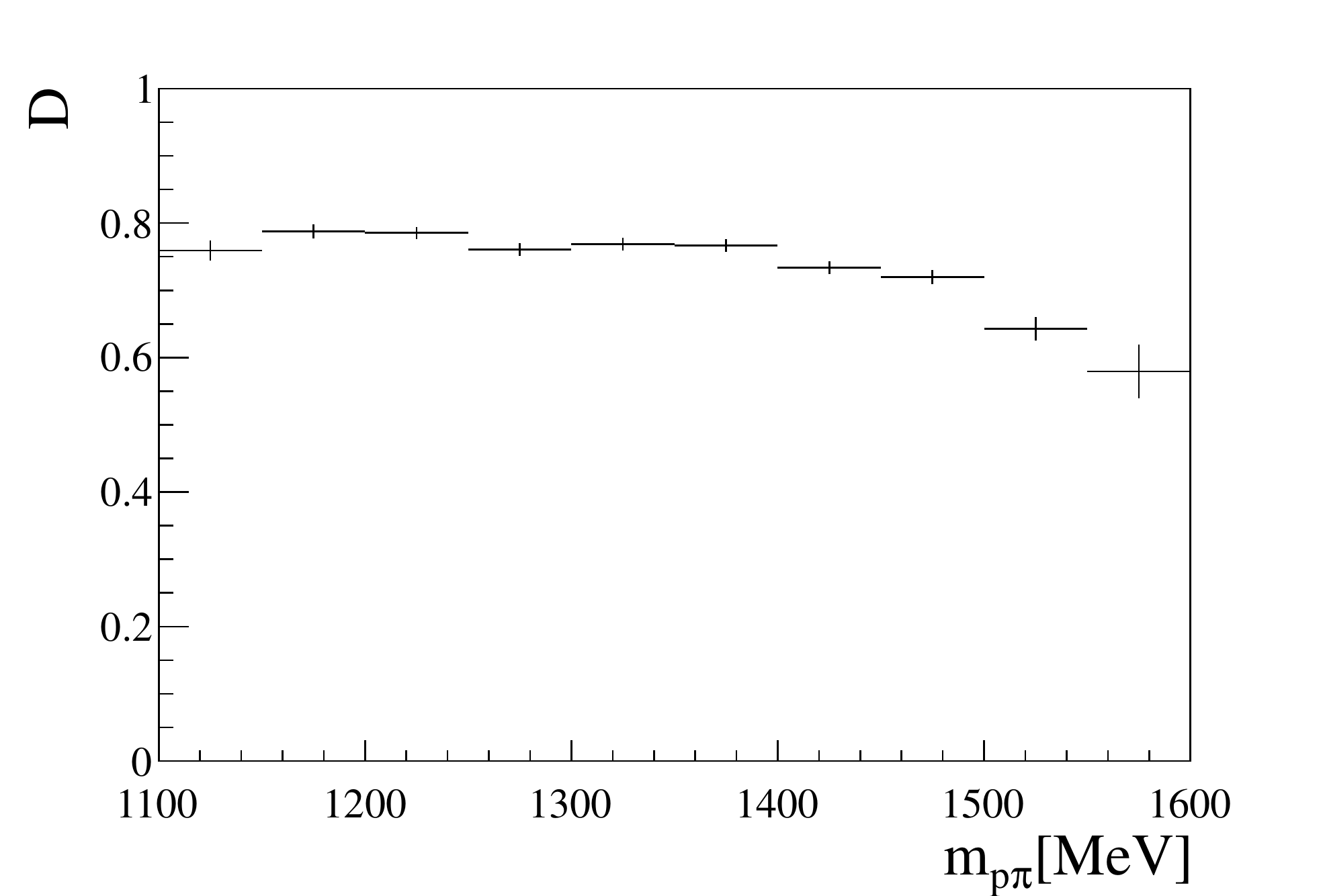}
      \put(78,52){(a)}
             \put(29,14){\large{\textcolor{gray}{preliminary}}}

      \end{overpic}
    \hspace*{-0.5cm}
      \begin{overpic}[width=0.27\textwidth]{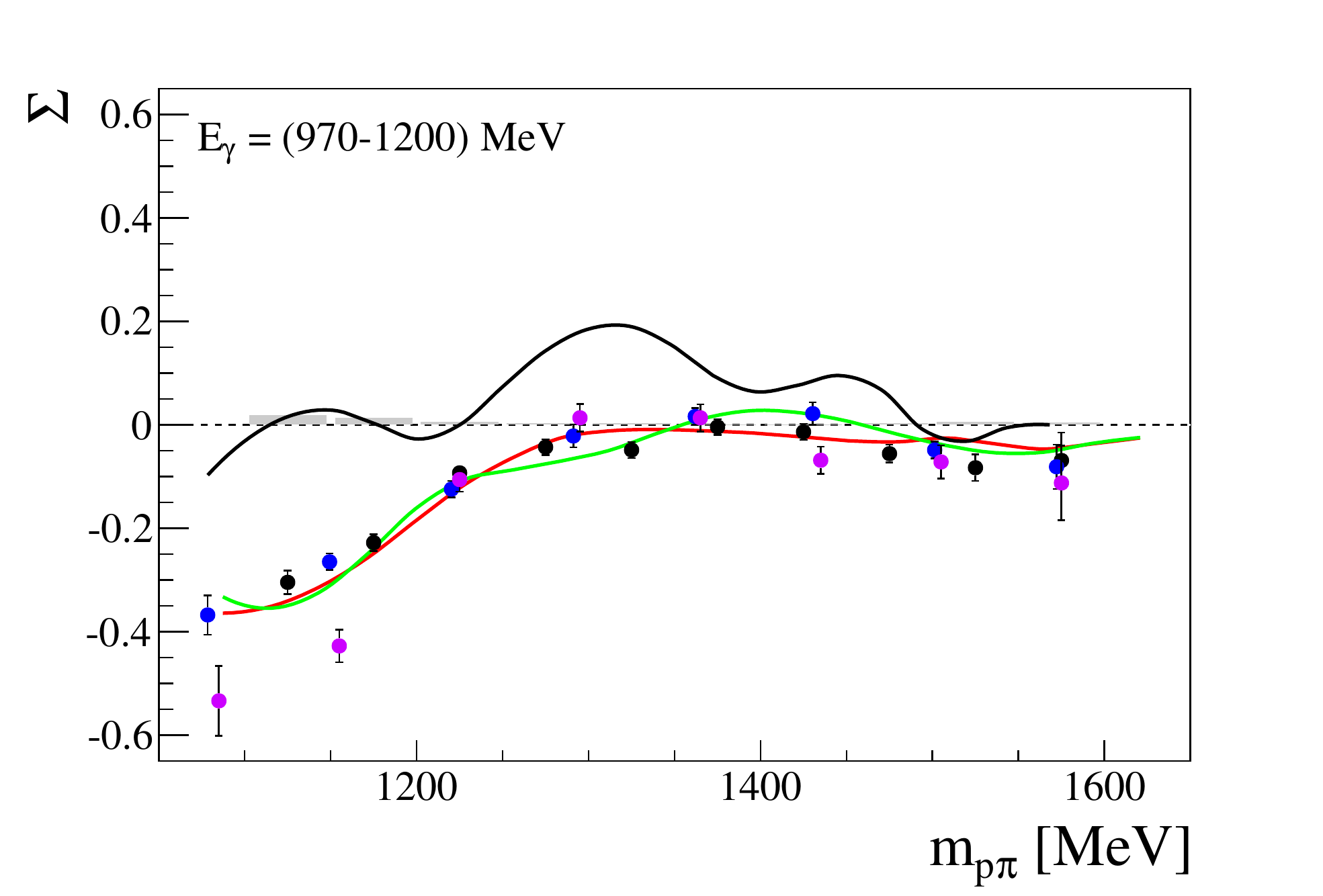}
      \put(78,52){(b)}
             \put(29,14){\large{\textcolor{gray}{preliminary}}}

      \end{overpic}
\hspace*{-0.5cm}
      \begin{overpic}[width=0.27\textwidth]{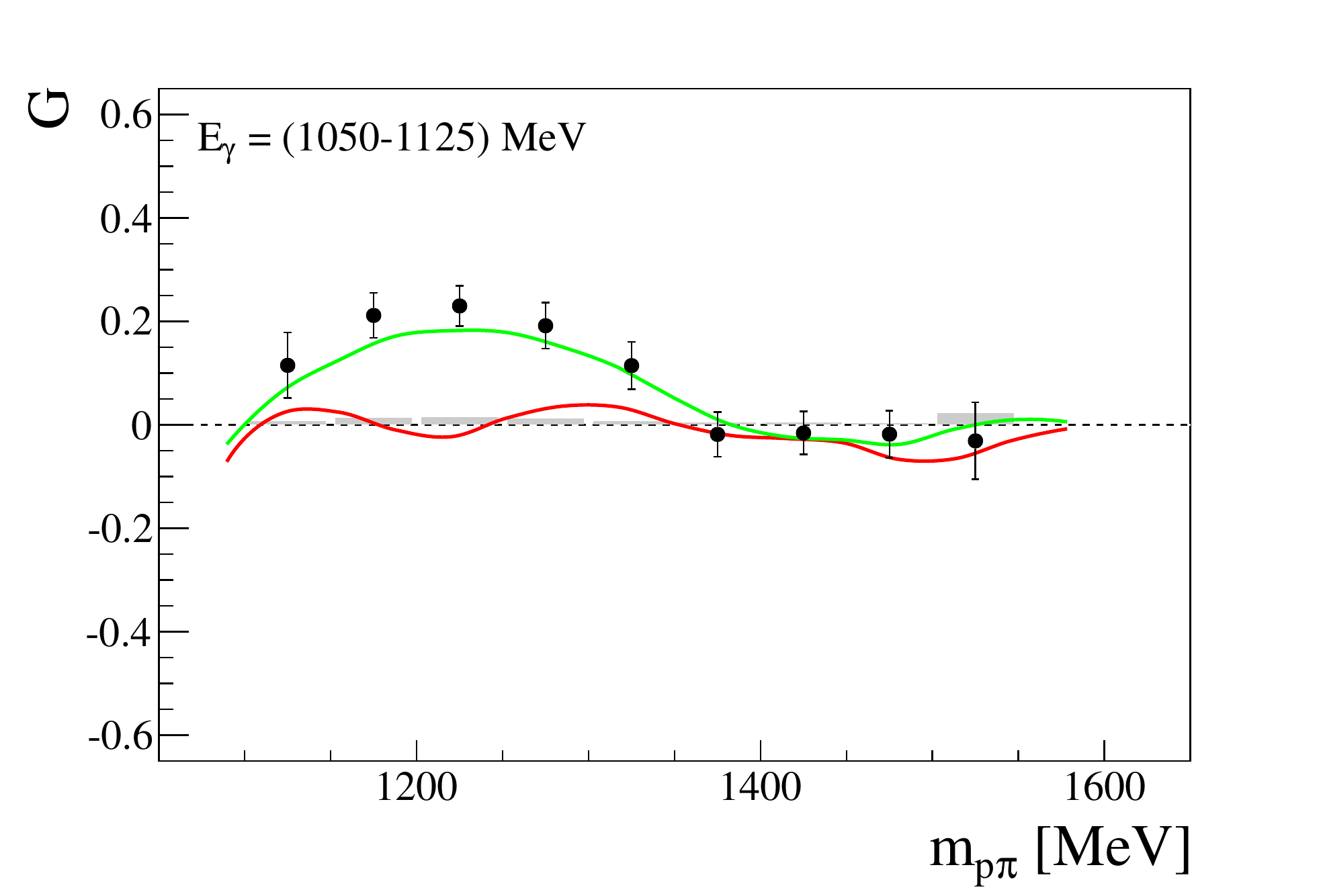}
      \put(78,52){(c)}

       \put(29,14){\large{\textcolor{gray}{preliminary}}}
      \end{overpic}
\hspace*{-0.5cm}
      \begin{overpic}[width=0.27\textwidth]{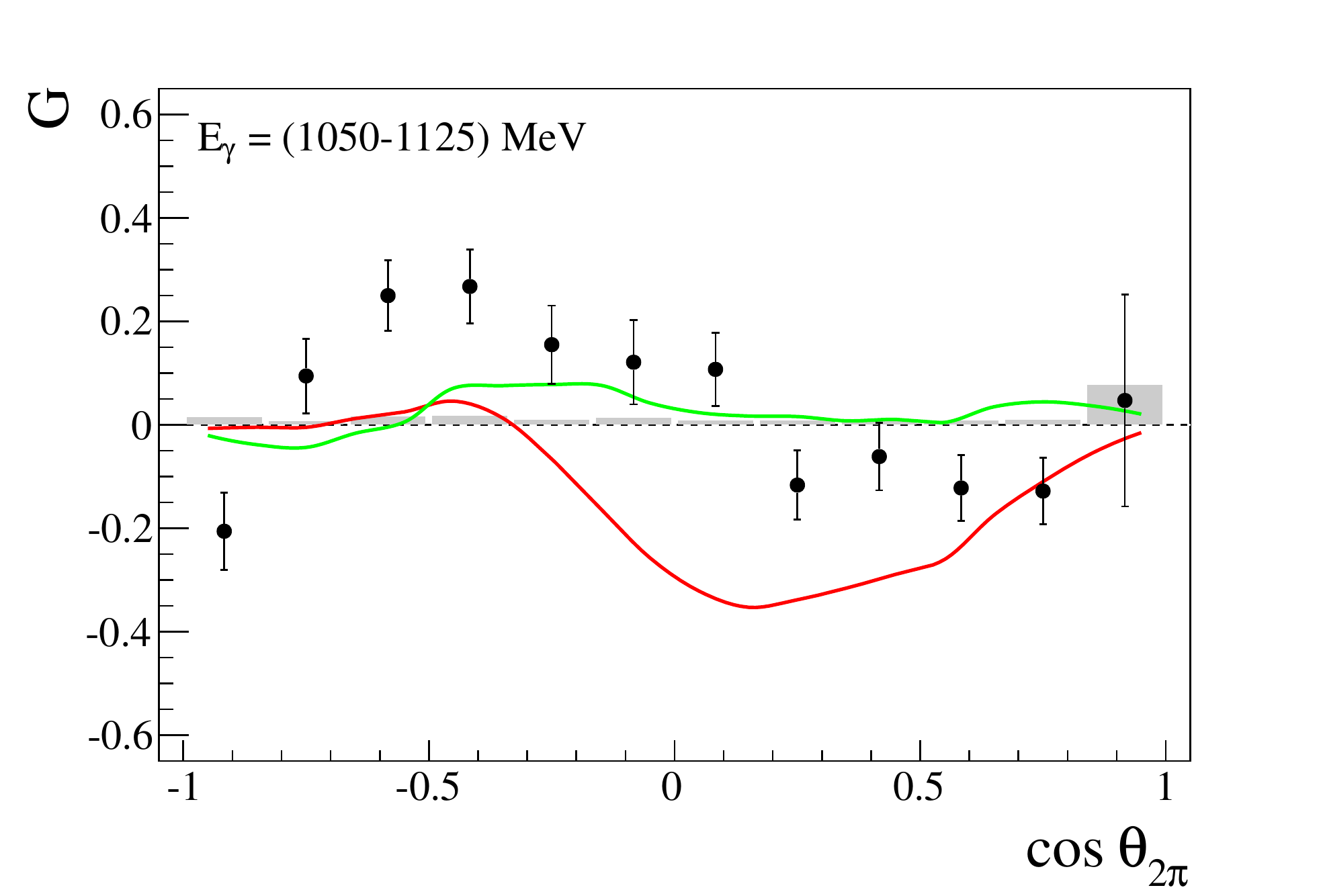}
            \put(78,52){(d)}

       \put(29,14){\large{\textcolor{gray}{preliminary}}}

	   \end{overpic}
\caption{Panel (a): Dilution factor D for $E_\gamma$ = 970--1200~MeV as a function of the kinematic variable $m_{p\pi}$. Panel (b): Beam asymmetry $\Sigma_{B}$ as a function of the kinematic variable $m_{p\pi}$. Black dots: this work, blue dots: GRAAL data \cite{graal}, violet dots: CBELSA/TAPS data \cite{vahe}, black curve: 2$\pi^0$-MAID \cite{maid}, green curve: BG2011-01 and red curve: BG2011-02 \cite{bg1,bg2}. Panel (c) and (d): Double polarization observable G as a function of the kinematic variables $m_{p\pi}$ and ${\cos \theta_{2\pi}}$.}
\vspace*{-0.2cm}
\label{fig:allgemein}
\end{figure} 
 The dilution factor indicates the amount of unbound, polarizable protons in the selected data of the butanol target (see Fig.\ref{fig:allgemein}.a).\\
The preliminary results for the observables in the reaction $\gamma p \rightarrow p ~ \pi^0 \pi^0$ have been determined in an energy range  $E_\gamma$~=~600--1275~MeV and as a function of several kinematic variables. The results for the beam asymmetry $\Sigma_B$ of the butanol target are compared with the predictions of the MAID \cite{maid} and the BnGA PWA \cite{bg1,bg2}, whereas the double polarization observable G is only compared with the BnGa PWA. The BnGa PWA provides two different solutions, namely BG2011-01 and BG2011-02. \\
The beam asymmetry $\Sigma_B$ in the energy range $E_\gamma$~=~970--1200~MeV shows similar shapes when compared to the GRAAL \cite{graal} and the recent CBELSA/TAPS data \cite{vahe}. However small deviations are also visible (see Fig.\ref{fig:allgemein}.b). They might occur due to the different acceptance and phase space coverage of the two experimental setups. In addition, the GRAAL and CBELSA/TAPS results are measured only on unbound protons whereas in the results of this work contributions of bound protons are present (cf. Fig.\ref{fig:allgemein}.a). Both BnGa solutions describe the data reasonably well while MAID does not reproduce both, the absolute value and the overall evolution of the observable. \\
For the first time, the double polarization observable G for the double $\pi^0$ photoproduction off the proton has been measured with the CBELSA/TAPS experiment. The extracted observable shows several accordances with the BnGa solutions but also significant
deviations (see Fig.\ref{fig:allgemein}.c,d). The observable G is therefore a further step to constrain different PWA solutions in the double $\pi^0$ photoproduction off the proton.

\newpage

\subsection{CMD-3 and SND Results at VEPP-2000}
\addtocontents{toc}{\hspace{2cm}{\sl S.~Eidelman}\par}

\vspace{5mm}

{\sl S.~Eidelman} \\for the CMD-3 and SND Collaborations

\vspace{5mm}

\noindent
Budker Institute of Nuclear Physics SB RAS and Novosibirsk State University, 
Novosibirsk, Russia \\

\vspace{5mm}

Since 2010 a new $e^+e^-$ collider VEPP-2000 has been running in the Budker 
Institute of Nuclear Physics in Novosibirsk based on
a new concept of round beams~\cite{Shatunov:2000zc}.
Its main parameters are shown in Table I in comparison with those of its 
predecessor, $e^+e^-$ collider VEPP-2M.\\

\begin{center}
Table I. Comparison of main parameters of VEPP-2M and VEPP-2000 \\
\begin{tabular}{|l|c|c|c|}
\hline 
Collider & Operation & $\sqrt{s}$, MeV &  
${\cal L}, 10^{30}{\rm cm}^{-2}{\rm s}^{-1}$ \\
\hline
VEPP-2M & 1975-2000 & [360,1400] & 3 \\
\hline
VEPP-2000 & 2010- & [$2m_\pi,2000$] & 100 \\
\hline
\end{tabular}
\end{center}

The maximum luminosity achieved is $2 \cdot 10^{31}$ cm$^{-1}$s$^{-1}$ 
at 1.7-1.8 GeV falling much slower with decreasing energy than before.
In 2013 the center-of-mass (c.m.) energy of 2~$\times$~160 MeV, the smallest 
$\sqrt{s}$ ever, was reached. 
At high energies luminosity is limited by a deficit of positrons
and maximum energy of the booster (825 MeV now).
A long shutdown until 2015 will be used to increase the booster energy 
to 1 GeV and commission the new injection complex to reach 
the designed luminosity. \\ 

Two detectors are installed at VEPP-2000: CMD-3,
a general-purpose magnetic (1.3 T) detector with three electromagnetic 
calorimeters (LXe, CsI, BGO)~\cite{Fedotovich:2006cq} and SND,
a high-resolution  NaI calorimeter with excellent tracking and 
PID~\cite{Achasov:2012nw}. Both detectors 
plan  to measure cross sections
of various exclusive final states with hadrons with high accuracy to
improve knowledge of the hadronic vacuum polarization needed for
the muon anomalous magnetic moment.
An important feature facilitating this ambitious goal is a possibility
of high-precision measurement of the absolute beam energy based on
the Compton laser backscattering~\cite{Abakumova:2013fsa}. 
In 2011-2013 both detectors collected about 60 pb$^{-1}$ each.\\ 


CMD-3 aims at measuring the pion form factor with 0.35\% relative accuracy 
near the $\rho$ meson peak. Particle identification at low energy is
performed using the information from the Drift Chamber allowing separation 
of the $\mu^+\mu^-$ process while at high energy one uses the energy 
deposition in the calorimeters~\cite{Logashenko:2014wxa}. For multihadronic
processes the goal is to reach a 3\% accuracy for the dominating reactions.
Luminosity measurements are currently based on $e^+e^- \to e^+e^-$ and 
$e^+e^- \to \gamma\gamma$ events and provide a
1\% precision~\cite{Akhmetshin:2014lea}. A number of the hadronic final 
states have already been studied by both 
detectors and preliminary results were reported 
in Ref.~\cite{Achasov:2014xsa}. \\

CMD-3 completed the analysis of
the process  $e^+e^- \to 3(\pi^+\pi^-)$, measured the cross section and
performed the first analysis of the process dynamics~\cite{Akhmetshin:2013xc}.
The cross section confirms a dip in the cross section near the threshold
of nucleon-antinucleon production earlier reported 
by BaBar~\cite{Aubert:2006jq}. It is interesting that a similar phenomenon 
is also observed by CMD-3 in the cross section of the process
 $e^+e^- \to 2(\pi^+\pi^-\pi^0)$~\cite{Achasov:2014xsa}. SND used high 
efficiency of photon detection to measure the cross section of the
process $e^+e^- \to \pi^0\pi^0\gamma$~\cite{Achasov:2013btb1} and for the 
first time at the energy above 1.4 GeV  
$e^+e^- \to \eta\gamma$~\cite{Achasov:2013eli}. \\

It is also worth mentioning two very recent results.  
The CMD-3 Collaboration performed a new search for the process  
$e^+e^- \to \eta^\prime(958) \to \eta\pi^+\pi^-,~\eta \to 2\gamma$  
using an integrated luminosity of 2.69 pb$^{-1}$ collected at the c.m. 
energy $\approx m_{\eta^\prime}=957.78\pm0.06 $ 
MeV/$c^2$~\cite{Akhmetshin:2014hxv}. 
From the absence of the signal they obtain
$\Gamma_{\eta^\prime\to e^+e^-}{\cal {B}}_{\eta^\prime\to\pi\pi\eta}
{\cal {B}}_{\eta\to\gamma\gamma}< 0.00041~{\rm eV~at~90\%~C.L.}$
or finally $\Gamma_{\eta^\prime\to e^+e^-}<0.0024~{\rm eV}$,
a factor of 25 more stringent than before, but 
still 300 times higher than the unitarity bound.\\

ND measured the cross section of the process $e^+e^- \to n\bar{n}$
near threshold~\cite{Achasov:2014ncd}, one of the very few measurements 
of this final state ever and with much higher precision than in the old 
measurement of FENICE~\cite{Antonelli:1998fv}. \\

The first results from VEPP-2000 are very promising and show that
already with the existing statistics the precision comparable or 
better than that of BaBar using ISR can be achieved.

\newpage

\subsection{Investigation of the charge symmetry breaking reaction ${dd \rightarrow ^{4}}$He$\pi^0$ with the WASA-at-COSY experiment}
\addtocontents{toc}{\hspace{2cm}{\sl M.~\.Zurek}\par}

\vspace{5mm}

{\sl M.~\.Zurek}\\for the WASA-at-COSY Collaboration

\vspace{5mm}

\noindent
Institut f\"ur Kernphysik, Forschungszentrum J\"ulich GmbH, Germany\\
Institut f\"ur Kernphysik, Universit\"at zu K\"oln, Germany

\vspace{5mm}

Isospin symmetry is one of the fundamental symmetries of Quantum Chromodynamics.
In the Standard Model, isospin symmetry is broken because of the electromagnetic interactions and the mass difference of the lightest quarks \cite{Gasser, Weinberg}. Studies of reactions in which these two isospin symmetry violating sources can be disentangled allow access to quark mass ratios \cite{Gasser, Miller, Leutwyler}. On the hadronic level, the isospin breaking observables are dominated by the pion mass difference, which is an almost purely electromagnetic effect. Therefore, in general, it is difficult to get access to the quark mass difference. 

For a special case of isospin symmetry breaking, namely charge symmetry breaking (CSB), the pion mass difference term does not contribute. Charge symmetry is a rotation of $180^{\circ}$ in isospin space, interchanging up and down quarks. First successful measurements of CSB observables were the forward-backward asymmetry in the $np \rightarrow d\pi^0$ reaction \cite{Opper} and the total cross section of the $\sign$ reaction close to threshold \cite{Stephenson}. Theoretical studies, based on Chiral Perturbation Theory, showed the importance of a consistent analysis of CSB in both reactions \cite{Miller06, Filin, Hanhart, Gardestig, Nogga, Lahde, Fonseca}. Especially, a measurement of higher partial waves in $dd -> ^{4}He\pi^0$ can provide a non-trivial test of our understanding of isospin symmetry breaking \cite{Nogga}. In order to provide the necessary experimental input a corresponding program was initiated using the WASA-at-COSY setup \cite{H.H.Adam, COSY}. 

Theoretical control over the initial state interactions in $\sign$ is one of the main challenges in the calculations of this reaction. Thus, high accuracy wave functions for $dd\rightarrow 4N$ at relatively high energy and in low partial waves are needed. The CSB program started with a measurement of the charge symmetry conserving reaction $\bg$ at $p_d = 1.2$~GeV/$c$, which shares some of the partial waves in the initial state with the signal reaction. To describe the obtained results, a two-fold model ansatz was used. The data were compared to a quasi-free reaction model $dd \to ^3 He \pi^0 + n_{\mathrm{spec}}$ and a partial-wave expansion for the three-body reaction limited to at most one $p$-wave in the system, both added incoherently \cite{3He}.

In a next step, first results for the $\sign$ reaction at a beam momentum of $p_d = 1.2$~GeV/$c$ were obtained. The extracted total cross section amounts to $\sigma_{\mathrm{tot}}=(118\pm18_{\mathrm{stat}}\pm13_{\mathrm{sys}}\pm8_{\mathrm{ext}})$ pb. While the differential cross section is consistent with $s$-wave pion production, due to the low statistics no stringent limit on higher partial waves could be extracted \cite{4He}. Thus, a new measurement aiming at higher statistics and increased sensitivity has been performed. The eight-week long experimental run took place at the beginning of 2014. A modified WASA detector setup was used, introducing the possibility to measure time-of-flight, in order to improve the $^{3}$He-$^{4}$He separation and the kinetic energy reconstruction. For this, most of the detector layers in the forward detector were removed, resulting in a free flight path of about 1.5 m. Data analysis is currently under way.

\newpage

\subsection{ Review on dibaryon resonances }
\addtocontents{toc}{\hspace{2cm}{\sl M.~Bashkanov}\par}

\vspace{5mm}

{\sl M.~Bashkanov}

\vspace{5mm}

\noindent
Physikalisches Institut, Eberhard--Karls--Universit\"at 
 T\"ubingen, Auf der Morgenstelle~14, 72076 T\"ubingen, Germany\\

\vspace{5mm}

Despite their long painful history~\cite{Seth1,Seth2} dibaryon
searches (where dibaryon means a baryon number $B = 2$ state independently on the internal structure (genuine six-quark state/baryonic-molecule)
have recently received new interest, in particular by the recognition
that there are more complex quark configurations than just the
familiar $q \bar q$ and $qqq$ systems. The "hidden color" aspect makes dibaryons
a particularly interesting object in QCD~\cite{BBC}

A resonance like structure recently observed in double-pionic fusion
to deuteron~\cite{mb,MB,MBC}, at $M = 2.38 GeV$ with $\Gamma = 70
MeV$ and $I(J^{p}) = 0(3^{+})$ meanwhile proved to be the so-called
inevitable dibaryon~\cite {Gold} $d^{*}(2380)$.

 To investigate its structure we have measured
its decay branches into the $d\pi^0 \pi^0$~\cite{mb, MB, MBC},
$d\pi^+\pi^-$~\cite{MBC}, $pp\pi^-\pi^0$~\cite{TS1}, $pn\pi^0\pi^0$
~\cite{TS2} and $pn$~\cite{MBE1,MBE2} channels by
$pd$ and $dp$ collisions in the quasi-free reaction mode, utilizing
the WASA detector setup at CELSIUS and COSY. Further information on
$pn\pi^+\pi^-$ decay branch is expected to come from HADES experiment
in near future.

The $pn$ decay channel was measured by use of polarized deuterons in
inverse kinematics~\cite{MBE1,MBE2}. These new $np$ analyzing power data
exhibit a pronounced resonance effect in their energy dependence. The
SAID partial-wave analysis with inclusion of these data reveals a pole
in the complex plane of the coupled $^3D_3 - ^3G_3$ partial waves at
$(2380\pm 10)MeV -i(40\pm 5)MeV$ in accordance with the $d^*$ resonance
hypothesis~\cite{MBE1,MBE2}. An effect of the resonance in the
$^3G_3$ partial wave might point out to a non-vanishing $D$-wave $\Delta\Delta$
component of the $d^*(2380)$ dibaryon recently predicted by Huang et
al~\cite{Huang}.

Since in the double-pionic fusion reactions to $^3$He~\cite{EP} and
$^4$He~\cite{AP} the signature of this resonance is observed too, it obviously is robust enough to survive even in a nuclear surrounding, which may have
interesting consequences for nuclear matter under extreme
conditions. It has been shown that $d^*$ resonance can explain some
dilepton yield~\cite{BC} in heavy-ion collisions ("DLS Puzzle"
~\cite{DLS1,DLS2}). Dibaryons are bosons, hence not Pauli-blocked and as such allow for higher densities of compressed
nuclear matter. The effect of dibaryons on the equation of state for nuclear matter has been
considered in various theoretical investigations, see
e.g. Refs. ~\cite{NM1,NM2,NM3,NM4}.

Various theoretical calculations on $d^*$ internal structure can be
found in Refs. ~\cite{AG1,AG2,Huang,FW}.

Further investigations on the internal structure of the $d^*$ dibaryon,
the SU(3) multiplet companions as well as the mirror partners are
expected to be done in near future by COSY, MAINZ, JLab and J-PARC facilities.

\newpage

\subsection{Baryon resonances in $\pi-p$ and $p-p$ collisions}
\addtocontents{toc}{\hspace{2cm}{\sl  A.~Sarantsev}\par}

\vspace{5mm}
{\sl A.~Sarantsev}

\vspace{5mm}

\noindent
Helmholtz-Institut f\"ur Strahlen- und Kernphysik,
Universit\"at Bonn, Germany\\
Petersburg Nuclear Physics Institute, Gatchina, Russia\\

\vspace{5mm}

The spectrum of the baryon states predicted by the quark model
\cite{Capstick:bm} is much reacher than that observed so far
experimentally. Moreover, a set of new models which reduce the
number of baryon excitations (see e.g.\cite{Forkel:2008un}) still
predict a number of "missing" experimentally states in the mass
region below 2 GeV.

Until recent time our knowledge of the baryon spectrum was based on
the partial wave analyses of the elastic $\pi N$ data performed
about 30 years ago \cite{Hohler:1979yr,Cutkosky:1980rh}.
However the most recent analysis of these data \cite{Arndt:2006bf}
which includes the new polarization information did not confirm a
set of the observed earlier states including a number of "well
established resonances".

However, states with a small branching ratio to the $\pi N$ channel
can escape the identification it in the analysis of the elastic
data. Therefore such states should be studied in inelastic $\pi N$
reactions or in reactions with another (than $\pi N$) initial and
final states. One of the source of such data is meson
photoproduction reactions, another one is the data on the
nucleon-nucleon collisions with production of kaons and hyperons in
the final state.

One of a clear examples of the state which can escape the
identification in the elastic data but is well seen in the inelastic
channel is $P_{11}(1710)$. The analysis of the $\pi p\to K\Lambda$
reaction shows that below 2 GeV the reaction cross section is
dominated by the production of the three partial waves, $S_{11}$,
$P_{11}$ and $P_{13}$. The $P_{11}$ wave shows a clear resonant
structure in the mass region around 1700 MeV (see left-side panel of
Fig.~\ref{inelastic}). The $P_{13}$ partial wave shows a relatively
broad resonant structure at 1900 MeV which is fully compatible with
the $P_{13}(1900)$ states found in the analysis of the $\gamma p\to
K\Lambda$ data. The $S_{11}$ partial wave also reveals a relatively
narrow structure in the region around 1900 MeV. The $S_{11}(1895)$
resonance also was firstly observed in the analysis of the
photoproduction data were it improved the description of the $\gamma
p\to K\Lambda$ reaction.

\begin{figure}[h]
\centerline{\epsfig{file=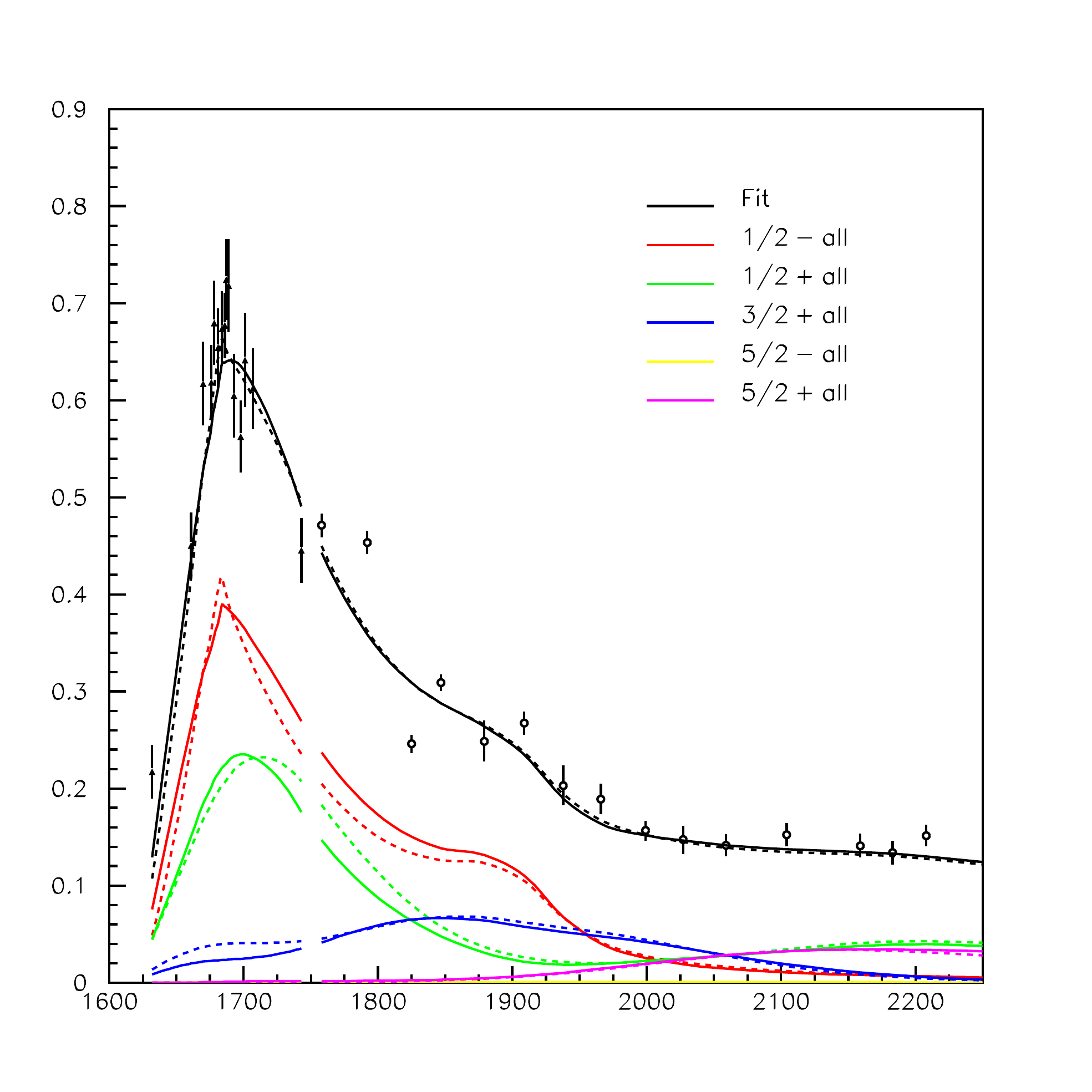,width=0.35\textwidth,height=0.35\textwidth}
\epsfig{file=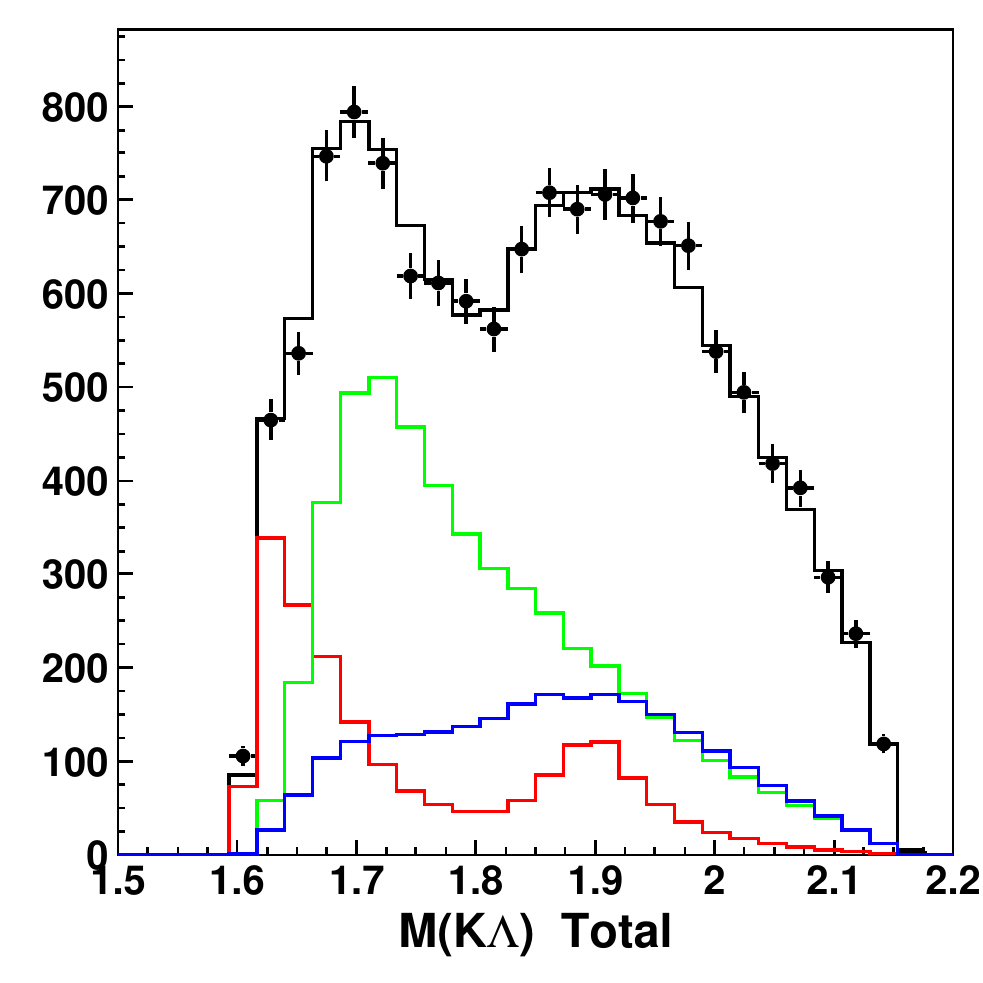,width=0.35\textwidth,,height=0.34\textwidth}}
\centerline{\epsfig{file=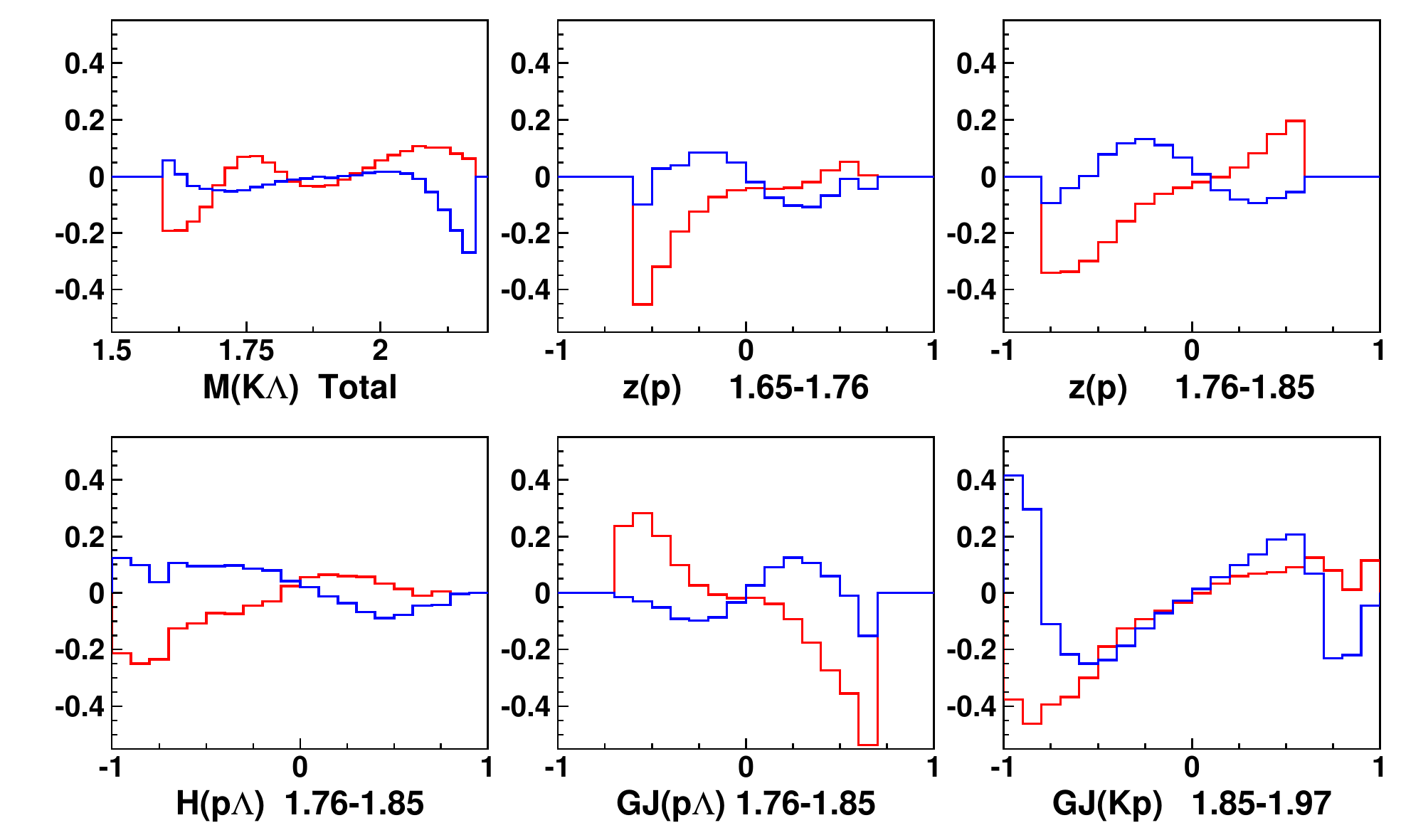,width=0.67\textwidth,
height=0.36\textwidth}}
 \caption{Contribution of the $J^P=\frac12^-$
(red), $\frac 12^+$ (green) and $\frac 32^+$ (blue) partial waves to
the $\pi^-p\to K^-\Lambda p$  cross section (left-side panel, mb)
and to the $K\Lambda$ mass projection from the $pp\to pK\Lambda$
HADES data (middle panel). The prediction of the recoil asymmetry
for the $pp\to K\Lambda p$ reaction for 2 solution (right-side
panel. The best solution is shown with red curves and the solution
without $P_{11}(1710)$ with blue curves}
\label{inelastic}
\end{figure}

The new HADES data on the $pp\to pK\Lambda$ reaction \cite{epple}
can provide a very strong confirmation for the existence of these
states. One of the best solutions found in the event-by-event
likelihood analysis shows a very similar structure in the partial
wave decomposition of these data. The $K\Lambda$ mass projection
(see middle panel of Fig.\ref{inelastic}) shows a clear signals from
the $P_{11}(1710)$, $P{13}(1900)$ and $S_{11}(1895)$ states. The
comparison of the resonance masses and widths extracted from the
combined analysis of the $\pi N$ and photoproduction data
\cite{Anisovich:2011fc} with parameters extracted from the fit of
the $pp\to pK\Lambda$ data alone is given in Table~\ref{resonances}.
It is seen that the parameters obtained in the analysis of different
reactions agree with a very good accuracy.

\begin{table}[ht]
\begin{center}
\renewcommand{\arraystretch}{1.0}
\begin{tabular}{|l|cc|cc|cc|}
\hline
 &\multicolumn{2}{|c|}{$P_{11}(1710)$} & \multicolumn{2}{|c|}{$S_{11}(1895)$}
 &\multicolumn{2}{|c|}{$P_{13}(1900)$}\\
 \hline
~ & M (MeV) &$\Gamma$ (MeV) & M (MeV) &$\Gamma$ (MeV) & M (MeV) &$\Gamma$ (MeV) \\
\hline
$\pi N+\gamma N$   & $1692\pm 9$  & $170\pm 20$ & $1907\pm 15$ &
$100^{+40}_{-15}$ & $1910\pm 30$ &$280\pm 50$ \\
$pp\to K^+\Lambda p$& $1690\pm 10$ & $168\pm 27$ & $1891\pm 7$ &
$84\pm 22$ &  $1906\pm 19$ & $290\pm 55$ \\
\hline
\end{tabular}
\renewcommand{\arraystretch}{1.0}
\end{center}
\vspace{-0.2cm} \caption{\label{resonances} Comparison of the
resonance parameters extracted from the combined analysis of the
$\pi N$ and photoproduction data with parameters extracted from the
analysis of the $pp\to p K\Lambda$ reaction.}
\end{table}

However the analysis of the $pp\to pK\Lambda$ data does not provide
a unique solution due to absence of the polarization information. We
also found a solution without the contribution from the
$P{11}(1710)$ state and a solution where the $P_{13}(1900)$ state
was replaced by a $D_{13}$ resonance. Although the first solution
has a problem in the description of the peak in the $K\Lambda$ mass
projection and the solution with $D_{13}$ state is rather unlikely:
we did not observe notable contributions from the $D_{13}$ partial
wave in the $\pi N\to K\Lambda$ and $\gamma N\to K\Lambda$
reactions, such solutions can not be ruled out from the analysis of
the $pp$ data alone. An extraction of the polarization information
from $\Lambda$ hyperon decay already would provide a strong tool for
reducing of the number of existing solutions. The prediction of the
recoil asymmetry from two solutions which have a compatible
description of the unpolarized angular distributions is shown in
Fig.~\ref{inelastic}. It is seen that measurement of these
observables would provide a significant impact for obtaining a
unique partial wave decomposition of the data.

\newpage

\subsection{HADES results in elementary reactions}
\addtocontents{toc}{\hspace{2cm}{\sl  B.~Ramstein}\par}

\vspace{5mm}
{\sl B.~Ramstein}  \\for the HADES Collaboration
\vspace{5mm}

\noindent
Institut de Physique Nucl\'{e}aire (UMR 8608), CNRS/IN2P3 - Universit\'{e} Paris Sud, F-91406~Orsay Cedex, France \\

\vspace{5mm}

The main goal of the High Acceptance Di-Electron experiment (HADES) \cite{Agakichiev09_techn} at GSI is the study of hadronic matter in the 1-3.5 GeV/nucleon incident energy range. The present interpretation of dilepton spectra measured in heavy-ion reactions at various energies is based on hadronic models which predict in-medium modifications of the $\rho$ meson spectral function due to its coupling to resonance-hole states \cite{Rapp99}.  In the energy range of the HADES experiments,  the $\rho$ meson is mainly produced in primary NN  or secondary $\pi$N collisions which opens the possibility to constrain the interpretation of medium effects by measuring dielectron emission in elementary reactions and better understand the relation between the couplings of the baryonic resonances to the $\rho$ meson and  the electromagnetic structure of the corresponding baryonic transitions.\par
 A first measurement of the \Del\ Dalitz decay could be achieved in the exclusive pp$\rightarrow$pp\epem\ channel at 1.25 GeV and was found consistent with the QED calculation, using constant electromagnetic form factors. 
 The exclusive one pion  (pp$\to$pp$\piz$ and pp$\to$pn$\pip$ ) and dielectron  (pp$\to$pp\epem ) production channels  were combined to provide an interpretation of the measurements in the pp reaction at 3.5 GeV including various baryonic resonances \cite{Agakishiev12_pp35,Agakishiev14_pp35_exclusive}. The effect of  the coupling of the $\rho$ meson to light baryonic resonances (N(1520, N(1535),...), was clearly observed.  It  induces a distortion of the $\rho$ meson spectral function in NN collision which  puts very strong constraint on the interpretation of medium effects \cite{Agakishiev12_pNb}.\par
  Although a definite explanation of the unexpectedly high dielectron yield measured by HADES in quasi-free pn reactions \cite{Agakishiev10_elem} does not exist yet, several rather successful ideas were presented \cite{Shyam10,Martemyanov11}. The most recent one \cite{Bashkanov14} is based on  virtual $\rho$ production in double \Del\ final state interaction. In order to constrain this process, the  double pion production is analyzed by HADES in different channels of the pp and np reactions. In addition, these analyses provide  independent checks on the existence of the d$^{\star}$ resonance (M=2.38 GeV/c$^2$) which seemed to be observed by the WASA collaboration \cite{Bashkanov14_bis}.\par
Moreover, HADES was able to  improve the upper limit on the kinetic mixing parameter of the dark photon in a range of mass from 0.02 to 0.1 GeV/c$^2$ \cite{Agakishiev14_darkphoton}. The  upper limits on the $\eta\to$\epem\ branching ratio could also be lowered twice in the last years by the HADES collaboration \cite{Agakishiev12_pp35,Agakishiev14_darkphoton}.\par
Very recently, the HADES collaboration took data using the GSI pion beam. Such a beam, covering the energy range of the second and third resonance region, is presently unique in the world and offers the possibility to improve the very scarce data base for pion-nucleon reactions.  The perspectives from this first experiment, which was focused  to the N(1520) region,  are very promising. The measurement of \epem\ production will bring completely new information on the Dalitz decay branching ratio of this resonance and will allow to check the contradictory predictions for \epem\ production at high invariant masses. The measurement of the pion production in an energy scan around the N(1520) resonance will allow for a better determination of the branching ratios of this resonance, with a special interest for  the  $\rho$N decay, since it has a direct impact  on the in-medium distortions of the $\rho$ meson spectral function, as discussed earlier. Pion beam experiments will be continued in GSI in the next years, then the HADES experimental program will be pursued using the proton and ion beams at FAIR.\\
\vskip 2 mm
\noindent
The HADES collaboration gratefully acknowledges the support by the grants:\\
PTDC/FIS/113339/2009 LIP Coimbra, NCN grant 2013/10/M/ST2/00042 SIP JUC Cracow, Helmholtz Alliance HA216/EMMI GSI Darmstadt, VH-NG-823, Helmholtz Alliance HA216/EMMI TU Darmstadt, 283286, 05P12CRGHE HZDR Dresden, Helmholtz Alliance HA216/EMMI, HIC for FAIR (LOEWE), GSI F$\&$E Goethe-University, Frankfurt VH-NG-330, BMBF 06MT7180 TU München, Garching BMBF:05P12RGGHM JLU Giessen, Giessen UCY/3411-23100, University Cyprus CNRS/IN2P3, IPN Orsay, Orsay MSMT LG 12007, AS CR M100481202, GACR 13-06759S NPI AS CR, Rez, EU Contract No. HP3-283286.

%


\newpage

\subsection{Luminosity determination via $dp$-elastic scattering at ANKE}
\addtocontents{toc}{\hspace{2cm}{\sl  C.~Fritzsch}\par}

\vspace{5mm}

{\sl C.~Fritzsch}\\for the ANKE Collaboration

\vspace{5mm}

\noindent
Institut f\"ur Kernphysik, Westf\"alische Wilhelms-Universit\"at M\"unster,
Germany\\

\vspace{5mm}

Studies on the total cross sections of the reaction $d+p
\rightarrow{}^3\text{He}+\eta$ are of special interest since they differ
strongly from a pure phase space behaviour near threshold \cite{Mers2007a,
Smyrski2007, Berger1988, Mayer1996, Adam2007}. This behaviour could be an
indication for a quasi bound state of the $\eta^3\text{He}$-system \cite{Wilken2007}. 
New high precision data from the ANKE experiment \cite{ANKE} at
the accelerator ring COSY at the Forschungszentrum J\"ulich allow the
extraction of precise absolute cross section values for the $\eta$ production up to an excess energy
of $Q=$~15~MeV. Therefore, a careful luminosity determination was realized via 
$dp$-elastic scattering ($d+p \rightarrow d+p$) for 18 beam
momenta in a range between $3120.17~\nicefrac{\text{MeV}}{c} \le p_d \le
3204.16~\nicefrac{\text{MeV}}{c}$.\\\\ 
The $dp$-elastic scattering is very well suited as normalization reaction. 
Its broad data base of available reference data and their high differential 
cross sections in the region of interest ensure an excellent signal-to-background ratio.\\
Most of the deuterons of the $dp$-elastic scattering cause only a low momentum
transfer on the target proton. Consequently, these deuterons have a momentum
close to the beam momentum, which means that the D2 magnet of ANKE will deflect them under small
laboratory scattering angles towards the forward detection system.
This detection system consists of a multiwire drift chamber and two multiwire proportional 
chambers used for track reconstruction and two layers of scintillation
hodoscopes for particle identification. The identification of this reaction is
ensued via the missing mass technique (see figure \ref{fig:lum} left).

\begin{figure}[h]
\centering
\includegraphics[width=0.5\textwidth]{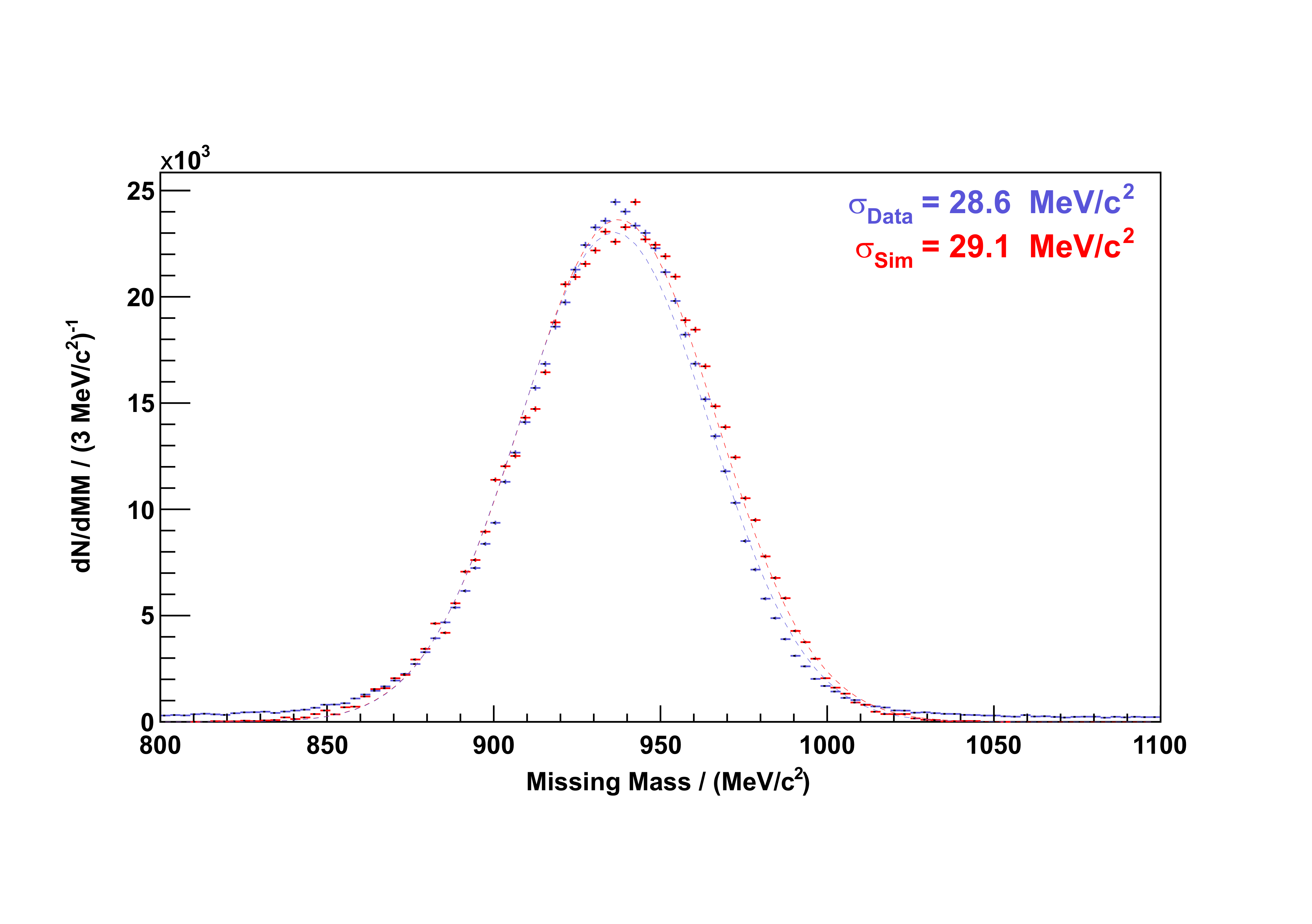}
\includegraphics[width=0.48\textwidth]{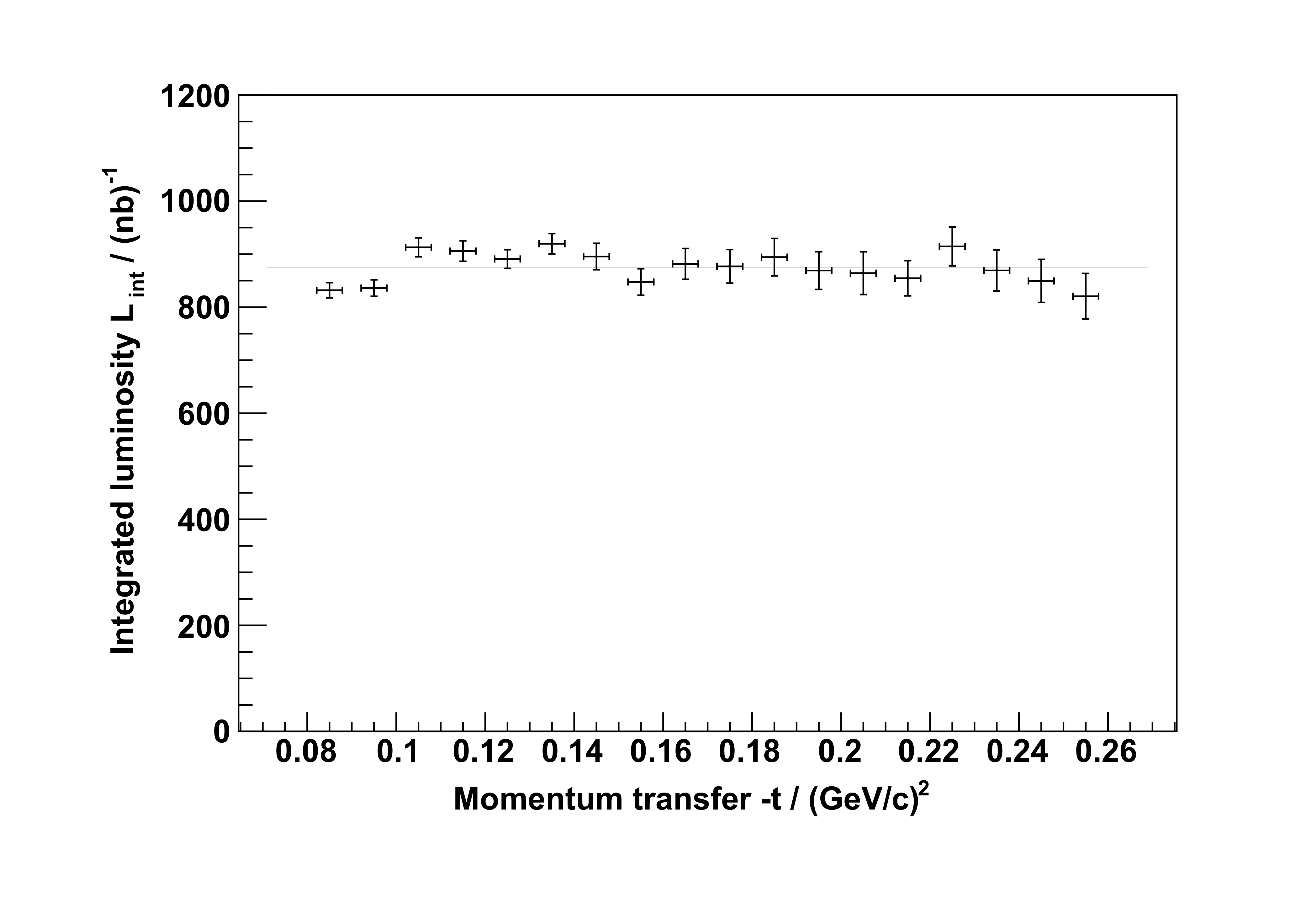}
 \caption{{\bf Left:} Missing-mass distribution of the identified deuteron (blue). The corresponding Monte Carlo simulation is the red distribution. {\bf Right:} Integrated luminosity for 18 momentum transfer bins for a beam momentum of $p_d = 3150.42~\nicefrac{\text{MeV}}{c}$.}
\label{fig:lum}
\end{figure}

The luminosity has to be independent of the momentum transfer (see figure
\ref{fig:lum} right) and for systematic tests the determination has been
performed for 18 momentum transfer bins for each of the 18 beam momenta. By this
luminosities could be extracted with high
\begin{wrapfigure}[22]{r}{7.5cm}
		\includegraphics[width=0.4\textwidth ,trim = -1.5cm 0.15cm 0mm
		0cm]{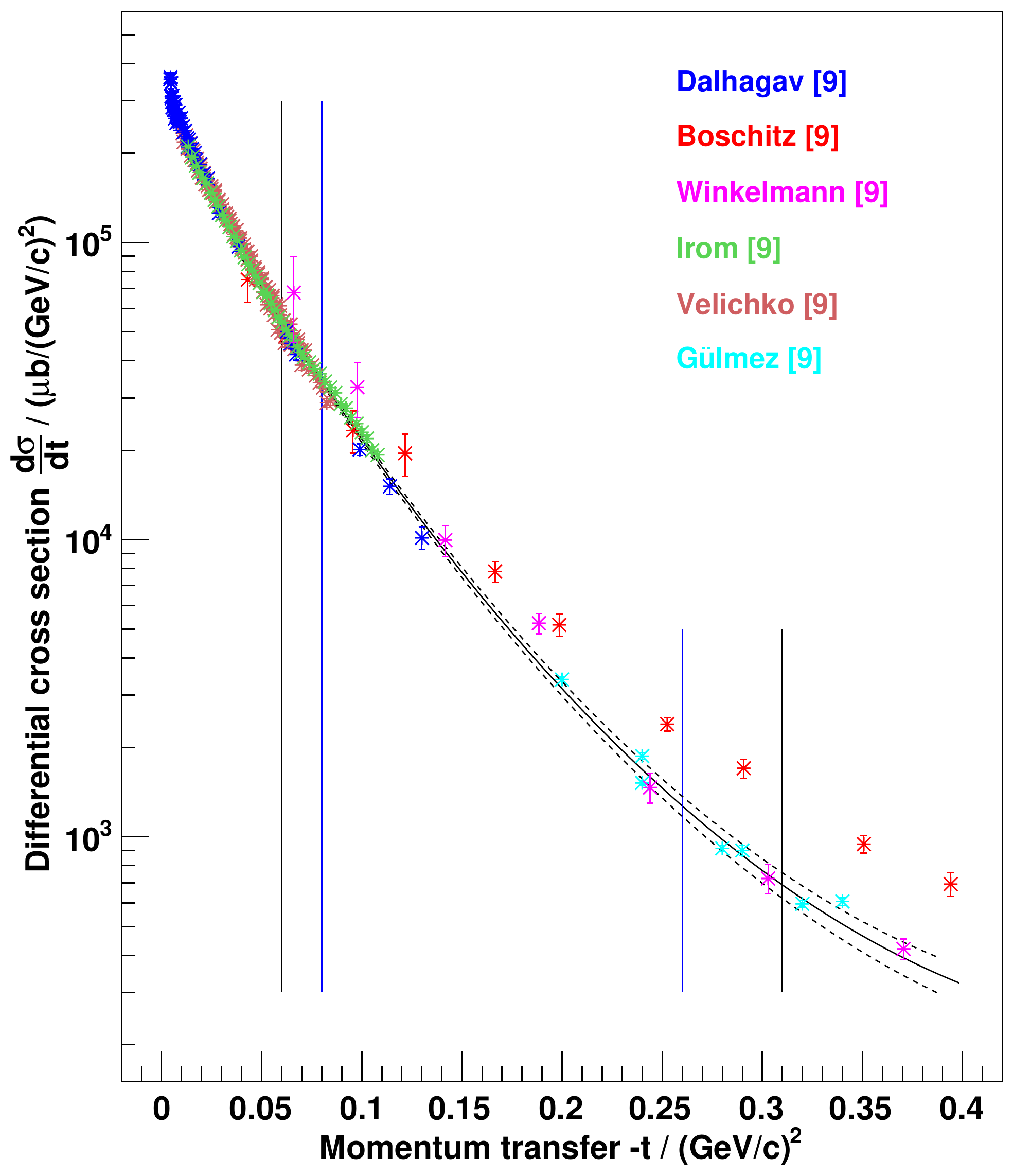} \caption{Reference cross sections
		for $dp$-elastic scattering as a function of momentum transfer $-t$. The black vertical lines tag the range of ANKE
	acceptance and the blue vertical lines the range which was used for the
	luminosity determination.}
	\label{fig:wq}
\end{wrapfigure}
precison ($\Delta_{\text{stat}} = 1\,\%$ and $\Delta_{\text{sys}} = 6\,\%$). Especially
the systematic uncertainties were improved by at least a factor of two compared
to previous determinations. These luminosities were already used to
determine differential and double differential cross sections for the reaction $d + p \rightarrow
{}^3\text{He} + \pi^+ \pi^-$ \cite{Mielke2014}.\\\\
Furthermore, at higher momentum transfers ($\geq 0.12~(\nicefrac{\text{GeV}}{c})^2$) the 
available reference data base shows a limited number of data
points and discrepancies between some measurement sets (see figure \ref{fig:wq}). Due to
the high quality and statistics of the ANKE data set on the $dp$-elastic
scattering in this momentum transfer region, new precision data can be provided.
For this purpose an independent absolute normalization is currently in progress.
First estimations show that a precision of the extracted
differential cross sections of $\Delta_{\text{stat}}$ = $1\,\% - 2\,\%$ and
$\Delta_{\text{sys}}$ = $2\,\% - 3\,\%$ can be achieved.\\\\
This work has been supported by the COSY-FFE program of the
Forschungszentrum J\"ulich.

\newpage

\subsection{ Investigating the $\text{pd}\rightarrow {}^3\text{He}\,\eta$ production cross section between $Q\approx\unit[13.6]{MeV}$ and $Q\approx\unit[80.9]{MeV}$}
\addtocontents{toc}{\hspace{2cm}{\sl N.~H\"usken}\par}

\vspace{5mm}
{\sl N.~H\"usken} \\ for the WASA-at-COSY Collaboration\vspace{5mm}

\noindent
Institut f\"ur Kernphysik, Westf\"alische Wilhelms-Universt\"at M\"unster, Germany\\

\vspace{5mm}

The production cross section of the $\text{pd}\rightarrow {}^3\text{He}\,\eta$ reaction has been studied in great detail in the near threshold region \cite{Mers2007,Smyrski2007a,Berger1988a,Mayer1996a,Adam2007a}, whereas at higher excess energies the amount of available data is limited \cite{Raus2009,Bilger2002,Bilger2004,Betigeri2000}. Moreover, while the data from ANKE and WASA/PROMICE expose a total cross section plateau, recent results from the WASA-at-COSY experiment \cite{Adlarson2014} suggest an unexpected narrow variation of the total cross section at $Q=\unit[48.8]{MeV}$ that is in tension with earlier results, as shown in Figure \ref{fig:totalcrossFB}. 

\begin{figure}[h]
	\centering	
	\includegraphics[width=1.00\textwidth]{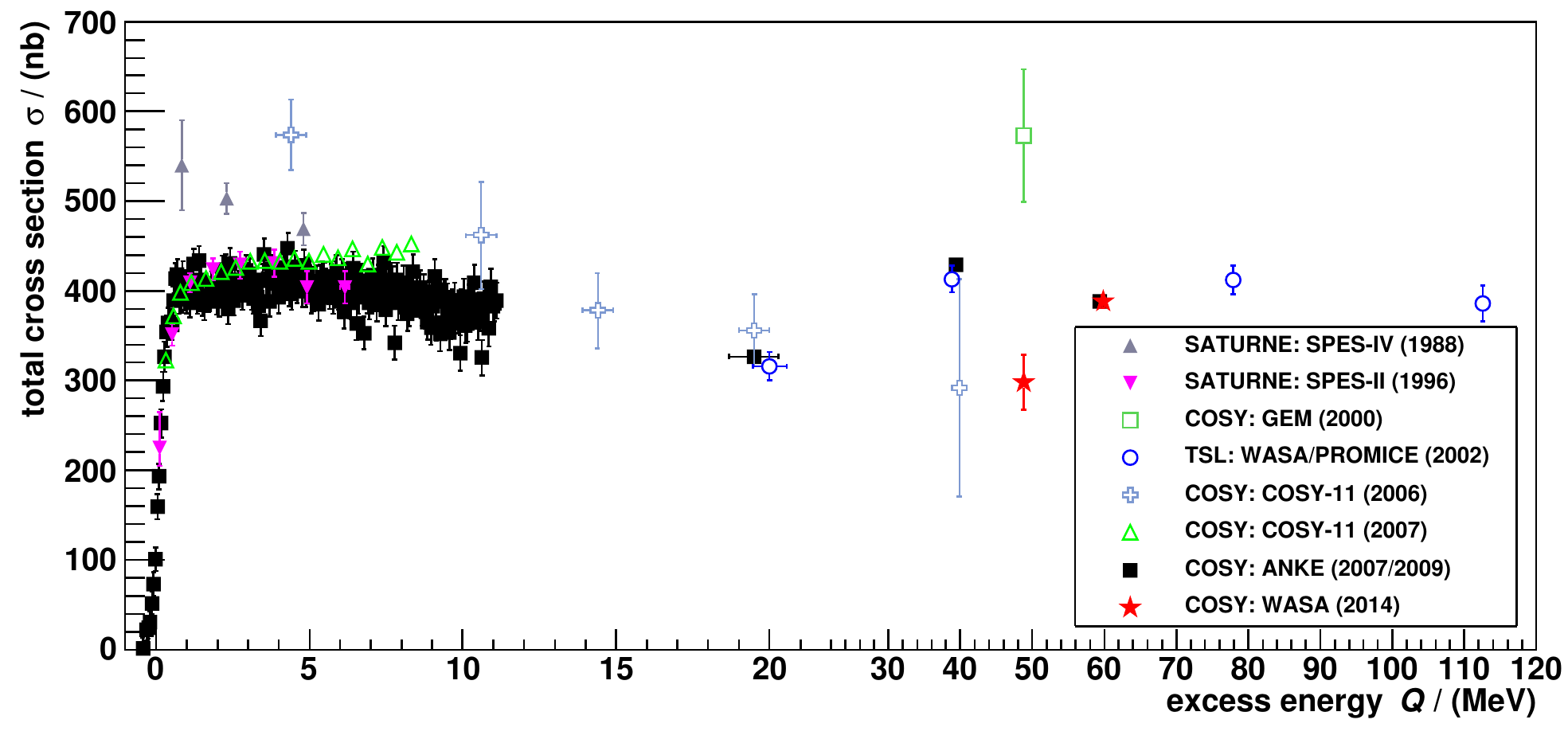}
	\caption{Total cross section data for the reaction $\text{pd}\rightarrow {}^3\text{He}\, \eta$ obtained by \cite{Mers2007,Smyrski2007a,Berger1988a,Mayer1996a,Adam2007a,Raus2009,Bilger2002,Bilger2004,Betigeri2000,Adlarson2014} as a function of the excess energy $Q$. The uncertainties shown do not include systematic uncertainties resulting from an absolute normalization. The data from WASA-at-COSY (red stars) are scaled to the ANKE data point at $Q=\unit[59.4]{MeV}$. Figure taken from \cite{Adlarson2014}.}
	\label{fig:totalcrossFB}
\end{figure}

The WASA-at-COSY experiment is perfectly suited to study the behaviour of the total cross section of the $\text{pd}\rightarrow {}^3\text{He}\, \eta$ reaction. In May 2014 a beam time was realized in order to investigate the excess energy region of interest. The COSY storage ring of the Forschungszentrum J\"ulich was used to scatter protons with beam momenta between $p_p=\unit[1.60]{GeV/c}$ and $p_p=\unit[1.74]{GeV/c}$ on a deuterium pellet target. The measurement covered 15 different beam momenta, resulting in a Q-value range between $Q\approx \unit[13.6]{MeV}$ and $Q\approx \unit[80.9]{MeV}$ with a stepsize of about $\unit[4.8]{MeV}$. 
The identification and reconstruction of light mesons and their decay particles can be realized with the WASA Central Detector, while the WASA Forward Detector is used to measure the full four-momenta of the produced ${}^3\text{He}$ ions. Therefore, both the ${}^3\text{He} \eta$ and the ${}^3\text{He}\pi^0$ final states can be reconstructed, with the latter one used for normalization purposes. \\
The ongoing analyses will result in angular distributions and differential cross sections for the $\text{pd}\rightarrow {}^3\text{He}\eta$ reaction for all 15 excess energies. The high statistics obtained of at least 75000 reconstructed $\eta$ mesons per excess energy yield an estimated point-to-point uncertainty of the order of $\unit[8]{\%}$ (not including an overall normalization uncertainty). In order to extract total cross sections a normalization will be done using the $\text{pd}\rightarrow {}^3\text{He}\pi^0$ reaction. \\
The new data will allow us to investigate possible cross section variations in great detail. In addition, precise total and differential cross section data will be provided, which are of high interest for the understanding of the underlying production processes and the development of new theoretical production models.
\vspace{5mm}

Supported by FFE program of the Forschungszentrum J\"ulich. The research leading to these results has received funding from the European Union Seventh Framework Programe (FP7/2007-2013) under grant agreement n° 283286.

\newpage

\section{List of participants}

\begin{flushleft}
\begin{itemize}
\item Patrik Adlarson, Universt\"at Mainz, {\tt adlarson@kph.uni-mainz.de}
\item Farah Noreen Afzal, University of Bonn, {\tt afzal@hiskp.uni-bonn.de}
\item Mikhail Bashkanov, Universit\"at T\"ubingen, {\tt bashkano@pit.physik.uni-tuebingen.de}
\item Marco Battaglieri, INFN-Genova, {\tt marco.battaglieri@ge.infn.it}
\item Marcin Berlowski, National Center for Nuclear Research, Warsaw, {\tt Marcin.Berlowski@fuw.edu.pl}
\item Johan Bijnens, Lund University, {\tt  bijnens@thep.lu.se}
\item Caterina Bloise, LNF, {\tt Caterina.Bloise@lnf.infn.it}
\item Fabio Bossi, LNF, {\tt Fabio.Bossi@lnf.infn.it}
\item Li Caldeira Balkestahl, Uppsala University, {\tt li.caldeira\_balkestahl@physics.uu.se}
\item Francesca Curciarello, University of Messina, {\tt curciarello@unime.it}
\item Veronica de Leo, University of Messina, {\tt deleo@unime.it}
\item Kay Demmich, University of M\"unster, {\tt demmich@wwu.de}
\item Gernot Eichmann, University of Giessen, {\tt Gernot.Eichmann@theo.physik.uni-giessen.de}
\item Simon Eidelman, Novosibirsk State University, {\tt eidelman@mail.cern.ch }
\item Shuangshi Fang, IHEP Beijing, {\tt fangss@ihep.ac.cn}
\item Christopher Fritzsch, University of M\"unster, {\tt c.fritzsch@wwu.de}
\item Aleksander Gajos,  Jagiellonian University, {\tt  aleksander.gajos@uj.edu.pl}
\item Paolo Gauzzi, Sapienza Universit\'a di Roma e INFN, {\tt paolo.gauzzi@roma1.infn.it}
\item Simona Giovannella, Laboratori Nazionali di Frascati, {\tt  simona.giovannella@lnf.infn.it}
\item Frank Goldenbaum, Forschungszentrum Juelich, {\tt  f.goldenbaum@fz-juelich.de}
\item Sergi Gonzalez-solis, IFAE and U.A. Barcelona, {\tt sgonzalez@ifae.es}
\item Evgueni Goudzovski, University of Birmingham, {\tt  eg@hep.ph.bham.ac.uk}
\item Dieter Grzonka, Forschungszentrum Juelich, {\tt  d.grzonka@fz-juelich.de}
\item Lena Heijkenskj\"old, Uppsala University, {\tt lena.heijkenskjold@physics.uu.se}
\item Martin Hoferichter, University of Bern, {\tt hoferichter@itp.unibe.ch}
\item Tom\'a\v s Husek, Charles University, {\tt Tomas.Husek@cern.ch}
\item Nils H\"usken, University of M\"unster, {\verb|n_hues02@uni-muenster.de|}
\item Tord Johansson, Uppsala University, {\tt tord.johansson@physics.uu.se}
\item Tom\'a\v s Kadav\'y, Charles University, {\tt tomas.kadavy@gmail.com}
\item Daria Kaminsla, Jagiellonian University, {\tt dk.dariakaminska@gmail.com}
\item Karol Kampf, Charles University, {\tt karol.kampf@mff.cuni.cz}
\item Marc Knecht, CNRS/cPT Marseille, {\tt marc.knecht@aeres-evaluation.fr}
\item Mari\'an Koles\'ar, Charles University, {\tt  kolesar@ipnp.troja.mff.cuni.cz}
\item Bernd Krusche, University of Basel, {\tt bernd.krusche@unibas.ch}
\item Wojciech Krzemien, Jagiellonian University, {\tt wojciech.krzemien@if.uj.edu.pl}
\item Bastian Kubis, University of Bonn, {\tt kubis@hiskp.uni-bonn.de}
\item Andrzej Kupsc, Uppsala University, {\tt Andrzej.Kupsc@physics.uu.se }
\item Matteo Mascolo, LNF, {\tt Matteo.Mascolo@lnf.infn.it}
\item Pere Masjuan, Universt\"at Mainz, {\tt masjuan@kph.uni-mainz.de}
\item Michael Papenbrock, Westf\"alische Wilhelms-Universit\"at M\"unster, {\tt  michaelp@uni-muenster.de}
\item Emilie Passemar, Indiana University/JLab, {\tt epassema@indiana.edu}
\item Elena Perez del Rio, LNF, {\tt elena.perezdelrio@lnf.infn.it}
\item Beatrice Ramstein, Institut de Physique Nucleaire Orsay, {\tt amstein@ipno.in2p3.fr}
\item Christoph Florian Redmer, Universit\"at Mainz, {\tt redmer@uni-mainz.de }
\item Johan Relefors, Lund University, {\tt johan.relefors@thep.lu.se}
\item Piotr Salabura, Jagiellonian University,  {\tt piotr.salabura@uj.edu.pl}
\item Pablo Sanchez-Puertas, University of Mainz, {\tt sanchezp@kph.uni-mainz.de}
\item Andrey Sarantsev, PNPI, {\tt andsar@hiskp.uni-bonn.de}
\item Ivano sarra,  Laboratori Nazionali di Frascati, {\tt  ivano.sarra@lnf.infn.it}
\item Diane Schott, The George Washington University, {\tt dschott@email.gwu.edu}
\item Olga Shekhovtsova, IFJ PAN, {\tt Olga.Shekhovtsova@lnf.infn.it}
\item Alexander Somov, Jefferson Lab, {\tt somov@jlab.org}
\item Karsten P. Spieker, HISKP Bonn, {\tt spieker@hiskp.uni-bonn.de}
\item Joanna Stepaniak, National Centre for Nuclear Research, Warsaw, {\tt joanna.stepaniak@fuw.edu.pl}
\item Adam Szczepaniak, Indiana University and Jefferson Lab, {\tt aszczepa@indiana.edu}
\item Sean Tulin, York University, {\tt stulin@yorku.ca}
\item Andrew Wilson. HISKP, Bonn, {\tt awilsonhiskp.uni-bonn.de}
\item Andreas Wirzba, Forschungszentrum Juelich, {\tt a.wirzba@fz-juelich.de}
\item Martin Zdr\'ahal, Charles University, {\tt  martin.zdrahal@mff.cuni.cz}
\item Maria Zurek, Forschungszentrum Juelich, {\tt m.zurek@fz-juelich.de}
\end{itemize}
\end{flushleft}

\end{document}